\documentstyle[aps,prb,twocolumn,epsf]{revtex}
\begin{document}

\tighten
\draft

\title{\centerline{Double-Layer Systems at Zero Magnetic Field}}
\author{C.~B.~Hanna, Dylan Haas\cite{haas}, and J.~C.~D\'{\i}az-V\'{e}lez}
\address{Department of Physics, Boise State University,
Boise, Idaho~~83725}

\date{\today}
\maketitle

\begin{abstract}

\vspace{-0.3in}

We investigate theoretically the effects of
intralayer and interlayer exchange
in biased double-layer electron and hole systems,
in the absence of a magnetic field.
We use a variational Hartree-Fock-like approximation to analyze the
effects of layer separation, layer density, tunneling, and applied gate
voltages on the layer densities and on interlayer phase coherence.
In agreement with earlier work, we find that for very low layer separations
and layer densities, an interlayer-correlated ground state possessing
spontaneous interlayer coherence (SILC) is obtained,
even in the absence of interlayer tunneling.
In contrast to earlier work, we find that as a function of total density,
there exist four, rather than three, distinct noncrystalline phases for
balanced double-layer systems without interlayer tunneling.
The newly identified phase exists for a narrow range of densities
and has three components and slightly unequal layer densities,
with one layer being spin-polarized and the other unpolarized.
An additional two-component phase is also possible in the presence of
sufficiently strong bias or tunneling.
The lowest-density SILC phase is the fully spin- and pseudospin-polarized
``one-component'' phase discussed by Zheng and co-workers
[Phys. Rev. B {\bf 55}, 4506 (1997)].
We argue that this phase will produce a finite interlayer Coulomb drag
at zero temperature due to the SILC.
We calculate the particle densities in each layer
as a function of the gate voltage and total particle density,
and find that interlayer exchange can reduce or prevent abrupt
transfers of charge between the two layers.
We also calculate the effect of interlayer exchange on the
interlayer capacitance.

\end{abstract}

\vspace{-0.1in}
\section{Introduction}
\label{sec:intro}

In the last several years, double-layer electron and hole systems
have provided an exceptionally useful tool for investigating
the effects of interparticle Coulomb interactions in two dimensions,
particularly at low particle densities where exchange and correlation effects
are significant.
This has been especially true in the quantum Hall regime,\cite{prange,daspin}
where the combination of a strong perpendicular magnetic field
(which quenches the kinetic energy)
and very small layer separation
(which enhances interlayer exchange)
stabilizes remarkable interlayer-coherent quantum Hall
states.\cite{dlexpt1,dlexpt2,murphy,yang,moon,bigyang,gmchap}
Even in the absence of any magnetic field, 
interlayer capacitance and drag measurements in
double-layer electron systems (2LES's) have provided quantitative measures
of the effects of electronic interactions on both the
thermodynamics\cite{jpe}
and transport (Coulomb drag\cite{drag})
of two-dimensional electron and hole systems.
Unless otherwise specified, we shall take the notation 2LES
to also include double-layer hole systems.

Our work is motivated by the rapid pace of advancements
in the engineering of double-layer semiconductor devices.
We expect that high-mobility double-layer devices will eventually be built
with both
(1) {\em separately contactable} layers, and
(2) layer separations and carrier densities so small that
the {\em interlayer} correlations between the carriers are substantial,
perhaps even without the aid of a strong quantizing magnetic field.
Such devices will allow direct measurements of the effects
interlayer many-body effects in double-layer systems.
As a starting point to analyze the zero magnetic field situation,
we have developed a simple mean-field model that incorporates both
intralayer and interlayer exchange in biased double-layer electron
and hole systems in the absence of a magnetic field.
We use the model to calculate theoretically the effect of interlayer
exchange on the layer densities and interlayer capacitance
as a function of layer spacing, particle density, and applied gate voltage.

Gated double-layer systems have the great advantage of allowing the
layer densities of electrons or holes to be varied by the application
of a bias (gate voltage).
At high densities, the kinetic energy per unit
area dominates the exchange and correlation energies,
and in a translationally invariant system,
the symmetry and properties of the ground state
of a many-body system are in one-to-one correspondence
with those of a free-electron gas.
At lower electron densities,
the Coulombic exchange and correlation energies can produce
qualitative changes in the nature of the many-particle ground state:
numerical work on two- and three-dimensional electron gases show that
a spin ferromagnetic state is obtained at low densities,
which is eventually supplanted by a Wigner crystal state
at the lowest densities.
The low-density ferromagnetic state of the interacting electron gas
was anticipated some 70 years ago by Bloch;\cite{bloch}
ferromagnetism can be found even within the Hartree-Fock approximation
when the exchange interaction energy, which favors occupation of
single-particle states of the same spin, dominates the kinetic energy,
which favors reducing the Fermi energy by equal occupation of both
spin states.

Multilayer semiconductor devices enhance the effects of
interparticle interactions through the combination of
reduced dimensionality and low particle density, and by
the presence of an additional electronic degree of freedom,
the layer index.\cite{ahmmulti}
In double-layer systems,
layer occupancy can be specified by a introducing a pseudospin variable
that points up for one layer and down for the other
layer.\cite{pseudo}
Extending the notion of an exchange-driven ferromagnetic transition
to double-layer systems suggests that at low enough densities,
the electronic ground state should be both spin and pseudospin polarized.
Such reasoning, supported by Hartree-Fock calculations, led
Ruden and Wu to propose that, at low enough electron densities and
for small enough layer separations, the electrons of a 2LES would
minimize their ground-state energy by having all electrons occupy
a single layer.\cite{ruden}
The Ruden-Wu scenario implies that as the total density of a balanced 2LES
is lowered, the electrons should eventually experience an interlayer charge
transfer instability that spontaneously empties out one of the two layers.
Ruden and Wu also suggested that the inherent bistability of the resulting
low-density 2LES (either layer could be the one to lose or gain particles)
would constitute an exchange-driven logic gate.
The possibility of an exchange-driven interlayer charge-transfer instability
has been a subject of both
theoretical\cite{ruden,staging,zheng,conti}
and experimental\cite{kata,ying,millard,papa}
interest.

Although pseudospin polarization at sufficiently low densities is
likely, it does not require unequal layer densities.
This has been demonstrated theoretically in great detail
for closely spaced double-layer systems in a strong magnetic field
at unit filling factor and appears to give a good explanation of
experimental results.\cite{yang,moon,bigyang}
The key point is that electrons in double-layer structures
are not restricted to occupying only one of two layer eigenstates:
quantum mechanics allows states that are superpositions of the two
layers.
The layer pseudospin must therefore be treated as a Heisenberg variable,
as was done in Refs. \onlinecite{yang} and \onlinecite{moon}
in the quantum Hall regime,
and by Zheng and co-workers in zero magnetic field,\cite{zheng} rather than
as an Ising variable (where only ``up'' and ``down'' allowed) as was done
by Ruden and Wu.
For example, when interlayer tunneling is present,
the single-particle eigenstates are symmetric and antisymmetric
combinations of layer states.
A major insight of Refs.~\onlinecite{yang,moon,bigyang}
was the concept of ``spontaneous interlayer coherence'' (SILC):
at sufficiently small layer separations,
electrons can spontaneously create and occupy linear combinations
of layer states in which each layer has the same average number of electrons,
{\em even without any interlayer tunneling}.
The spontaneous formation of superposed layer states
in double-layer quantum Hall (2LQH) sytems
can be accomplished by the interlayer exchange interaction alone.
SILC in balanced 2LQH systems corresponds to $XY$ pseudospin ferromagnetism
in which the pseudospins spontaneously magnetize,
but do not point either up or down, since neither layer has
(on average) more particles than the other.\cite{yang}

The application of SILC to the zero magnetic field case was first made
by Zheng and co-workers,\cite{zheng} who considered the same model system as
Ruden and Wu\cite{ruden} --
electrostatically balanced zero-thickness layers
of interacting electrons without interlayer tunneling --
but came to a very different conclusion.
They proved within the Hartree-Fock approximation (HFA) that
(1) the 2LES becomes spin ferromagnetic before it becomes
pseudospin ferromagnetic for any finite layer separation
(this possibility was not considered by Ruden and Wu), and
(2) at low enough densities and small enough layer separations,
the pseudospin ferromagnetic state possesses SILC,
with all electrons occupying one subband composed
of a superposition of layer states with equal average density in each layer.
Conti and Senatore have also
argued that for electrostatically balanced layers,
the 2LES ground state cannot be one in which all electrons
are eigenstates of the same single layer\cite{conti}.
SILC at sufficiently small layer separations
should also follow from earlier calculations presented in
Ref.~\onlinecite{swier},
although no explicit inference of SILC was made in that work.
Recent work that goes beyond the HFA and includes correlation effects
within the STLS approximation\cite{stls} also finds that SILC
is favored over single-layer occupancy for balanced layers that are
sufficiently close together.\cite{lerwen}

When studying a 2LES using a density-functional approach,
it is important to note that SILC is a nonlocal effect.
Calculations based on local-density approximations
will not find SILC if they treat the pseudospin
as an Ising-like variable.\cite{zheng}
The same caveat applies to the work of Ruden and Wu,
who used a restricted HFA that excluded the possibility of SILC.
Once the intralayer separation between electrons becomes comparable
to  the interlayer separation between layers, the possibility of
interlayer correlations (such as SILC) must be considered.
The spontaneous charge-transfer state predicted by Ruden and Wu
follows quite generally (even beyond the HFA) from the fact that
when interlayer correlations are ignored,
the negative compressibility of the electron gas guarantees
an interlayer charge-transfer instability at sufficiently small
layer spacing.
But it is precisely at small layer spacings that interlayer correlations
become important, and so their effects must be included to obtain
physically meaningful results.

We have extended the study the effects of Coulombic exchange
in double-layer electron and hole systems
to include an applied bias due to front and back gate voltages,
while allowing for the possibility interlayer exchange.
In the balanced case, we have found that the four- to two-component
transition is always interrupted by the presence of a three-component
phase with slightly unequal layer densities.
There are therefore four possible noncrystalline phases for a 2LES
with balanced gates.
Under bias or tunneling a second (pseudospin-polarized) two-component
state is also possible.
We have enumerated the transitions between the five allowed states
in the presence of bias,
and explored the effects of bias on the one-component state.

The rest of this paper is organized as follows:
In Sec.~\ref{sec:model},
we introduce a simplified model for double-layer systems,
review the concept of interlayer capacitance,
and give a general criterion for stability against
spontaneous interlayer charge transfer.
In Sec.~\ref{sec:mfa}, we develop a mean-field approximation for biased
double-layer systems that allows for the possibility of
interlayer coherence.
In Sec.~\ref{sec:balance},
we examine the special case of electrostatically balanced layers,
enumerate the resulting four possible noncrystalline phases,
and explore the onset of interlayer coherence and its effect
on the size of the subband splitting.
We also obtain a phase diagram for the balanced case, and perform
an alternate calculation for the onset of the one-component phase.
In Sec.~\ref{sec:bias}, we explore the effect of bias
on the layer occupancies and interlayer capacitance
for large, intermediate, and small layer separations.
We develop simple models capable of closely fitting experimental
layer-occupancy data, and expore the transitions
between different phases induced by layer imbalance.
In Sec.~\ref{sec:onecomp}, we analyze the onset and properties
of the one-component state under bias.
We summarize our findings and speculate on the possible relevance
of these results to the strong magnetic-field regime
in Sec.~\ref{sec:conclude}.

\vspace{-0.1in}
\section{Double-Layer Model and Interlayer Capacitance}
\label{sec:model}
\vspace{-0.1in}

In this section, we introduce an idealized model of double-layer
systems.
We review the condition for thermodynamic equilibrium between the
inner layers, obtain a necessary condition for stability against
interlayer charge transfer, and review an experimentally useful measure
of the interlayer capacitance, the Eisenstein ratio.\cite{jpe}

\begin{figure}[h]
\epsfxsize3.5in
\centerline{\epsffile{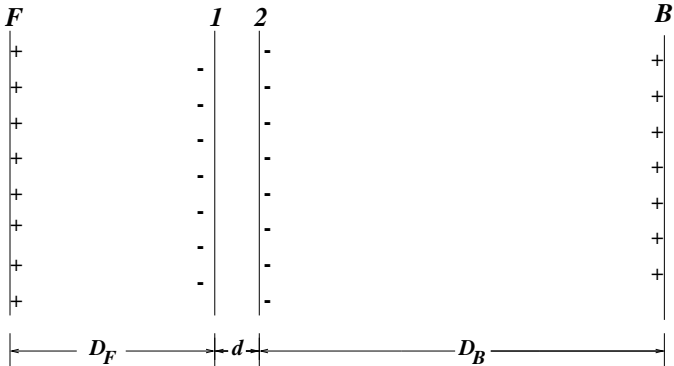}}
\caption{
Schematic figure of a double-layer device with
front-gate surface charge density $ep_F$ at left,
areal charge densities $-en_1$ and $-en_2$ in the quantum wells,
and back-gate surface charge density $ep_B$ at right.
}
\label{fig:device}
\end{figure}

Figure~\ref{fig:device} illustrates schematically the
geometry of the 2LES device.
We treat the quantum wells as zero-thickness layers
sandwiched between two plates of neutralizing charge,
which represent the effects of the front and back gates.
The distance between the front gate (at far left) and the first layer
(layer 1) is $D_F$; that between the back gate (at far right)
and the second layer (layer 2) is $D_B$;
the interlayer separation is denoted by $d$.
Typically, $d\sim 10$ nm, $D_F\sim 1 \mu$m, and $D_B\sim 1$ mm, so that
$d \ll D_F \ll D_B$; thus, Fig.~\ref{fig:device} is not at all
to scale.

The two inner layers are assumed to be in thermodynamic equilibrium
with each other,
and the voltage of the front (back) gate relative to the common chemical
potential of the inner layers is denoted by $V_F$ ($V_B$).
It is also assumed that small changes $\delta V_\alpha$ in the gate voltages
$V_\alpha$ ($\alpha={F,B}$) produce small changes in surface charge densities
only at the gates ($e\delta p_\alpha$) and in the layers
($-e\delta n_i$, $i=1,2$).
Overall charge neutrality requires that the total charge density vanish:
\begin{equation}
\label{eq:elecneut}
ep_F + ep_B - en_1 - en_2 = 0 .
\end{equation}
(Strictly speaking, we only require that the {\em change}
in the total charge density vanish:
$\delta ep_F + \delta ep_B - \delta en_1 - \delta en_2 = 0$.)
In writing Eq.~(\ref{eq:elecneut}), we have assumed that any stray charge
not included in Eq.~(\ref{eq:elecneut}) is unchanged when the gate voltage
are varied.
This implies that the only significant effect of the stray charges is
to shift the gate voltages by constant (empirically determined) amounts.
Sheet charge densities on the gates and inner layers
produce electric fields between the double-layer and the gates
($E_\alpha$) and between the two layers ($E_{12}$) according to Gauss's law,
\begin{eqnarray}
\label{eq:gauss}
E_\alpha &=& (e/\epsilon) p_\alpha , \\ \nonumber
E_{12} &=& (e/\epsilon) (p_F - n_1)
              = (e/\epsilon) (n_2 - p_B) .
\end{eqnarray}

We now obtain the conditions for thermodynamic equilibrium and
stability between the layers.
Regarding the gate charge densities $p_\alpha$ as fixed quantities,
we seek the values of the particle density in the inner layers
that minimize the total energy per unit area ${\cal{E}}_0/L_xL_y$.
Following Ref.~\onlinecite{jungwirth}, we separate the electrostatic
part of the total energy per unit area from the rest:
\begin{equation}
\label{eq:uovera}
\frac{{\cal{E}}_0}{L_xL_y} = \frac{\epsilon}{2} E_{12}^2 d +
                   \varepsilon(n_1,n_2) ,
\end{equation}
up to an irrelevant constant, where the first term is the
interlayer electrostatic energy density,
with $E_{12}$ given by Eq.~(\ref{eq:gauss}).
The quantity $\varepsilon(n_1,n_2)$ represents the total energy per unit area
for a fully interacting double-layer system in which each layer contains
a uniform neutralizing charge density, for particle densities
$n_1$ and $n_2$ in layers 1 and 2, respectively.
The Fermi, exchange, and correlation energies,
both intralayer and interlayer,
are contained in $\varepsilon(n_1,n_2)$, and it is this quantity
that is calculated using many-body techniques.\cite{jungwirth}
In the next section, we make an approximate calculation of
$\varepsilon(n_1,n_2)$ which includes the effects of
interlayer and intralayer exchange.

To obtain the condition for thermodynamic equilibrium between the
inner layers, we note that for fixed external charge densities
$p_\alpha$, the constraint of overall charge neutrality implies
that the particle density in one layer (e.g, $n_1$) is determined
by that in the other layer (e.g., $n_2$):
\mbox{$n_1 = p_F + p_B - n_2$}.
We may thus regard the total energy per unit area ${\cal{E}}_0/L_xL_y$
as a function of the layer density $n_2$, and
extremize ${\cal{E}}_0/L_xL_y$ with respect to $n_2$
at fixed $p_F$ and $p_B$
to obtain\cite{jpe,jungwirth}
\begin{equation}
\label{eq:equilib}
\mu_1 - \mu_2 = e E_{12} d ,
\end{equation}
where 
\begin{equation}
\label{eq:mui}
\mu_i \equiv \partial\varepsilon(n_1,n_2)/\partial n_i
\end{equation}
is the chemical potential measured relative to the energy minimum
of layer $i$.
Equation~(\ref{eq:equilib}) states that the difference in the
layer values of the chemical potential is equal to
the electrostatic potential energy difference between the layers.
If the equation of state determining $\mu_i(n_1,n_2)$ were known,
then Eqs. (\ref{eq:gauss}) and (\ref{eq:equilib}) would together determine
the values of layer densities $n_1$ and $n_2$ for which the
total energy ${\cal{E}}_0/L_xL_y$ is an extremum.

We now examine a necessary condition for interlayer thermodynamic stability
(i.e., for the local extremum to be a local minimum of the energy).
First we follow Ref.~\onlinecite{jpe} and introduce a set of lengths
that describe the dependence of the layer chemical potentials
$\mu_i$ on the layer densities $n_j$,
\begin{equation}
\label{eq:sij}
s_{ij} \equiv \frac{\epsilon}{e^2}
              \frac{\partial \mu_i}{\partial n_j}
            = \frac{\epsilon}{e^2}
              \frac{\partial^2 \varepsilon}{\partial n_j\partial n_i} .
\end{equation}
For the extremum condition in Eq.~(\ref{eq:equilib}) to represent
a local minimum of the total energy per unit area,
we require that the second derivative of ${\cal{E}}_0/L_xL_y$
with respect to $n_2$ be positive.
(We again regard $n_1$ as being determined by $n_2$ for
fixed $p_\alpha$ by the requirement for overall charge neutrality.)
This gives a necessary condition for stability:
\begin{equation}
\label{eq:stability}
d + s_1 + s_2 > 0,
\end{equation}
where
\begin{equation}
\label{eq:s1s2}
s_1 \equiv s_{11} - s_{12}, \qquad
s_2 \equiv s_{22} - s_{21} .
\end{equation}
The above inequality guarantees that the 2LES is at least metastable.
If Eq.~(\ref{eq:stability}) is violated, then the layer densities constitute
a local energetic maximum rather than a local minimum, and
there will be an interlayer charge instability that causes charge
to flow between the layers until a new energetic minimum is reached.

In the absence of interlayer correlations (the case considered in
Refs. \onlinecite{ruden} and \onlinecite{jpe}),
$s_{12}=s_{21}=0$, and the length $s_{ii}$
is directly related to the electronic compressibility $\kappa_i$ in
layer $i$ according to\cite{jungwirth}
\begin{equation}
\label{eq:sii}
s_{ii} = \frac{\epsilon}{e^2 n_i^2 \kappa_i} .
\end{equation}
Equations~(\ref{eq:stability}) and (\ref{eq:sii}) together with the
experimentally measured negative compressibility ($\kappa<0$) of the electrons
imply that when interlayer correlations are ignored ($s_{12}=0$),
sufficiently small interlayer separation $d$ will always
lead to a charge-transfer instability when the density of either (or both)
layer is sufficiently small.
This result is true even when intralayer correlations are included
beyond the HFA calculation of Ruden and Wu; only interlayer correlations
that produce sufficiently negative values of $s_{12}$
to satisfy Eq.~(\ref{eq:stability})
can suppress interlayer charge-transfer instabilities
at very small interlayer separations.

If the two layers of the 2LES sample can be contacted separately,
then the Eisenstein ratio $R_E$
provides a sensitive measure of the interlayer capacitance
that avoids the large gate-distance factors that dominate
the gate capacitances per unit area.\cite{jpe,jungwirth}
The Eisenstein ratio is defined as the ratio of the differential change
in the electric field $E_{12}$ between the inner layers
to that of the electric field $E_F$
between the front gate and the inner layers:
\begin{equation}
\label{eq:re}
R_E \equiv \frac{\delta E_{12}}{\delta E_F}
= 1 - \delta n_1/ \delta p_F ,
\end{equation}
where we have made use of Gauss's law, Eq.~(\ref{eq:gauss}).
In the following sections, we calculate the layer occupancies
$n_i$ as a function of the gate charges $p_\alpha$;
we then use Eq.~(\ref{eq:re}) to obtain $R_E$ by computing the
derivative of $n_1$ with respect to $p_F$.
In the classical limit (corresponding to large enough particle densities
and layer separations so that only the electrostatic energies are relevant),
$n_1=p_F$ (for $p_F>0$), so that by Eq.~(\ref{eq:re}), $R_E=0$.
Note also that if $n_1=0$ (e.g., due to $p_F<0$), then $R_E=1$.
By using Eqs. (\ref{eq:equilib}) and (\ref{eq:sij}) to express
differential changes in $E_{12}$ in terms of the electronic lengths
$s_{ij}$ and using Gauss's law,
the Eisenstein ratio may be expressed as
\begin{equation}
\label{eq:regen}
R_E = \frac{s_1 - s_2\delta E_B/\delta E_F}{d + s_1 + s_2} .
\end{equation}

The Eisenstein ratio
has an especially simple form for fixed total density
since then $\delta E_B=-\delta E_F$; from Eq.~(\ref{eq:regen}),
$R_E=s/(d+s)$, where $s\equiv s_1+s_2$.
However, most experiments fix the back-gate voltage $V_B$
rather than the total density,
and sweep the front-gate voltage $V_F$.
Because of the relatively large size of the back-gate distance $D_B$,
the constraint of fixed back-gate voltage ($\delta V_B = 0$)
is very nearly equivalent to that of fixed back-gate sheet charge density
($\delta p_B=0$ or, by Gauss's law, $\delta E_B=0$).
For $\delta E_B=0$, Eq.~(\ref{eq:regen}) shows that
the Eisenstein ratio is very nearly
\begin{equation}
\label{eq:repb}
R_E \approx \left(\delta E_{12}/\delta E_F\right)_{p_B}
= \frac{s_1}{d + s_1 + s_2} .
\end{equation}

The advantage of measuring $R_E$ rather than the usual gate
capacitances per unit area (such as $e\delta p_F/\delta V_F$)
is that $R_E$ depends only on the electronic lengths $s_i$ and
the interlayer distance $d$, not on the much larger gate distances
$D_\alpha$.
The difficulty in measuring $R_E$ is that (at least one of)
layers 1 and 2 must be separately contactable, which becomes increasingly
difficult as the layer separation $d$ becomes very small.
Nonetheless, measurements of the Eisenstein ratio have been used to
demonstrate the negative compressibility of the electron gas,\cite{jpe}
and it is to be expected that devices with separately contactable
layers will be built with increasingly narrow layer separations.
Note also that $R_E$ is very sensitive to charge-transfer instabilities.
In fact, from Eq.~(\ref{eq:stability}), the condition for the onset of an
instability to interlayer charge transfer, \mbox{$d+s_1+s_2=0$},
shows that $R_E$ formally diverges at the instability.
This can also be seen from the relation $R_E=1-\delta n_1/\delta p_F$,
since a charge transfer instability would produce an abrupt change
in the layer density $n_1$ in response to a small change in the 
gate density $p_F$.
Although evidence for abrupt interlayer charge transfers has been
reported based on Shubnikov-de Haas (SdH) measurements,\cite{kata}
the Eisenstein ratio would be a far more sensitive measure of abrupt interlayer
charge transfers.

In subsequent sections, we calculate the layer densities $n_i$
as a function of the front-gate particle density $p_F$,
for fixed back-gate particle density $p_B$.
Fortunately, $p_B$ may be found experimentally from SdH measurements as
the value of the layer densities when the system is balanced: i.e.,
for equal layer densities ($n_1=n_2$) and minimum subband separation
\mbox{$(n_a-n_b)$}.
Once $p_B$ (which we assume is very nearly constant) is known,
$p_F$ may be determined from charge conservation by measuring
the total layer density, either by SdH (as the sum of the
subband densities) or by Hall effect measurements.
It is therefore possible to determine $p_F$ experimentally,
without recourse to the gate voltages.
Of course, experimentally, it is $V_F$ that is varied directly
while $V_B$ is kept fixed, and $p_F$ changes in response to $V_F$
(while $p_B$ changes very little for large $D_B$).
We describe how the gate voltages may be determined from a knowledge
of $p_F, p_B$, and $\mu_i$ in Sec. \ref{app:gatevolt} of the Appendix.

\vspace{-0.1in}
\section{Mean-Field Approximation}
\label{sec:mfa}
\vspace{-0.1in}

In this section, we introduce the microscopic Hamiltonian
for the gated double-layer system and make a variational
approximation for the ground-state wave function that allows
for interlayer coherence.
The resulting approximate ground-state energy per unit area
depends on both intralayer and interlayer exchange.

We now consider the microscopic Hamiltonian for the double-layer
system illustrated in Fig.~\ref{fig:device}.
We idealize the inner layers as being two-dimensional,
and treat interlayer tunneling in the tight-binding approximation.
The Hamiltonian for the interacting system is then
\begin{eqnarray}
{\cal{H}} &=&
\sum_{j{\bf k}s} \varepsilon_k
                  c_{j{\bf k}s}^\dagger c_{j{\bf k}s}
- t \sum_{{\bf k}s} (c_{1{\bf k}s}^\dagger c_{2{\bf k}s}
                   +   c_{2{\bf k}s}^\dagger c_{1{\bf k}s}) \\ \nonumber
&+& \frac{1}{2L_xL_y} \sum_{\bf q} \sum_{j_1{\bf k}_1s_1}
                        \sum_{j_2{\bf k}_2s_2} V_{j_1j_2}(q) \\ \nonumber
&& \qquad \quad \times c_{j_1{\bf k_1+q}s_1}^\dagger
                        c_{j_2{\bf k_2-q}s_2}^\dagger
                        c_{j_2{\bf k_2}s_2} c_{j_1{\bf k_1}s_1} \\ \nonumber
&-& \sum_{j{\bf k}s} \sum_{\alpha} V_{j\alpha}(q=0)  p_\alpha
c_{j{\bf k}s}^\dagger c_{j{\bf k}s} \\ \nonumber
&+& \frac{L_xL_y}{2}\sum_{\alpha\beta} V_{\alpha\beta}(q=0)
                                       p_\alpha p_\beta ,
\label{eq:edensity}
\end{eqnarray}
where
$c_{j{\bf k}s}$ ($c_{j{\bf k}s}^\dagger)$ denotes the second-quantized
destruction (creation) operator for an electron or hole in layer
$j$ with momentum $\hbar{\bf k}$ and spin $s$.
Here
$\varepsilon_k = \hbar^2 k^2/2m^*$ is the kinetic energy in the
effective mass ($m^*$) approximation,
$t$ is the interlayer tunneling amplitude, and
the Fourier-transformed Coulomb potential is given by
\begin{equation}\label{eq:vj1j2}
V_{ij}(q) = \frac{e^2}{2\epsilon q} e^{-qd_{ij}} ,
\end{equation}
where $d_{ij}$ is the distance between layer $i$ and layer $j$.
The indices $\alpha,\beta$ in Eq.~(\ref{eq:edensity}) take
the values $F$ (front gate) or $B$ (back gate),
$V_{\alpha\beta}$ is the direct Coulomb interaction between gates
$\alpha$ and $\beta$, and
$V_{\alpha j}$ is the direct Coulomb interaction between gate
$\alpha$ and layer $j$.
The last two terms in Eq.~(\ref{eq:edensity}) represent
the direct interaction between the layers and gates, and
between  front and back gates, respectively.

We use a mean-field approximation (MFA) that is variationally based and
that reduces to the Hartree-Fock approximation for the balanced case
of equal layer densities.
This approximation includes the effects of interlayer correlations
in the simplest possible way.
The variational ground-state wave function is composed of two subbands,
$a$ and $b$, containing spin up ($\uparrow$) and
spin down ($\downarrow$) electrons:
\begin{equation}
|\Psi_0\rangle =
\prod_{{\bf k_4}}^{k_4\le k_{b\downarrow}} b_{{\bf k_4}\downarrow}^\dagger
\prod_{{\bf k_3}}^{k_3\le k_{b\uparrow}} b_{{\bf k_3}\uparrow}^\dagger
\prod_{{\bf k_2}}^{k_2\le k_{a\downarrow}} a_{{\bf k_2}\downarrow}^\dagger
\prod_{{\bf k_1}}^{k_1\le k_{a\uparrow}} a_{{\bf k_1}\uparrow}^\dagger
|0\rangle ,
\label{eq:psimfa}
\end{equation}
where $k_{as}$ and $k_{bs}$ denote the Fermi wave vectors for electrons
or holes of spin $s$ in subbands $a$ and $b$.
The creation operators for the subbands are related to
the layer creation operators by a canonical transformation
that we take to be of the form
\begin{eqnarray}
a_{{\bf k}s}^\dagger &=&
\cos(\theta/2) c_{1{\bf k}s}^\dagger +
\sin(\theta/2) e^{i\phi} c_{2{\bf k}s}^\dagger \\ \nonumber
b_{{\bf k}s}^\dagger &=&
-\sin(\theta/2) e^{-i\phi} c_{1{\bf k}s}^\dagger +
\cos(\theta/2) c_{2{\bf k}s}^\dagger
\label{eq:canon}
\end{eqnarray}
When $\theta=\pi/2$ and $\phi=0$, subband $a$ is the symmetric subband and
subband $b$ is the antisymmetric subband.
In the language of pseudospin, the superposition of layer states
in Eq.~(\ref{eq:canon}) corresponds to treating the layer pseudospin
as a Heisenberg, rather than an Ising, spin variable.\cite{zheng}
The form of the canonical transformation in Eq.~(\ref{eq:canon})
is not completely equivalent to a fully self-consistent
Hartree-Fock calculation because we have taken $\theta$ and $\phi$
to be independent of the wave vector ${\bf k}$ and spin $s$.
It would be interesting to explore the effects of
including the ${\bf k}$ and $s$ dependence of $\theta$ and $\phi$
in a future calculation.
Our simpler variational calculation,
which is equivalent to Ref.~\onlinecite{zheng} for the special
case of balanced layers, offers a reasonable starting point,
which is probably qualitatively correct over a large range of
layer densities.
It certainly gives layer densities that are in close agreement
with experimental values obtained from SdH measurements, as we
shall show.

The layer occupation numbers may be expressed in terms of
the subband occupation numbers by using Eq.~(\ref{eq:canon}):
\begin{eqnarray}
\langle c_{1{\bf k}s}^\dagger c_{1{\bf k}s} \rangle &=&
\cos^2(\theta/2) \langle a_{{\bf k}s}^\dagger a_{{\bf k}s} \rangle +
\sin^2(\theta/2) \langle b_{{\bf k}s}^\dagger b_{{\bf k}s} \rangle
\\ \nonumber
\langle c_{2{\bf k}s}^\dagger c_{2{\bf k}s} \rangle &=&
\sin^2(\theta/2) \langle a_{{\bf k}s}^\dagger a_{{\bf k}s} \rangle +
\cos^2(\theta/2) \langle b_{{\bf k}s}^\dagger b_{{\bf k}s} \rangle
\\ \nonumber
\langle c_{1{\bf k}s}^\dagger c_{2{\bf k}s} \rangle &=&
\sin(\theta/2)\cos(\theta/2)e^{i\phi}
(\langle a_{{\bf k}s}^\dagger a_{{\bf k}s} \rangle -
\langle b_{{\bf k}s}^\dagger b_{{\bf k}s} \rangle)
\\ \nonumber
\langle c_{2{\bf k}s}^\dagger c_{1{\bf k}s} \rangle &=&
\langle c_{1{\bf k}s}^\dagger c_{2{\bf k}s} \rangle^* ,
\label{eq:ccvals}
\end{eqnarray}
where the asterisk denotes complex conjugation,
and we have used Eq.~(\ref{eq:psimfa}) to eliminate cross terms such as
$\langle a_{{\bf k}s}^\dagger b_{{\bf k}s} \rangle$.
SILC occurs when 
\begin{equation}
\label{eq:sipcdef}
\langle c_{1{\bf k}s}^\dagger c_{2{\bf k}s} \rangle \ne 0
\end{equation}
in the absence of interlayer tunneling.
According to Eqs. (\ref{eq:ccvals}) and (\ref{eq:sipcdef}),
SILC requires that the following occurs:
(1) $\theta\ne 0,\pi$ so that the subband densities are different
from the layer densities.
SILC is therefore excluded when the pseudospin is treated
as an Ising variable.
(2) $n_a\ne n_b$ so that the subband densities are not equal
(nonzero pseudospin polarization).
When $n_{as}=n_{bs}$ (completely unpolarized pseudospin),
the MFA ground state can be expressed as
the product of two uncorrelated single-layer wavefunctions
by performing a global pseudospin rotation.

Although it is straightforward to generalize our approach to
finite temperature,
we calculate numerical results in the limit of zero temperature ($T=0$),
both for the sake of simplicity and because measurements can (and have)
been made on double-layer systems at low temperatures, even
down to millikelvin temperatures in the quantum Hall regime.\cite{murphy}
For the zero-magnetic-field case treated here, we expect that
finite temperature will not produce signifcant qualitative changes
in the layer densities if the temperature $T$ is below a fraction of the
Fermi temperature $T_F\equiv E_F/k_B$, where $E_F$ is Fermi energy
and $k_B$ is Boltzmann's constant.
For a layer density of $n=10^{10} $cm$^{-2}$, $T_F$ is roughly
4 K for $n$-type GaAs, and 1K for $p$-type GaAs.
The scale of the Hartree charge transfer energy, $e^2dn/2\epsilon$,
is larger than the Fermi energy except for ultrasmall layer separations
($d<5$ nm for $n$-type GaAs and $d<1$ nm for $p$-type GaAs.)

The other reason we work at zero temperature is to address matters
of principle, such as whether an interlayer charge transfer instability
can occur when the layers are very close together; finite temperatures
would presumably smear out such a transfer, if it could occur.
In the limit of zero temperature,
Eq.~(\ref{eq:psimfa}) implies that
\begin{equation}
\label{eq:occups}
\langle a_{{\bf k}s}^\dagger a_{{\bf k}s} \rangle =
\Theta(k_{as}-k) , \quad
\langle b_{{\bf k}s}^\dagger b_{{\bf k}s} \rangle =
\Theta(k_{bs}-k) ,
\end{equation}
where the subband Fermi wave vectors and number densities are related
through
\begin{equation}
\label{eq:kasbs}
k_{as} = \sqrt{4\pi n_{as}} , \qquad k_{bs} = \sqrt{4\pi n_{bs}} .
\end{equation}
Summing Equations (\ref{eq:ccvals}) over wave vector ${\bf k}$
relates the number densities of the layers to those of the subbands:
\begin{eqnarray}
n_{1s} + n_{2s} &=& n_{as} + n_{bs} , \\ \nonumber
n_{1s} - n_{2s} &=& (n_{as} - n_{bs})\cos\theta .
\label{eq:n1n2}
\end{eqnarray}

We may use the preceding equations to express
the ground-state energy per unit area in terms of
the subband occupancies $n_{\alpha s}$ and the angle $\theta$:
\begin{eqnarray}
\frac{{\cal{E}}_0}{L_xL_y} &=&
\frac{1}{\nu_0} \sum_{\alpha s} n_{\alpha s}^2 -
t(n_a-n_b)\sin\theta\cos\phi \\ \nonumber &+&
\frac{e^2d}{8\epsilon} \left[ (n_a-n_b)\cos\theta - (p_F-p_B) \right]^2
\\ \nonumber &-&
\frac{1}{2L_xL_y} \sum_{{\bf q}\alpha s} V_{11}(q) I_{\alpha\alpha s}(q)
\\ \nonumber &+&
\frac{\sin^2\!\theta}{4L_xL_y} \sum_{{\bf q}s}
       \left[ V_{11}(q)-V_{12}(q) \right]
\\ \nonumber &&\qquad\qquad\quad
\times \left[ I_{aas}(q)+I_{bbs}(q)-2I_{abs}(q) \right] ,
\label{eq:etheta}
\end{eqnarray}
where $\nu_0 = m^*/(\pi\hbar^2)$ is the density of states per unit area
for noninteracting spin-1/2 particles in two dimensions,
\mbox{${\bf q}={\bf k}_1-{\bf k}_2$}, and
\begin{eqnarray}
I_{\alpha\beta s}(q) \equiv
\frac{1}{L_xL_y} \sum_{\bf K} &&
\Theta(k_{\alpha s}-|{\bf K+q}/2|) \\ \nonumber
&& \times \Theta(k_{\beta s}-|{\bf K-q}/2|) .
\label{eq:iab}
\end{eqnarray}
Here ${\bf K}=({\bf k}_1+{\bf k}_2)/2$,
and the subband indices $\alpha$ and $\beta$ can be either $a$ or $b$.
Equation (\ref{eq:iab}) says that $I_{\alpha\beta s}(q)$ is
$1/(2\pi)^2$ times the common area of two circles of radii
$k_{\alpha s}$ and $k_{\beta s}$ whose centers are separated by $q$.
When $\beta=\alpha$, then $k_{\beta s}=k_{\alpha s}$,
and the first exchange integral in Eq.~(\ref{eq:etheta})
may be carried out explicitly:
\begin{equation}
\label{eq:v11term}
-\frac{1}{2L_xL_y} \sum_{\bf q} V_{11}(q) I_{\alpha\alpha s}(q) =
-\frac{8}{3\sqrt{\pi}} \frac{e^2}{4\pi\epsilon} n_{\alpha s}^{3/2} .
\end{equation}
Equation~(\ref{eq:v11term}) is just the exchange energy per unit area
for a uniform single-layer spin-polarized two-dimensional electron gas
of areal density $n_{\alpha s}$.

The last term in Eq.~(\ref{eq:etheta}), which contains the interlayer
exchange contribution, may be conveniently expressed as
\begin{equation}
\frac{\sin^2\theta}{4} \frac{e^2d}{2\epsilon} (n_a-n_b)^2 \Gamma ,
\end{equation}
where $\Gamma$ is the interlayer exchange parameter, given by
\begin{eqnarray}
\Gamma &\equiv& \frac{1}{L_xL_y} \sum_{{\bf q}s}
\left[ \frac{V_{11}(q)-V_{12}(q)}{e^2d/2\epsilon} \right] \\ \nonumber
&& \qquad \qquad \times
\left[ \frac{I_{aas}(q)+I_{bbs}(q)-2I_{abs}(q)}{(n_a-n_b)^2} \right] .
\label{eq:gamma}
\end{eqnarray}
The properties of $\Gamma$ are described in
Sec. \ref{app:sinteg} of the Appendix.
The last term in Eq.~(\ref{eq:etheta}) must in general be evaluated
numerically, although it vanishes at $d=0$ or when $n_{as}=n_{bs}$.
It also vanishes if $\theta$ is $0$ or $\pi$, in which
case subband $a$ is the layer (1 or 2) with most particles, while
subband $b$ is the layer with the fewest particles.
Ruden and Wu implicitly treated the layer pseudospin as an Ising-like
variable with $0$ and $\pi$ as the only allowed values for $\theta$;
in their approximation, the last term in Eq.~(\ref{eq:etheta}) vanishes,
and the interlayer effects we shall discuss here do not appear.

For definiteness, we take $n_a\ge n_b$,
$n_{\alpha\uparrow}\ge n_{\alpha\downarrow}$,
and $0\le\theta\le\pi$.
Our procedure consists of finding the values of
$n_{\alpha s}$ and $\theta$ that minimize the
expected energy per unit area, Eq.~(\ref{eq:etheta}).
Within our variational approximation, we find (as in\cite{zheng} the HFA)
that the spins in a given subband are always either completely
polarized (ferromagnetic at sufficiently low subband densities)
or completely unpolarized (paramagnetic at higher densities).
Real systems probably possess intermediate polarization for
a range of low densities.\cite{ortiz,rapi}
For finite $t>0$,
the ground-state energy per unit area, Eq. (\ref{eq:etheta}),
is minimized for $\phi=0$.
In the absence of interlayer tunneling, the ground-state energy per unit
area is independent of $\phi$, provided that $\phi$ is constant;
for convenience we set $\phi=0$.
The layer densities $n_{1s}$ and $n_{2s}$ may be obtained from
$n_{\alpha s}$ and $\theta$ via Eq.~(\ref{eq:n1n2}).
We begin our calculations in the next section, by considering the
case of electrostatically balanced gates.

\vspace{-0.1in}
\section{Balanced Gates}
\label{sec:balance}
\vspace{-0.1in}

In this section, we consider the case of electrostatically balanced
gates ($p_F=p_B$), beginning with zero interlayer tunneling.
This was the situation originally considered by Ruden and Wu,\cite{ruden}
and more recently in Refs. \onlinecite{zheng,conti}, 
and \onlinecite{swier}.
For balanced gates, our approximation is equivalent to the unrestricted
HFA of Zheng and co-workers,\cite{zheng} and
except for our analysis of the three-component phase,
most of our results agree with theirs.

\vspace{-0.1in}
\subsection{Zero tunneling}
\label{sec:zerotun}
\vspace{-0.1in}

The balanced case raises an important question of principle:
can exchange and correlation effects alone, unaided by applied
gate biases and unhindered by interlayer tunneling, ever
produce a ground state in which the densities of the inner layers
are not equal?
Based on a restricted HFA (which did not allow for SILC)
Ruden and Wu proposed that for small enough layer densities
and layer separations, the answer is yes.
Zheng and co-workers argued recently that an unrestricted HFA (which allows for,
but does not mandate, SILC) gives the opposite answer.\cite{zheng}
We find that, except for a small region in density which supports a
three-component phase that has a slight layer imbalance, the layer densities
are equal when the gates are balanced.

\subsubsection{Zero layer separation}

In order to classify the four noncrystalline phases that we find
for gate-balanced double-layer systems, it is useful to begin with
the idealized case of zero layer separation ($d=0$).
For $d=0$, the Hamiltonian is invariant under spin rotation,
pseudospin rotation, and the interchange of spin and pseudospin.
It is the
same as the Hamiltonian for a four-layer system of spinless fermions
with zero separation between all the layers: layers, subbands, and spins
become interchangeable labels for the four components.
The $d=0$ double-layer system is therefore equivalent to a
single-layer two-dimensional system of fermions with $CP(3)$ symmetry.
At $d=0$, the interlayer Hartree energy is zero
and the interlayer ($V_{12}$) and intralayer ($V_{11}$)
Coulomb interactions are equal.
As a consequence, the variational energy in Eq.~(\ref{eq:etheta}) is
independent of $\theta$ and (for $t=0$) of $\phi$ when $d=0$,
\begin{equation}
\label{eq:edzero}
\frac{{\cal{E}}_0}{L_xL_y} = \sum_{\alpha s} \left[
\frac{n_{\alpha s}^2}{\nu_0} -
\frac{8}{3\sqrt{\pi}} \frac{e^2}{4\pi\epsilon} n_{\alpha s}^{3/2} \right]
\equiv \sum_{\alpha s} \varepsilon(n_{\alpha s}) ,
\end{equation}
and the MFA is equivalent to the HFA.

For generality, we first
consider an $N$-component system, where $N$ is twice the number of layers:
$N=4$ for double-layer spin-1/2 systems.
At $d=0$, the MFA lacks intercomponent correlations; thus,
the total energy of the system is just the sum of the
individual energies $\varepsilon(n_{\alpha s})$ of each component.
We can investigate the distribution of component densities in the
MFA ground state in an $N$-component system
by taking all but two of component densities to be fixed,
and minimizing the energy of the remaining two-component system.
If we label the two components we seek to minimize as $1$ and $2$,
then according to Sec.~\ref{sec:model},
the condition for stable equilibrium (local minimum of the total energy) is
\begin{equation}
\mu(n_1)=\mu(n_2), \qquad s(n_1)+s(n_2)>0 ,
\end{equation}
where $\mu(n_j)=\partial\varepsilon(n_j)/\partial n_j$ is the chemical
potential relative to the minimum energy of component $j$ and 
$s(n_j)=\partial\mu(n_j)/\partial n_j$
is inversely proportional to the compressibility of component $j$.
When $\varepsilon(n_j)$ is the sum of the kinetic and exchange energies
as in Eq.~(\ref{eq:edzero}), then the MFA energy is minimized only if
(1) both layer densities are equal ($n_1=n_2$), or
(2) one or both layers are empty.
There are no intermediate possibilities in the MFA.
This $d=0$ result for $N=2$ gives the results found by Ruden and Wu:
when $n_1+n_2$ is sufficiently large, the component densities are equal;
when $n_1+n_2$ is sufficiently small, there is an exchange-driven
intercomponent charge instability that empties out one of the components.
In the absence of intercomponent correlations, we expect that at
low enough densities the component compressibilities will be negative
and that therefore one of the components will empty out, even
if intracomponent correlations are included.
In the $N$-component $d=0$ MFA ground state,
any pair of layers either has equal density or has at least one
of the layers empty.

There are therefore $N$ possible MFA ground states
in an $N$-component system at $d=0$,
characterized by the number of components $p$ that have nonzero
and equal densities.
The remaining $N-p$ components have zero density.
Defining the dimensionless average interparticle spacing
{\em per component} by \mbox{$r_s=1/\sqrt{\pi(n_T/N)a_0^2}$}
for an $N$-component system,
where $a_0=4\pi\epsilon\hbar^2/m^*e^2$ is the effective Bohr radius,
we may write the energy per unit area for the ``$p$-component''
MFA ground state as
\begin{equation}
\label{eq:ep}
\frac{{\cal{E}}_p}{L_xL_y} =
\frac{e^2}{4\pi\epsilon a_0^3} p \left[
\pi\left(\frac{N}{\pi r_s^2p}\right)^2 -
\frac{8}{3\sqrt{\pi}} \left(\frac{N}{\pi r_s^2p}\right)^{3/2}
\right] .
\end{equation}
In the limit $N\rightarrow\infty$, $p$ becomes a continuous variable
that minimizes the energy (\ref{eq:ep}) for $p=p_\infty$, where
\begin{equation}
\label{eq:pinfty}
p_\infty = N\left(\frac{3\pi}{4r_s}\right)^2
= \left(\frac{3\pi}{4}\right)^2 \pi a_0^2 n_T ,
\end{equation}
where we have assumed that $r_s \ge (3\pi/4)$;
otherwise, $p_\infty=N$.
Equation (\ref{eq:pinfty}) shows that in the limit $N\rightarrow\infty$,
$p$ is proportional to $n_T$ for $r_s \ge (3\pi/4)$,
and equal to $N$ otherwise.
As expected, the number $p$ of equally occupied components drops as
$n_T$ is reduced.
Equation (\ref{eq:pinfty}) is equivalent to saying the dimensionless
interparticle separation per component for which $p=p_\infty$ is
\begin{equation}
\label{eq:rspinfty}
r_s(p_\infty) = \frac{3\pi}{4} \sqrt{\frac{N}{p_\infty}}
\end{equation}
when $p_\infty/N\le 1$.
The equilibrium value of the \mbox{$N\rightarrow\infty$} MFA energy
corresponding to Eq.~(\ref{eq:pinfty}) is
\begin{equation}
\varepsilon_\infty=-\left(\frac{4}{3\pi}\right)^2
\frac{e^2}{4\pi\epsilon a_0}n_T .
\end{equation}

For arbitrary finite $N$,
Eq.~(\ref{eq:ep}) can be used to find the interparticle spacing
per component $r_s^{(0)}(p,p+1)$ at which the $p$- and the $(p+1)$-component 
phases have the same energies at $d=0$:
\begin{equation}
\label{eq:rspp1}
r_s^{(0)}(p,p+1) = \frac{3\pi}{8} \left(
\sqrt{\frac{N}{p}} + \sqrt{\frac{N}{p+1}} \right) .
\end{equation}
This is the interparticle spacing per component
for the MFA transition between the $p$- and $(p+1)$-component phases.
It is interesting to compare this result to Eq.~(\ref{eq:rspinfty})
and to note that
\begin{equation}
r_s^{(0)}(p,p+1) < r_s(p_\infty) <  r_s^{(0)}(p-1,p) ,
\end{equation}
so that even for finite $N$,
$r_s(p_\infty)$ always gives a value for the interparticle spacing
that is in the $p$-component MFA phase at $d=0$.

For systems of physical interest containing $N/2$ layers of
spin-1/2 particles, it is convenient to work with
the interparticle spacing {\em per layer}
(rather than the spacing {\em per component}).
This is accomplished by dividing Eq.~(\ref{eq:rspp1}) by $\sqrt{2}$.
For double-layer systems of spin-1/2 particles ($N=4$),
\begin{eqnarray}
\label{eq:rszero}
r_s^{(0)}(1,2) &=& \frac{3\pi}{8} (\sqrt{2}+1) \approx 2.844 \\ \nonumber,
r_s^{(0)}(2,3) &=& \frac{3\pi}{8} (1+\sqrt{2/3}) \approx 2.140 \\ \nonumber,
r_s^{(0)}(3,4) &=& \frac{3\pi}{8} (\sqrt{2/3}+\sqrt{1/2}) \approx 1.795,
\end{eqnarray}
where the superscript $(0)$ denotes zero layer separation.
Note that the direct four- to two-component MFA transition
predicted to occur at
\begin{equation}
r_s(2,4) = \frac{3\pi}{8} (1+\sqrt{1/2}) \approx 2.011
\end{equation}
does not exist.
In fact, we find that when $r_s=r_s(2,4)$,
the gate-balanced ($p_F=p_B$) double-layer system is always in the
three-component MFA phase, regardless of the layer separation.
The three-component system, which at $d=0$ has
$n_{a\uparrow}=n_{a\downarrow}=n_{b\uparrow}=n_T/3$,
has a spin-unpolarized subband ($n_{a\uparrow}=n_{a\downarrow}$)
with greater density than the other subband ($n_a>n_b$),
which is spin-polarized.
These features of the three-component phase persist at finite layer
separations, although the layer imbalance is greatly reduced.

\subsubsection{Finite layer separation}

Classically (when only the electrostatic energies are considered),
balanced layers ($n_1=n_2$) are obtained when $p_F=p_B$, in order
to make the electric field $E_{12}$ between the inner layers vanish.
This result
gives the asymptotically correct behavior for high layer densities
and large layer separations.
At sufficiently low densities and layer separations, the exchange energy
can dominate the kinetic and electrostatic energies, so that the
possibility of strong intralayer exchange leading to an interlayer
charge-transfer instability must be considered.
However,
within the MFA, it can be proved that the inner layer densities
are always equal, except in the three-component phase.

If the subband densities are equal
($n_a=n_b$, the case of ``pseudospin paramagnetism''),
then Eq.~(\ref{eq:n1n2}) shows that $n_1=n_2$.
Thus the four-component ($n_{\alpha s}=n_T/4$) and two-component
(with $n_{a\uparrow}=n_{b\uparrow}=n_T/2$) phases have balanced layers.
This is because the MFA state constructed by occupying equally
the single-particle subband states $a$ and $b$
is equivalent (up to a global pseudospin rotation)
to the MFA state constructed by occupying equally the
single-particle layer states $1$ and $2$.
The fact that the ground-state energy in Eq.~(\ref{eq:etheta}) is
independent of $\theta$ and $\phi$ when $n_a=n_b$ is due to
the invariance of the ground-state energy under global pseudospin
rotation.

If the subband densities are not equal ($n_a>n_b$),
extremizing the total energy per unit area in Eq.~(\ref{eq:etheta})
with respect to $\theta$ for $p_F=p_B$ and $t=0$
gives the condition $\sin(2\theta)=0$.
The requirement that the extremum be a minimum (i.e., that the
second derivative of the total energy per unit area with respect
to $\theta$ be positive) gives
\begin{equation}
\sin\theta = \left\lbrace
\begin{array}{ll}
1  & \qquad \mbox{if $\Gamma<1$} \\
0  & \qquad \mbox{if $\Gamma>1$}
\end{array}
\right.
\end{equation}
where the interlayer exchange parameter $\Gamma$
is defined in Eq.~(\ref{eq:gamma}).
The properties of $\Gamma$ are described in
Sec. \ref{app:sinteg} of the Appendix.
Using the inequality
\begin{equation}
\label{eq:inequal}
e^2d/2\epsilon>V_{11}(q)-V_{12}(q) ,
\end{equation}
which is true for $d>0$,
it follows from Eq.~(\ref{eq:gamma}) that for $d>0$,
\begin{eqnarray}
\label{eq:gineq}
\Gamma &<& \frac{1}{L_xL_y} \sum_{{\bf q}s}
\left[ \frac{I_{aas}(q)+I_{bbs}(q)-2I_{abs}(q)}{(n_a-n_b)^2} \right]
\\ \nonumber && \qquad =
\sum_s \frac{(n_{as}-n_{bs})^2}{(n_a-n_b)^2} .
\end{eqnarray}
Thus the condition
\begin{equation}
\label{eq:pi2cond}
(n_{a\uparrow}-n_{b\uparrow})(n_{a\downarrow}-n_{b\downarrow}) \ge 0
\end{equation}
is sufficient to guarantee that $\Gamma<1$ for $d>0$,
so that $\theta=\pi/2$, which balances the layers.
The one-component phase ($n_{a\uparrow}=n_T$) satisfies
Eq.~(\ref{eq:pi2cond}), so $\theta=\pi/2$ and the layers are balanced.
No interlayer charge transfer is obtained in the one-component phase
because the combined effects of electrostatics and interlayer exchange,
which favor balanced layers,
always dominate the unbalancing influence of intralayer exchange:
\mbox{$e^2d/2\epsilon+V_{12}(q)>V_{11}(q)$}.

Within the MFA, the spins in subbands $a$ and $b$
are either completely unpolarized (at higher densities) or fully polarized
(at sufficiently low densities).
Therefore the only possible MFA configurations of spin and pseudospin
which could have unbalanced layers ($n_a>n_b$, $\theta\ne\pi/2$)
when $p_F=p_B$ would be three-component states with
\begin{equation}
n_{a\uparrow}=n_{a\downarrow}<n_{b\uparrow} , \quad n_{b\downarrow}=0 .
\end{equation}
We find numerically that such states always have $\Gamma>1$ when $t=0$.
so that $\sin\theta=0$.
Hence, there is no interlayer phase coherence
and the pseudospins are Ising-like.
States with $n_a>n_b$ and $\sin\theta=0$ have partially unbalanced
layers, with the lower-density layer being spin-polarized and the
higher-density layer being spin-unpolarized,
even for balanced gates ($p_F=p_B$).
This is the behavior we find for the three-component MFA phase.
For infinitesimal $d$, one (spin-unpolarized) layer has twice the density
of the other (spin-polarized) layer, and the phase exists in the range
$r_s^{(0)}(3,4)<r_s<r_s^{(0)}(2,3)$.
At finite $d$, the three-component phase has only a slight layer imbalance
and exists only in a narrow region of average interparticle spacing around
$r_s\approx r_s(2,4)$.

The equality of the inner layer densities in the balanced case
(except for the three-component phase)
has also been shown to be true for the one-component phase
when intralayer and interlayer correlations
are included within the STLS approximation.\cite{lerwen}
We note that if interlayer exchange were omitted from
the total energy per unit area
by setting $V_{12}(q)=0$ in Eq.~(\ref{eq:etheta}),
then four MFA ground states would still be obtained,
and the four-, three-, and two-component states would be unchanged.
However, Eq.~(\ref{eq:inequal}) would not be satisfied at small $q$,
and at large interparticle distances (at small values of $k_F$,
or low densities) the vanishing of the second derivative with respect to
$\theta$ would give the condition $\cos(2\theta)>0$, so that
$\cos(\theta)=\pm 1$.
This would produce the interlayer charge instability
of Ruden and Wu for the one-component phase.\cite{ruden}
The fact that the one-component phase has equal densities is due to
the effects of interlayer exchange.

Before obtaining the MFA phase diagram for double-layer systems,
it is convenient to express lengths and energies as dimensionless
quantities.
We therefore express the layer separation $d$
and the average density per layer $n_T/2$
in terms of the effective Bohr radius of the sample,
\mbox{$a_0=4\pi\epsilon\hbar^2/m^*e^2$},
where $\epsilon$ is the dielectric
constant in the material, and $m^*$ is the effective mass.
For GaAs, the dielectric constant is $\epsilon\approx 13\epsilon_0$.
For $n$-type GaAs, $m^*\approx 0.07m_e$ so that
$a_0\approx 9.8$ nm,
while for $p$-type GaAs, $m^*\approx 0.3m_e$ so that
$a_0\approx 2.3$ nm.
The average density per layer can be expressed in terms of
dimensionless ratio $r_s=r_0/a_0$ of the interparticle spacing $r_0$
for a single layer of averaged density $n_T/2$
(defined through $\pi r_0^2 n_T/2=1$) to the effective Bohr radius
$a_0$.
Thus, for the same total density, $p$-type GaAs will have a value for $r_s$
that is about $9.8/2.3\approx 4.3$ times larger than for $n$-type GaAs.
We define the Fermi wave vector $k_F$ in terms of the total density
$n_T$ as
\begin{equation}
\label{eq:kf}
k_F = \sqrt{4\pi n_T/p} =
\sqrt{\frac{2}{p}}\frac{2}{r_s a_0}
\end{equation}
for $p=1,2,4$.
For the spin-polarized $p=2a$ state and and for the completely unpolirized 
$p=4$, $k_F$ is equal to the Fermi wave vector corresponding to
the average density per layer ($n_T/2$).
The energy scale associated with the effective Bohr radius is
\mbox{$v_0 = e^2/4\pi\epsilon a_0 = \hbar^2/m^*a_0^2$},
which gives $v_0\approx 11$ meV for $n$-type GaAs,
and $v_0\approx 48$ meV for $p$-type GaAs.

The HFA phase diagram for balanced double-layers without tunneling was
obtained by Zheng and co-workers,\cite{zheng}
except for the three-component phase.
Like them, we find that within the MFA, three of the stable phases
have equal average inner layer densities.
Only the three-component phase has unequal layer densities.
To understand the origin of the MFA phases, it useful to consider
the five terms that contribute to the total energy per unit area in
Eq.~(\ref{eq:etheta}).
The first term of Eq.~(\ref{eq:etheta}) is
the kinetic (Fermi) energy, which favors distributing
the particles equally among the subbands and spins.
At the highest densities, the kinetic energy term dominates, and
the double-layer system is a four-component spin and pseudospin paramagnet:
$n_{a\uparrow}=n_{a\downarrow}=n_{b\uparrow}=n_{b\downarrow}=n_T/4$.
The second term of Eq.~(\ref{eq:etheta}) is
the tunneling energy, which we take to be zero for now.
In general, it favors $n_a>n_b$ (pseudospin polarization),
without regard to the real spin.
The third term of Eq.~(\ref{eq:etheta}) is
the electrostatic energy, which vanishes when
the gates and inner layers are balanced.
In general, the electrostatic term favors complete screening, which would make
$n_1=p_F$ and $n_2=p_B$, without regard to the real spin.

The fourth term in Eq.~(\ref{eq:etheta}) is
an intrasubband exchange term that dominates at the smallest densities
and layer separations.
It has the opposite effect of the kinetic energy, eventually producing
a one-component spin and pseudospin ferromagnet at very low densities
and small layer separations.
The last term (containing the interlayer exchange) favors pseudospin
paramagnetism ($n_{as}=n_{bs}$), but is indifferent to the polarization
of the real spin, so long as it is the same in both subbands.
Thus when $p_F=p_B$ and $d>0$, the last term and the electrostatic
terms are responsible for producing a two-component phase that is
ferromagnetic in real spin rather than in pseudospin:
\mbox{$n_{a\uparrow}=n_{b\uparrow}=n_T/2$}.
(At $d=0$, two-component states that are ferromagnetic in either
the spin or pseudospin are degenerate.
For $d>0$, the pseudospin ferromagnetic $p=2b$ state is favored only for
substantial tunneling $t$ and/or layer imbalance $|p_F-p_B|$.)
In the absence of the kinetic energy term (e.g., in the limit
$m^*\rightarrow\infty$), the real spin is always polarized due to
intralayer exchange.
However, because of the interlayer exchange,
the two-component state is always obtained at high densities or large
layer separations, while the one-component state is favored only for
low densities and small layer separations.

Finite layer separation ($d>0$) differentiates between spin and pseudospin,
so that the symmetry of the problem becomes $SU(2)\times U(1)$
rather than the $CP(3)$ symmetry at $d=0$.
At finite layer separation, \mbox{$V_{12}(q)<V_{11}(q)$},
and the last term in the energy per unit area in Eq.~(\ref{eq:etheta})
is minimized by equal subband densities, rather than by equal
real spin densities.
This does not effect the $p=1$ (fully spin- and pseudospin-polarized)
or $p=4$ (completely spin and pseudospin unpolarized) phases.
However, when $t=0$ and $p_F=p_B$, the two-component ($p=2$) MFA phase
has its real spin fully polarized and its pseudospin completely unpolarized.

Finite layer separation also changes the densities at which the
transitions between neighboring $p$-component phases occur.
The densities at which the transitions between the phases occur
in the MFA can be determined by comparing MFA energies.
This task is made easier by the fact that
the MFA always makes the real spins in a given subband either completely
unpolarized (paramagnetic) of fully polarized (maximally ferromagnetic).
This would probably not be true if correlation-energy effects were
properly included,\cite{ortiz}
and it is likely that in real double-layer systems, states with
partial polarization may be stable in some regions of density.
The same MFA behavior (the restriction to the two extremes
of either zero or full polarization)
is also found for the pseudospin when the layers are balanced
and the interlayer tunneling is zero, except for the three-component
phase, which has partial pseudospin polarization.

The four-component and two-component phases both have equal subband densities,
so their MFA energies are independent of the layer separation $d$.
If there were a direct transition between these two phases,
it would be simply a spin paramagnetic
to ferromagnetic transition in each subband or layer.
Therefore, within the MFA, such a four-component to two-component transition
would occur at the same layer density as the spin-polarization transition
for a single-layer system with a layer density equal
to the subband densities: i.e., for
\mbox{$r_s(2,4)=3\pi(1+\sqrt{1/2})/8\approx 2.011$},
independent of the layer separation.
As we discuss below, the direct four- to two-component transition
is interrupted by a three-component phase, which has one subband
(layer) spin-polarized and the other spin-unpolarized.
So it is still true that MFA spin polarization transitions occur
near $r_s\approx r_s(2,4)$.
However, the actual value of $r_s$ needed to obtain spin-polarization
in a real sample is likely to be significantly higher.
For single-layer systems,
diffusion Monte Carlo simulations show that the low-density ferromagnetic
state predicted by the HFA does occur; however, correlation-energy effects
move the transition to densities that are probably 100 times lower,
to $r_s\approx 20$.\cite{rapi}
Such high values of $r_s$ have been achieved in low-density $p$-type GaAs
samples, which possess a larger effective mass (and therefore larger $r_s$)
than $n$-type samples.\cite{shapira}
Large values of the effective mass will favor the existence of the
lower-component ($p<4$) described here, in that they increase $r_s$.

\begin{figure}[h]
\epsfxsize3.5in
\centerline{\epsffile{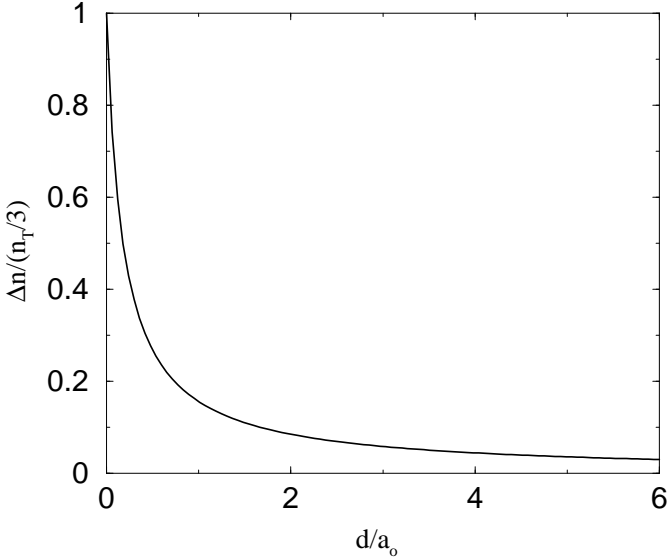}}
\caption{Layer imbalance ratio
$\Delta n/(n_T/3)=1-x$ in the three-component phase
for $r_s=r_s(2,4)\approx 2.011$. The imbalance decreases with increasing
layer separation.}
\label{fig:xplot}
\end{figure}

We find empirically that the three-component MFA phase has $\sin\theta=0$.
In order to analyze this phase, consider a MFA ground state with
\begin{eqnarray}
n_{b\uparrow} &=& (n_T/3)(1+x/2), \quad n_{b\downarrow}=0 , \\ \nonumber
n_{a\uparrow} &=& n_{a\downarrow} = (n_T/3)(1-x/4) ,
\end{eqnarray}
and $\sin\theta=0$, for $0 \le x \le 1$.
Note that the layer imbalance is given by
\begin{equation}
\Delta n \equiv n_a-n_b = (n_T/3) (1-x) .
\end{equation}
When $d=0$, the three-component phase distributes the
densities equally between the three components (but {\em not}
between the two layers), so that $x=0$.
As $d\rightarrow\infty$, the Coulombic cost of layer imbalance becomes
prohibitive, and $x\rightarrow 1$ so that $\Delta n \rightarrow 0$.
We plot the layer imbalance ratio
$\Delta n/(n_T/3) = (1-x)$
for $r_s=r_s(2,4)\approx 2.011$ in Fig.~\ref{fig:xplot}.

As $d\rightarrow 0$,
\begin{equation}
x \approx \frac{(4/3)(d/a_0)}{1-r_s/(\pi\sqrt{2/3})}
  \rightarrow 6.174\frac{d}{a_0} ,
\end{equation}
to linear order in $d/a_0$, where the last expression in the equation above
is for $r_s=r_s(2,4)$,
which is the only value of $r_s$ for which the double-layer system is
in the three-component phase for arbitrary layer separation.
As $d\rightarrow\infty$,
\begin{equation}
\label{eq:xdinf}
\frac{\Delta n}{n_T/3} =
1-x \approx \frac{a_0}{d}\frac{3}{4} \left[
\frac{r_s}{(1+1/\sqrt{2})\pi/2} \right] 
  \rightarrow \frac{3}{16}\frac{a_0}{d} ,
\end{equation}
to linear order in $a_0/d$, where the last expression in the equation above
is for $r_s=r_s(2,4)$.
Equation~(\ref{eq:xdinf}) says that the layer imbalance $\Delta n$
in the three-component phase is inversely proportional to $d/a_0$
for large values of $d/a_0$.

For infinitesimal $d$, the energy per unit area of three-component phase
increases by the amount $(e^2d/8\epsilon)(n_T/3)^2$ to linear order in $d$,
while that of the four- and two-component phases are unchanged.
Equating the three-component energy to the four- and to the two-component
energies gives
\begin{eqnarray}
r_s(2,3) &\approx& r_s^{(0)}(2,3) \left(1 - \frac{1}{3}\frac{d}{a_0}\right) ,
\\ \nonumber,
r_s(3,4) &\approx& r_s^{(0)}(3,4) \left(1 + \frac{2}{3}\frac{d}{a_0}\right) ,
\end{eqnarray}
to first order in $d/a_0$ as $d/a_0\rightarrow 0$, and
\begin{eqnarray}
r_s(2,3) &\approx& r_s^{(0)}(2,4) \left(1+\frac{1}{16}\frac{a_0}{d}\right) ,
\\ \nonumber
r_s(3,4) &\approx& r_s^{(0)}(2,4) \left(1-\frac{1}{16}\frac{a_0}{d}\right) ,
\end{eqnarray}
to first order in $a_0/d$ as $d/a_0\rightarrow\infty$.
Note that both $r_s(2,3)$ and $r_s(3,4)$ approach $r_s(2,4)$
in the limit $d\rightarrow\infty$.
This is because as $d\rightarrow\infty$,
the double-layer system consists of two independent layers, and
$r_s(2,4)$ is the interparticle spacing at which the
spin-polarized and spin-unpolarized energies are equal
in a single-layer system.
Thus, as $d\rightarrow\infty$, the energies of the
four-, three-, and two-component phases are all equal
at $r_s=r_s(2,4)$.

\begin{figure}[h]
\epsfxsize3.5in
\centerline{\epsffile{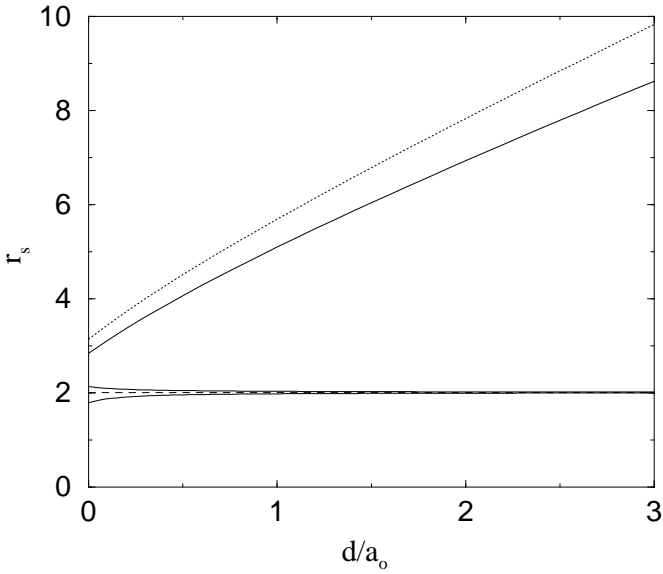}}
\caption{MFA (solid lines) phase diagram
for a gate-balanced ($p_F=p_B$) double-layer system with $t=0$.
At the highest densities (smallest $r_s$, bottom of figure),
the four-component state, which is both spin and pseudospin unpolarized,
is energetically favored.
In a narrow region around $r_s(2,4)\approx 2.011$ (dotted line),
the three-component state has the lowest energy.
At lower densitities, a two-component state which is spin
polarized but pseudospin unpolarized is favored.
At the lowest densities (large $r_s$, top of figure) a one-component
state which is both spin and pseudospin polarized is favored.
The dotted line at the top is a GRPA estimate for the onset of the
one-component phase.
}
\label{fig:phdiag}
\end{figure}

We now consider the two-component to one-component transition in the MFA.
For $d=0$, the MFA transition to pseudospin ferromagnetism is equivalent to
a real-spin paramagnetic to ferromagnetic transition in a single layer
having total density $n_T$ (rather than $n_T/2$).
Thus for $d=0$, the MFA critical density per layer
for the two- to one-component transition is exactly half the critical density
for the four- to two-component transition, so that
$r_s^{(0)}(1,2) = 3\pi(\sqrt{2}+1)/8\approx2.844$.
By equating the one- and two-component phase energies per area,
the critical density for the one- to two-component transition may
be obtained:
\begin{eqnarray}
\label{eq:1to2}
\frac{r_s^{(0)}(1,2)}{r_s(1,2)} &=&
1 - (1+1/\sqrt{2})\frac{3\pi z}{32}\Gamma_1(z)
\\ \nonumber &=&
1 - (1+1/\sqrt{2})\left[1-S(z)\right] ,
\end{eqnarray}
where
\begin{eqnarray}
\label{eq:z}
z &=& 2k_Fd=2d\sqrt{4\pi n_T}=4\sqrt{2}d/(r_sa_0) \\ \nonumber
  &=& (2-\sqrt{2}) \frac{32}{3\pi} \frac{d}{a_0}
      \frac{r_s^{(0)}(1,2)}{r_s} ,
\end{eqnarray}
$\Gamma_1 = \Gamma(n_{a\uparrow}=n_T)$, and
\begin{eqnarray}
\label{eq:sintegral}
S(z) &=& \frac{3}{2} \int_0^1 dx e^{-zx}
                     \left[\arccos(x) - x\sqrt{1-x^2}\right]
\\ \nonumber &=&
\frac{3\pi}{4z} \left\{ 1 - \frac{2}{z} \left[I_1(z)-L_1(z)\right] \right\}
\\ \nonumber &\rightarrow& \left\{
\begin{array}{ll}
1 - (3\pi/32)z + (1/5)z^2 , & z \rightarrow 0 \\
3\pi/4z , & z \rightarrow \infty
\end{array}
\right.
\end{eqnarray}
Here $I_1$ and $L_1$ are modified Bessel and modified Struve functions
of the first kind, respectively.
The derivation of the above formula is discussed in
Sec. \ref{app:onecomp} of the Appendix.
Equations (\ref{eq:1to2}) and (\ref{eq:z}) determine $r_s(1,2)$,
which we have plotted as the upper solid line in Fig.~\ref{fig:phdiag}.
It follows from Eq.~(\ref{eq:sintegral}) that
$r_s(1,2)/r_s^{(0)}(1,2)\rightarrow (1+d/a_0)$
in the limit $d/a_0\rightarrow 0$.
The value of $r_s(1,2)$ at very large layer separations
($d/a_0\rightarrow\infty$) can be obtained by setting
$r_s(1,2)/r_s^{(0)}(1,2)=0$ in the left-hand side of Eq.~(\ref{eq:1to2})
and then solving the resulting equation numerically for $z$.
The result is $z\rightarrow 4.015$, or
$r_s(1,2)/r_s^{(0)}(1,2)\rightarrow 0.495d/a_0$,
so that $r_s(1,2)\rightarrow 1.409d/a_0$.
We note that if interlayer exchange were ignored and the pseudospins
treated as Ising variables ($\sin\theta=0$), then the two- to one-component
MFA transition would occur at $r_s^{(0)}(1,2)(1+d/a_0)$
(for all values of $d/a_0$)
and would put all (spin-polarized) particles in a single layer.
Interlayer exchange makes the layer densities equal in the one-component
MFA state, and causes the two- to one-component MFA transition to occur
at a somewhat lower value of $r_s$ than would be predicted
using Ising pseudospins.

\vspace{-0.1in}
\subsection{Infinitesimal tunneling}
\label{sec:fintun}
\vspace{-0.1in}

In this subsection, we discuss the effects of very small interlayer tunneling.
The contribution of interlayer tunneling to the MFA energy per unit area
is given by $-2t(n_a-n_b)\sin\theta$.
Finite interlayer tunneling thus has two important effects.
First, it removes the Ising character of the spins by making
$\sin\theta>0$.
We discuss this in more detail in the next subsection.
Second, it always produces some degree of pseudospin polarization ($n_a-n_b>0$).
We have parametrized the dependence of the subband splitting $(n_a-n_b)$
on $t$ in terms of the pseudospin Stoner interaction parameter,
which we calculate in the second subsection below.

\vspace{-0.1in}
\subsubsection{Effect of tunneling on phase angle}
\label{sec:efftunnel}
\vspace{-0.1in}

We first consider the effect of interlayer tunneling on the interlayer
phase angle $\theta$.
Extremizing the energy per unit area [Eq.~(\ref{eq:etheta})] with respect
to $\theta$ for $t>0$ gives two possible solutions.
The solution
\begin{equation}
\cos\theta=0
\end{equation}
is always an extremum, and is a minimum whenever $\Gamma<1$,
which is true for all phases except $p=3$ near balance ($p_F\approx p_B$).
Of course, if $t$ were large enough to cause full pseudospin polarization
($n_a=n_T$), then $\cos\theta=0$.

If $\Gamma>1$ (three-component phase) then the solution
\begin{equation}
\sin\theta = \frac{t}{(\Gamma-1)e^2d(n_a-n_b)/2\epsilon}
\end{equation}
gives the correct minimum, provided that the right-hand side
is positive and does not exceed unity.
Even in the three-component phase, $t$ will increase $(n_a-n_b)$,
but since $(n_a-n_b)$ is already finite at $t=0$, $\sin\theta$
will be proportional to $t$, to first order in $t$.

The effect of including a small amount of interlayer tunneling
can often be described perturbatively, and produces a smooth increase
in the pseudospin polarization $(n_a-n_b)$ that is proportional to $t$.
However, the results obtained in the combined limits $t\rightarrow 0$
and $d\rightarrow 0$ can depend on the order of these limits.
For example, at $d=0$, the lowest-energy two-component phase has
$n_{a\uparrow}=n_{a\downarrow}=n_T/2$
(pseudospin-polarized but spin-unpolarized)
for arbitrarily small but finite $t$.
However, at $t=0$, the lowest-energy two-component phase has
$n_{a\uparrow}=n_{b\uparrow}=n_T/2$
(spin-polarized but pseudospin-unpolarized)
for arbitrarily small but finite $d$.
By comparing the MFA energies per unit area of two competing $p=2$ ground
states with $\cos\theta=0$ (spin-polarized $p=2a$ versus
pseudospin-polarized $p=2b$),
it can be shown that the pseudospin-polarized two-component ($p=2b$)
ground state requires
\begin{eqnarray}
\label{eq:tcrit}
t &>& \frac{e^2dn_T}{8\epsilon}\Gamma(n_{a\uparrow}=n_{a\downarrow}=n_T/2)
    = \frac{e^2dn_T}{8\epsilon}\Gamma_1(p=2)
\\ \nonumber &=& \frac{e^2dn_T}{8\epsilon}
\frac{16}{3\pi z}\left[1-S(z)\right]
\\ \nonumber && \quad \rightarrow \left\{
\begin{array}{ll}
e^2dn_T/16\epsilon , & k_Fd\rightarrow 0 \\
(4/3\sqrt{2\pi}) e^2\sqrt{n_T}/4\pi\epsilon , & k_Fd\rightarrow\infty .
\end{array}
\right.
\end{eqnarray}
Here $z=2k_Fd=2d\sqrt{2\pi n_T}$, and $S(z)$ is defined in
Eq.~(\ref{eq:sintegral}).
Therefore it is the size of $t$ relative to $e^2dn_T/16\epsilon$ that must
be considered as both $t$ and $d$ approach zero in the two-component phase.
A rough estimate of the minimum tunneling energy $t_c$ necessary to obtain
the pseudospin-polarized two-component phase may be made by calculating
$e^2dn_T/16\epsilon$ for the smallest value of $n_T$ that still gives
the two-component phase: i.e., for $r_s\sim r_s^{(0)}(1,2)\approx 2.844$.
For $d=10$ nm, this gives $t_c\sim 0.7$ meV for $n$-type GaAs and
$t_c\sim 13$ meV for $p$-type GaAs.
The differences between these two values of $t_c$ arise from the fact
that $n$-type GaAs spin-polarizes at a much lower density than $p$-GaAs,
due to the differences in the effective mass (and therefore in $r_s$.)
We stress that there exist two $p=2$ two-component states, which can
be either spin- or pseudospin-polarized.
The $p=2$ state that we focus on most will be the spin-polarized state
($p=2a$).

\vspace{-0.1in}
\subsubsection{Pseudsopin Stoner parameter}
\label{sec:stoner}
\vspace{-0.1in}

The difference $\Delta n \equiv n_a-n_b$ between the subband densities
(obtained from SdH measurements) of a balanced double-layer system
is often used as a measure of the size of the interlayer tuneling matrix
element $t$ by applying the formula \mbox{$n_a-n_b=(p/2)t\nu_0$},
which is valid for noninteracting particles.\cite{swier}
Here $p=4$ for spin-unpolarized particles,
$p=2$ for spin-polarized particles,
and $\nu_0=m^*/(\pi/\hbar^2)$ is the two-dimensional
density of states per unit area.
For interacting electrons or holes, the pseudospin
Stoner interaction parameter $I$ is defined through\cite{swier}
\begin{equation}
\label{eq:istoner}
n_a-n_b = \frac{(p/2)t\nu_0}{1-I} .
\end{equation}
For noninteracting particles, $I=0$.
For interacting particles, $I$ is a function of $r_s$ and $d/a_0$;
in general, it also a function of $t$.
We do not consider $p=3$ or $p=1$, since both the three- and the
one-component phases have $n_a>n_b$ at $t=0$, corresponding to
$I=1$.
The onset of SILC ($n_a>n_b$ and $\sin\theta>0$ even when $t=0$)
occurs at the two- to one-component transition, and
corresponds to $I\rightarrow 1$ for $p=2$.

The Stoner interaction parameter $I$ can be calculated analytically
in the limit of vanishing interlayer tunneling ($t\rightarrow 0$),
as we show in Sec. \ref{app:stoner} of the Appendix.
(For finite $t$, we calculate $I$ numerically.)
The basic idea is to start with equal subband densities ($n_a=n_b$)
and infinitesimally small $t$, and then calculate the change in energy
due to moving an infinitesimal amount of charge from subband $b$
to subband $a$.
Minimizing the change in energy with respect to the amount of charge
transferred between the subbands gives the linear response to small
interlayer tunneling and yields the following expression for
the $t\rightarrow 0$ limit of the Stoner interaction parameter:
\begin{eqnarray}
\label{eq:stonerfac}
I &=& \frac{\nu_0e^2}{2\pi\epsilon k_F} \left( 1 -
p \frac{\pi}{4} k_Fd \Gamma_0 \right)
\\ \nonumber
&=& \frac{\nu_0}{\pi} \left\lbrace 2V_{11}(2k_F) -
\int_0^1 dx \frac{[V_{11}(2k_Fx)-V_{12}(2k_Fx)]}{\sqrt{1-x^2}}
\right\rbrace \\ \nonumber
&=& \frac{r_s}{\pi} \sqrt{\frac{p}{2}}
\left[ 1 - \frac{1}{2} \int_0^{\pi/2} d\theta
\frac{\left(1 - e^{-2k_Fd\sin\theta}\right)}{\sin\theta} \right] ,
\end{eqnarray}
where $k_F=\sqrt{4\pi n_T/p}$ for $p=2,4$, and
$\Gamma_0\equiv \Gamma(n_a\rightarrow n_b)$.
Equation~(\ref{eq:stonerfac}) is equivalent to the generalized 
random-phase approximation (GRPA) result
in Eq.~(14) of Ref.~\onlinecite{swier}.

Note that $I=1$ (when $p=1$) corresponds to SILC, and that the GRPA result
for the phase boundary, which is shown as the dashed sloped line in
Fig.~\ref{fig:phdiag}, is different (has larger $r_s$) than the MFA result.
That the GRPA gives a higher value for $r_s(1,2)$ is not surprising,
given that the GRPA goes beyond MFA and contains correlation effects
in an approximate way.
To lowest order in $d$, \mbox{$I(d)/I(0)=(1-d/a_0)$}, so that
for $d\rightarrow 0$, the GRPA gives a higher critical value
of the interparticle spacing for SILC than the MFA:
$r_s(1,2)\approx\pi(1+d/a_0)$.
As expected,
a similar calculation of the linear response of the real spins to a weak
Zeeman field shows that a hypothetical four- to two-component GRPA transition
would occur at twice the density of the $d=0$ two- to one-component
transition, i.e., at $r_s(1,2)=\pi/\sqrt{2}$.
However, as with the MFA, we expect that a three-component GRPA phase
preempts any direct four- to two-component GRPA transition, and that
a three-component is always obtained within the GRPA at $r_s=\pi/\sqrt{2}$.

As pointed out in Ref.~\onlinecite{swier}, interactions enhance
the subband splitting ($I>0$) for $k_Fd<1.13$, but reduce the
splitting ($I<0$) for $k_Fd>1.13$.
The critical value for $r_s$ at large $d$ that separates the
one- and two-component phases is determined by
solving Eq.~(\ref{eq:stonerfac}) for $I=1$ and $r_s\rightarrow\infty$.
Asymptotically (for $r_s,d/a_0\rightarrow\infty$), it occurs at the same
value of $r_s$ that has $I=0$  i.e.,
$k_Fd\rightarrow 1.134$ or $r_s(1,2)/r_s^{(0)}(1,2)\rightarrow 0.620d/a_0$,
which gives $r_s(1,2)\rightarrow 1.764 d/a_0$.
Within the GRPA, SILC occurs once the interparticle
spacing is roughly twice the interlayer spacing.
In the limit $k_Fd\rightarrow\infty$,
\mbox{$I(d)/I(0)=1-(1/2)[\ln(4k_Fd)+\gamma]$},
where $\gamma\approx0.5772$ is Euler's constant.
In Ref.~\onlinecite{swier} it is argued that although
$I(d)$ is large and negative as $d\rightarrow\infty$,
its apparently divergent behavior is an unphysical artifact of the GRPA.

\begin{figure}[h]
\epsfxsize3.5in
\centerline{\epsffile{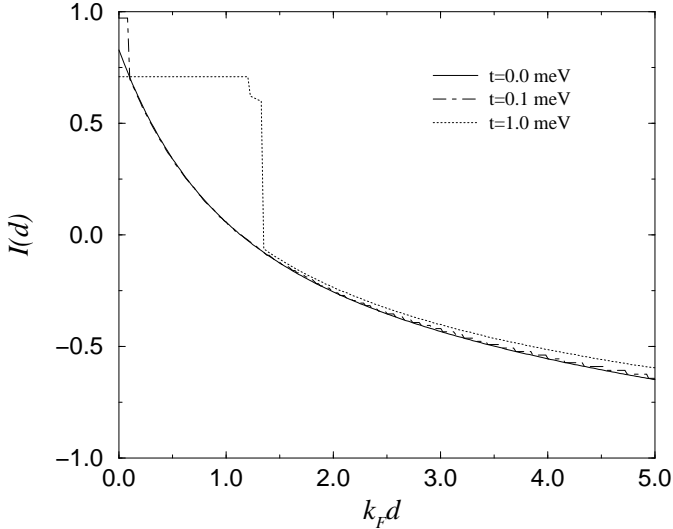}}
\caption{
Stoner interaction parameter $I=1-2t\nu_0/(n_a-n_b)$ versus $k_Fd$
for $n$-type GaAs at total density $n_T=10^{11}$cm$^{-2}$
for $t=0, 0.1, 1.0$ meV.
}
\label{fig:stoner07}
\end{figure}

\begin{figure}[h]
\epsfxsize3.5in
\centerline{\epsffile{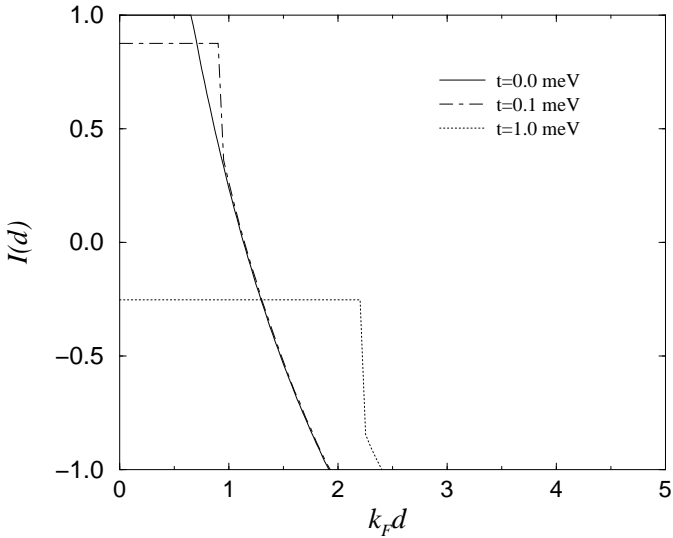}}
\caption{
Stoner interaction parameter $I=1-2t\nu_0/(n_a-n_b)$ versus $k_Fd$
for $p$-type GaAs at total density $n_T=10^{11}$cm$^{-2}$
for $t=0, 0.1, 1.0$ meV.
}
\label{fig:stoner30}
\end{figure}

We have calculated the Stoner interaction parameter $I$ for a
few values of the interlayer tunneling $t$ in
Figs. \ref{fig:stoner07} ($n$-type GaAs) and \ref{fig:stoner30} ($p$-type GaAs),
for a hypothetical sample with total density $n_T=10^{11} $cm$^{-2}$.
The complete polarization of the pseudospin ($n_a=n_T$) is indicated
by the mesa-like regions where $I$ becomes flat:
$n_{a\uparrow}=n_{a\downarrow}=n_T/2$ for $n$-type GaAs and
$n_{a_\uparrow}=n_T$ for $p$-type GaAs.
Increasing the size of $t$ favors pseudospin polarization
and allows it to persist to larger values of $k_Fd$.
Note that at fixed $t$, the Stoner interaction is equal to
$I_{\rm max}=1-2t\nu_0/n_T$ as $k_Fd\rightarrow 0$, so that
$I_{\rm max}$ decreases with increasing $t$.
The fact that $I_{\rm max}(t=0.1)>I_{\rm max}(t=0)$ is an artifact
of the order in which the limits $t\rightarrow 0$ and $d\rightarrow 0$
are taken: for any finite $t$, the two-component phase will have
$n_{a\uparrow}=n_{a\downarrow}=n_T/2$ as $d\rightarrow 0$, but
for any finite $d$, the two-component phase will have
$n_{a\uparrow}=n_{b\uparrow}=n_T/2$ as $t\rightarrow 0$.
For $n$-type GaAs with $t=1$ meV (dotted line in Fig.~\ref{fig:stoner07}),
the system goes through three phases as a function of $k_Fd$:
(1) a pseudospin-polarized two-component phase
($n_{a\uparrow}=n_{a\downarrow}=n_T/2$) for $k_Fd\rightarrow 0$,
(2) a three-component phase 
($n_{a\uparrow}=n_{a\downarrow}<n_{b\uparrow}$) for intermediate
values of $k_Fd$, indicated in by the ``missing piece'' on the
right side of the mesa in Fig.~\ref{fig:stoner07},
(3) a real-spin-polarized two-component phase
($n_{a\uparrow}=n_{b\uparrow}=n_T/2$) for larger $k_Fd\rightarrow\infty$.

SILC is indicated by having $I(t=0)=1$ and occurs only for $p$-type GaAs
(Fig.~\ref{fig:stoner30}, for $k_Fd<0.7$):
$I(t=0)$ is directly proportional to $\nu_0$,
and therefore to the effective mass of the particles, so that SILC
is more likely to be observed in $p$-type ($m^*/m_e\approx 0.3$) rather
than $n$-type ($m^*/m_e\approx 0.07$) GaAs.

\vspace{-0.1in}
\section{Effect of Bias}
\label{sec:bias}
\vspace{-0.1in}

In this section, we study the effect of bias ($p_F \ne p_B$,
due to applied gate voltages) on the subband and layer densities.
The classical results
\begin{equation}
n_1 = p_F\Theta(p_F) , \qquad n_2 = p_B\Theta(p_B) ,
\end{equation}
and
\begin{equation}
\label{eq:vfsimple}
V_F = eD_FE_F + edE_{12} + eV_F^{(0)}
\end{equation}
give the asymptotically correct behavior for high layer densities
and large layer separations.
Double-layer systems at low densities and small layer separations
show measureable deviations from the classical behavior, most notably
because of quantum-mechanical exchange.
We shall find it convenient to study the effects of layer imbalance
by fixing the total density ($n_T=p_F+p_B$) and then varying
the gate-imbalance parameter $\zeta$, defined by
\begin{equation}
\label{eq:zeta}
\zeta \equiv \frac{p_F-p_B}{p_F+p_B} = (p_F-p_B)/n_T .
\end{equation}
The case of balanced gates ($p_F=p_B$) corresponds to $\zeta=0$.

In the presence of bias and/or tunneling, there are five possible
noncrystalline MFA ground states, which we write in order of
increasing $r_s$ in Table~\ref{table:phases}.
Only the last ($p=1$) phase can exhibit SILC.
The last two phases are pseudospin-polarized and can have
$0 \le \sin\theta \le 1$, with the value of $\theta$ being determined by
the layer separation, density, bias, and tunneling.
The first three phases are Ising-like ($\sin\theta=0$) when $t=0$,
and therefore do not involve interlayer exchange, in the absence
of interlayer tunneling.
Note that there are two $p=2$ phases, one that is pseudospin-polarized
but spin-unpolarized ($p=2b$), and another that is spin-polarized but
pseudospin unpolarized ($p=2a$).
The pseudospin-polarized $p=2b$ requires bias and/or tunneling.
For all MFA phases, the real spin is either fully polarized or completely
unpolarized.
\begin{table}
\caption{Subband occupancies of the five possible non-crystalline MFA
ground states for a double layer system with gate imbalance
($|p_F-p_B|>0$) or tunneling ($t>0$).}
\label{table:phases}
\begin{tabular}{|l|l|l|} \hline
$p=4$ & $n_{a\uparrow}=n_{a\downarrow}\ge n_{b\uparrow}=n_{b\downarrow} > 0$ &
             \\ \hline
$p=3$ & $n_{a\uparrow}=n_{a\downarrow}\ge n_{b\uparrow} > 0$ &
             $n_{b\downarrow}=0$ \\ \hline
$p=2a$ & $n_{a\uparrow} \ge n_{b\uparrow} > 0$ &
             $n_{a\downarrow}=n_{b\downarrow}=0$ \\ \hline
$p=2b$ & $n_{a\uparrow}=n_{a\downarrow} > 0$ &
             $n_{b\uparrow}=n_{b\downarrow}=0$ \\ \hline
$p=1$ & $n_{a\uparrow}=n_T$ &
             $n_{a\downarrow}=n_{b\uparrow}=n_{b\downarrow}=0$ \\ \hline
\end{tabular}
\end{table}

We begin this section by studying the case of vanishing interlayer exchange,
which is the relevant situation for the majority of double-layer samples
that have been studied experimentally, except in the
quantum Hall regime.\cite{murphy}
This is the case when layer separations are sufficiently far apart
that interlayer exchange is negligible in zero or weak magnetic fields.
A model without interlayer exchange is able to fit existing double-layer
data well, and obtains the correct four-, three-, and two-component phases,
although it fails to properly describe the one-component phase.
We then include the effects of interlayer exchange and solve
for the subband and layer densities using our MFA, which allows for
the possibility of interlayer exchange, even in the absence of
interlayer tunneling.
Finally, we give a full treatment of the one-component phase within
the MFA, including bias and tunneling.

\vspace{-0.1in}
\subsection{No interlayer exchange}
\label{sec:nox}
\vspace{-0.1in}

It is simplest to begin our study of the effects of layer imbalance
($p_F\ne p_B$) by first considering the limit of vanishing interlayer exchange.
This limit is relevant to most of double-layer samples that have been studied
experimentally, except in the quantum Hall regime.\cite{murphy}
It corresponds to layer separations that are sufficiently far apart
that interlayer exchange is negligible in zero or weak magnetic fields.
We shall also demonstrate that a very simple model which assumes that
the particles are always spin-unpolarized gives a good fit to existing data
on the subband occupancies of double-layer systems, except at low densities.

\vspace{-0.1in}
\subsubsection{No tunneling}
\label{sec:noxnotun}
\vspace{-0.1in}

Interlayer exchange is negligible when the layer separation is
large compared to the interparticle spacing.
This condition may be expressed in various ways, e.g.,
$k_Fd\gg 1$ or $r_s \ll d/a_0$, and is satisfied for most samples.
In this limit, we ignore interlayer tunneling and correlations and
write the exchange-correlation energy in Eq.~(\ref{eq:uovera}) as
\begin{equation}
\label{eq:esep}
\varepsilon(n_1,n_2) \approx \varepsilon(n_1) + \varepsilon(n_2) ,
\end{equation}
where $\varepsilon(n)$ is the sum of the kinetic (Fermi), exchange,
and correlation (but {\em not} the electrostatic) energies,
for a single-layer two-dimensional electron gas (2DEG) of density $n$.
In the absence of interlayer exchange and tunneling, we can
work directly with the layer densities rather than the subband
densities because the pseudospins are Ising variables:
\begin{eqnarray}
n_a &=& \max(n_1,n_2) , \quad n_b = \min(n_1,n_2) , \\ \nonumber
\theta &=& \pi \Theta(p_B-p_F) =
\left\{
\begin{array}{ll}
\pi , & p_F<p_B \\
0 , & p_F>p_B
\end{array}
\right.
\end{eqnarray}
so that $\sin\theta=0$.

No interlayer exchange is required to correctly describe
the four-, three-, and two-component MFA phases at $t=0$,
since all have $\sin\theta=0$.
Our analysis of these phases therefore proceeds as before.
It is nonetheless useful to look at the phases of system in terms of 
the equilibrium and stability conditions of Sec.~\ref{sec:model},
which we do below.

Although it would be straightforward to include intralayer correlation-energy
effects in $\varepsilon(n)$, we do not do so here for the sake of simplicity.
Even so,
the resulting approximate model does a very good job of fitting SdH data
until the layer densities get so low that they violate
our initial assumption that $k_Fd$ is large.
We therefore begin with
\begin{eqnarray}
\label{eq:varepsn}
\varepsilon(n) &\approx&
\sum_s \left[\frac{n_s^2}{\nu_0}
-\frac{8}{3\sqrt{\pi}} \frac{e^2}{4\pi\epsilon} n_s^{3/2} \right]
\\ \nonumber &=&
\left\{
\begin{array}{ll}
n^2/2\nu_0
-(8/3\sqrt{2\pi}) (e^2/4\pi\epsilon) n^{3/2} , & n>n_c \\
n^2/\nu_0
-(8/(3\sqrt{\pi}) (e^2/4\pi\epsilon) n^{3/2} , & n<n_c
\end{array}
\right. ,
\end{eqnarray}
where $n_c$ is the critical density for the MFA spin-polarization transition
in a single-layer system.
The condition $n>n_c$ corresponds to spin-unpolarized
electrons and occurs at higher densities, whereas
$n<n_c$ corresponds to spin-polarized electrons and occurs
at lower densities.
When $\sin\theta=0$ (which is the case we are considering in this section),
$n_c$ is the critical density of the MFA (or the HFA) spin-polarization
transition for a single layer, which occurs when the single-layer $r_s$
has the value $r_s(2,4)=(3\pi/8)(1+\sqrt{1/2})\approx 2.011$,
shown as the dashed line in Fig.~\ref{fig:phdiag}.
Note that $r_s\equiv 1/\sqrt{\pi na_0^2}$  for a single-layer system
of number density $n$, which gives
$n_ca_0^2=(2/3)(4/\pi)^3(1-2\sqrt{2}/3)\approx 0.078 70$.
In the MFA, which is equivalent to the HFA for balanced layers,
the spin polarization is either completely unpolarized (at higher densities)
or completely polarized (at lower densities).
Correlation-energy effects probably produce a range of intermediate
spin polarizations.

The chemical potential measured relative to layer $i$ is $\mu_i=\mu(n_i)$,
where
\begin{eqnarray}
\label{eq:easymu}
\mu(n) &=& \partial\varepsilon(n)/\partial n \\ \nonumber &=&
\left\{
\begin{array}{ll}
n/\nu_0 - (4/\sqrt{2\pi}) (e^2/4\pi\epsilon) \sqrt{n} ,
& n>n_c \\
2n/\nu_0 - (4/\sqrt{\pi}) (e^2/4\pi\epsilon) \sqrt{n} ,
& n<n_c
\end{array}
\right.
\end{eqnarray}
where we have used Eqs. (\ref{eq:mui}) and (\ref{eq:varepsn}).
The values of the layer densities can be determined by using
Eq.~(\ref{eq:easymu}) in the equilibrium condition of Eq.~(\ref{eq:equilib}),
\mbox{$\mu_1-\mu_2=eE_{12}d$}.

The electronic lengths $s_{ij}$ that determine the Eisenstein ratio $R_E$
[Eq.~(\ref{eq:regen})]
and the condition for stability against interlayer charge transfer
[Eq.~(\ref{eq:stability})]
can be calculated from Eqs. (\ref{eq:sij}) and (\ref{eq:easymu}).
Ignoring interlayer correlations as in Eq.~(\ref{eq:esep}), gives
$s_{ij}=0$ for $i\ne j$, and $s_i=s_{ii}=s(n_i)$, where
\begin{eqnarray}
\label{eq:sn}
\frac{s(n)}{a_0} = \frac{\epsilon}{e^2a_0}
\frac{\partial\mu(n)}{\partial n} &=&
\left\{
\begin{array}{ll}
1/4 - (\sqrt{2/\pi})/(4\pi a_0\sqrt{n}) , & n>n_c \\
1/2 - (2/\sqrt{\pi})/(4\pi a_0\sqrt{n}) , & n<n_c
\end{array}
\right.
\\ \nonumber &=&
\left\{
\begin{array}{ll}
1/4 - (\sqrt{2}/4\pi)r_s , & n>n_c \\
1/2 - (1/2\pi)r_s , & n<n_c
\end{array}
\right. .
\end{eqnarray}
Note that the MFA compressibility $\kappa$ can be calculated from the above
result using Eq.~(\ref{eq:sii}), and that $s(n)$ and therefore $\kappa$
are negative at sufficiently low densities.
The length $s(n)$ and the compressibility $\kappa$ jump discontinuously
at $n=n_c$, where the ground state changes abruptly from spin-unpolarized
to spin-polarized.
For densities just above $n_c$, $s(n_c^+)/a_0\approx 0.0237$, while
for densities just below $n_c$, $s(n_c^-)/a_0\approx 0.1799$.
Sensitive measurements of the interlayer capacitance
(e.g., the Eisenstein ratio $R_E$) could detect the exchange-driven
spin polarization of a 2DEG through its effect on the compressibility,
especially in $p$-type GaAs samples when the density in a
layer could be made small enough to polarize the holes in
that layer.

It is straightforward to calculate the effect of spin polarization
in a low-density layer on $R_E$ using Eqs. (\ref{eq:repb}) and (\ref{eq:sn}).
For the usual case in which the interlayer separation $d$
is substantially larger than the electronic lengths $s(n_i)$
(i.e., for $d/a_0\gg 0.2$), the MFA gives an abrupt jump in $R_E$
by almost a factor of 8, from approximately $0.0237a_0/d$
for densities just above $n_c$, to approximately $0.1799a_0/d$
for densities just below $n_c$.
Of course, the MFA overestimates the size of the jump, but it is nonetheless
plausible that measurements of $R_E$ could detect changes in the
compressibility of a low-density layer due to the exchange-driven
polarization of the spins.

\vspace{-0.1in}
\subsubsection{Unpolarized spins}
\label{sec:unpol}
\vspace{-0.1in}

\begin{figure}[h]
\epsfxsize3.5in
\centerline{\epsffile{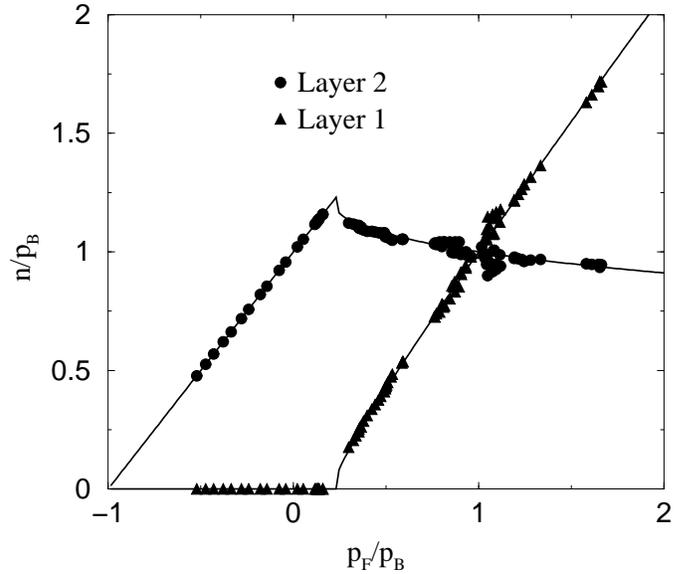}}
\caption{
Plot of the experimental (circles and triangles) and calculated (solid curves)
layer densities $n_1$ and $n_2$ versus front-gate number density $p_F$.
A simple model which neglects interlayer exchange, intralayer correlations,
and spin polarization and assumes zero layer thickness yields a good fit
to the experimental densities, except when $n_1$ becomes small.
}
\label{fig:finfig}
\end{figure}

Ruden and Wu assumed not only that the pseudospins were Ising-like
($\sin\theta=0$), but also that the real spins
were always unpolarized.\cite{ruden}
This limited the phases they found at $p_F=p_B$ to two:
pseudospin-unpolarized ($p=4$) at high density and pseudospin-polarized
($p=2b$) for low density.
It is straightforward to compare the energy of the four-component
phase with that of the hypothetical pseudospin-polarized ($p=2b$) phase of
Ruden and Wu and show that they are equal when
$r_s/r_s^{(0)}(1,2)=1+2d/a_0$.
Although neither assumption was, strictly speaking, correct, it is
an interesting and useful fact that making such assumptions can yield
a simple model that fits experimental data for layer densities versus
gate bias quite well, except at the lowest densities.
Figure~\ref{fig:finfig} shows experimental SdH data\cite{hamilton}
and a theoretical fit from a simple theory that ignores interlayer
exchange and takes the spins to be unpolarized.
The value of the interlayer separation $d$ used in the model is taken
to be a fitting parameter.
The values of $d$ that we obtain with this simplified model
always locate the idealized
two-dimensional electrons layers inside the confining quantum wells,
although $d$ always seems to be somewhat larger than the
midwell to midwell distance.
Note that the experimental data fits very well almost everywhere,
except where the density $n_1$ is very small when layer 1 is near depletion.
Here the simplest theory (which omits the possibilities of spin and
pseudospin polarization) erroneously predicts an interlayer charge transfer
instability.
As we discussed in Sec.~\ref{sec:model}, such an instability is unavoidable
when interlayer correlations ($s_{12}$) are neglected.
According to the stability criterion of Eq.~(\ref{eq:stability}),
the interlayer charge-transfer instability should occur in the
simplest theory (spin-unpolarized, no interlayer exchange) when
\begin{equation}
\label{eq:jcnew}
r_{s1} \ge \pi\sqrt{2}(1+2d/a_0) - r_{s2} ,
\end{equation}
where $r_{si}=1/\sqrt{\pi n_ia_0^2}$
is the dimensionless interparticle separation in layer $i$,
and we have taken $n_1\le n_2$ (or \mbox{$r_{s1} \ge r_{s2}$}).
At balance, setting $r_{s1}=r_{s2}$ in Eq. (\ref{eq:jcnew})
gives $r_s=(\pi/\sqrt{2})(1+2d/a_0)$
as the  GRPA version of the critical particle separation for Ruden and Wu's
hypothetical pseudospin-polarization ($p=4\rightarrow p=2b$) transition.
[In the MFA, $r_s(2b,4)=r_s^{(0)}(2,4)(1+2d/a_0)$ gives the Ruden-Wu
hypothetical pseudospin-polarization transition at balance.]

Note that even in the limit of equal layer densities and zero layer separation,
\mbox{$r_{s1}\ge\sqrt{\pi}/2$}, which is the GRPA value of the particle
separation for a single layer to spin polarize.
Thus even in a theory that neglects interlayer correlations, 
the particles in the lower-density layer (or both layers, if they
are balanced) spin-polarize before the layer empties out.
As noted in Ref.~\onlinecite{zheng}, the spin polarization of the electrons
predicted by the HFA was ignored by Ruden and Wu.\cite{ruden}
However, including the spin polarization does not eliminate the
interlayer charge transfer instability, which according to
Eqs. (\ref{eq:stability}) and (\ref{eq:sn}) would occur at
\begin{equation}
r_{s1} > 2\pi(3/4+d/a_0) - r_{s2}/\sqrt{2} ,
\end{equation}
when layer 1 is spin-polarized but layer 2 is not, and at
\begin{equation}
r_{s1} \ge 2\pi(1+d/a_0) - r_{s2} ,
\end{equation}
when both layers are spin polarized, if interlayer correlations
could be ignored.

\vspace{-0.1in}
\subsubsection{LDF model}
\label{sec:noxyest}
\vspace{-0.1in}

We now introduce a tight-binding local density functional (LDF) model,
which includes the effects of interlayer tunneling in a simple way.
We shall follow the previous section and ignore for now the effects
of interlayer exchange, and we shall even take the electron densities
to be always spin-unpolarized.
Such an elementary  model is capable of fitting experimental data quite
well, despite its simplicity.

The Kohn-Sham single-particle equations for our tight-binding LDF model
is conveniently expressed as a $2 \times 2$ matrix equation:
\begin{equation}
\label{eq:hmatrix}
\left(
\begin{array}{cc}
\epsilon_1 & -t \\
-t & \epsilon_2
\end{array}
\right)
\left(
\begin{array}{c}
z_1^{(\lambda)} \\
z_2^{(\lambda)}
\end{array}
\right)
= E_\lambda
\left(
\begin{array}{c}
z_1^{(\lambda)} \\
z_2^{(\lambda)}
\end{array}
\right) ,
\end{equation}
where
\begin{equation}
\label{eq:epsj}
\epsilon_j = (-1)^j\frac{1}{2}eE_{12}d + \mu_{xc}(n_j) 
\end{equation}
represents the ``on-site'' energy of layer $j$,
and the tunneling matrix element $-t$ is off-diagonal.
The amplitude of the wave function for subband $\lambda$
($\lambda=a,b$) in layer $j$ is $z_j^{(\lambda)}$,
and the subband energy is $E_\lambda$.
The Hartree contribution to the on-site energy enters via the
interlayer electric field $E_{12}$, as shown in Eq.~(\ref{eq:gauss}).
The intralayer exchange and correlation contributions to the
on-site energy are given by the exchange-correlation potential $\mu_{xc}$.
In LDF theory, $\mu_{xc}(n)$ is equal to the derivative, with respect
to density, of the exchange plus correlation energies per unit area
of a two-dimensional single-layer system of uniform areal density $n$.
Equivalently, $\mu_{xc}(n_j)$ is equal to $\mu_j$ [see Eq.~(\ref{eq:mui})]
minus the kinetic energy contribution to $\mu_j$.
For simplicity, we do not include intralayer correlation energy contributions
to $\mu_{xc}$, so we write
\begin{equation}
\label{eq:muxc}
\mu_{xc}(n) \approx -\frac{4}{\sqrt{\pi}} \frac{e^2}{4\pi\epsilon}\sqrt{n} .
\end{equation}

The density in layer $j$ is given by
\begin{equation}
\label{eq:nj}
n_j = \sum_{\lambda=1}^2 N_\lambda |z_j^{(\lambda)}|^2 ,
\end{equation}
where
\begin{equation}
\label{eq:nlambda}
N_\lambda\equiv (E_F-E_\lambda) \nu_0 \Theta(E_F-E_\lambda)
\end{equation}
is the areal-density contribution from subband $\lambda$,
and $\nu_0=m^*/\pi\hbar^2$ is the two-dimensional density of states
for noninteracting particles.
The self-consistency of the Kohn-Sham equations enters via
Eqs. (\ref{eq:nj}) and (\ref{eq:nlambda}), since the layer densities
$n_j$, together with the gate densities $p_\alpha$, determine the
interlayer electric field $E_{12}$ appearing in Eq.~(\ref{eq:epsj}).
The Fermi energy $E_F$ is chosen so that the sum of the subband densities
$N_\lambda$ is equal to the total density $p_F+p_B$.

\begin{figure}[h]
\epsfxsize3.5in
\centerline{\epsffile{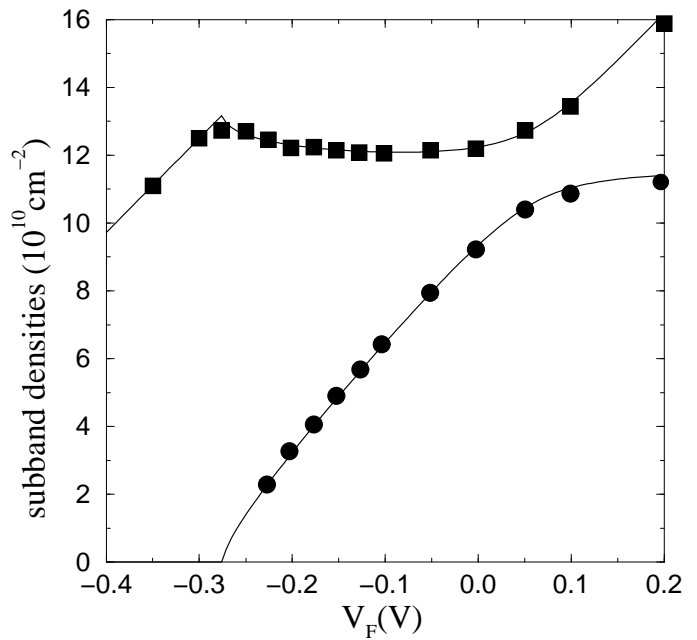}}
\caption{Experimental subband densities for $n_a$ (squares)
and $n_b$ (circles) versus front-gate voltage,
together with fit (solid curves) from tight-binding LDF model.
}
\label{fig:yinga}
\end{figure}

This simple LDF model, which takes the layers to have zero thickness
and assumes that the real spins are unpolarized,
is capable of fitting the experimental layer density data closely.
This is illustrated in Fig.~\ref{fig:yinga},
which shows SdH data taken from sample A of Ref.~\onlinecite{ying}.
The front-gate voltages used in Fig.~\ref{fig:yinga} were calculated
using Eq.~(\ref{eq:vfsimple}).

\vspace{-0.1in}
\subsection{Interlayer exchange}
\label{sec:incx}
\vspace{-0.1in}

We now allow for the possibility of interlayer exchange in biased systems.
We found that for balanced systems, interlayer exchange becomes
important only at low densities and small layer separations,
occuring only in the one-component phase.
In this section, we explore the effect of interlayer exchange
on biased double layers, and find that it can reduce or suppress interlayer
(although not intersubband) charge transfers.
We find that bias always increases pseudospin polarization and
sometimes reduces the total density required to achieve SILC.

\vspace{-0.1in}
\subsubsection{Interlayer phase angle}
\label{sec:xnotun}
\vspace{-0.1in}

When the interlayer tunneling $t$ is zero, it is possible to determine
the equilibrium value of $\theta$ that minimizes the total energy
per unit area, in terms of the equilibrium subband densities.
Extremizing Eq.~(\ref{eq:etheta}) with respect to $\theta$
for $n_a>n_b$ gives
\begin{eqnarray}
\label{eq:thsoln}
&& \cos\theta =
\left\lbrace \begin{array}{cc}
-1, & X \le -1 \\
X, & -1 \le X \le 1  \\
1, & X \ge 1 \\
\end{array} \right.
\\ \nonumber
&& X =
\frac{(p_F-p_B)/(n_a-n_b)}{1 - \Gamma} ,
\end{eqnarray}
where $\Gamma$ is defined in Eq.~(\ref{eq:gamma}).
In the four-component phase,
the kinetic energy dominates over the exchange energy,
and $(n_a-n_b)<|p_F-p_B|$, so that $\sin\theta=0$.
In practice, the only time that we find $|\cos\theta|<1$ (for $t=0$)
is in the one-component phase.
For Eq.~(\ref{eq:thsoln}) to minimize the energy with respect to $\theta$,
the second derivative of the energy with respect to $\theta$ must be
positive, which is equivalent to requiring that $\Gamma<1$.
If $\Gamma>1$ (which is true for the $p=3$ phase near balance),
then $\cos\theta=\pm 1$.

It follows from Eqs. (\ref{eq:n1n2}) and (\ref{eq:thsoln}) that
\begin{eqnarray}
\label{eq:nsoln}
(n_1-n_2) &=&
\left\lbrace \begin{array}{cc}
-(n_a-n_b), & X \le -1 \\
(p_F-p_B)/(1-\Gamma), & -1 \le X \le 1  \\
(n_a-n_b), & X \ge 1 \\
\end{array} \right.
\end{eqnarray}
In the special case of interest where $|X|<1$
(which requires $p=1$),
we can calculate the Eisenstein ratio $R_E$ at the point
where the layers are balanced:
\begin{equation}
\label{eq:retzero}
R_E(p_F=p_B) = -\frac{1}{2} (1-\delta p_B/\delta p_F)
\frac{\Gamma}{1-\Gamma} ,
\end{equation}
where we have used Eqs. (\ref{eq:re}) and (\ref{eq:nsoln}).
(Recall that $\delta p_B/\delta p_F\approx 0$ if the
back-gate voltage $V_B$ is kept fixed, but that
$\delta p_B/\delta p_F=-1$ if the total density is kept fixed.)

Within the tight-binding model of tunneling, the MFA model shows that
any finite value of the tunneling matrix element $t$ prohibits
either layer from completely emptying out, regardless of the gate
charges $(p_F,p_B)$.
Extremizing the energy per unit area [Eq.~(\ref{eq:etheta})] with
respect to $\theta$ for $t>0$ and $p_F\ne p_B$ shows that neither
$\sin\theta=0$ nor $\cos\theta=0$ are ever local extrema.
When $t$ is small and $p_F<<p_B$, including any negative values of $p_F$
(i.e., for $\zeta\le -1$, roughly speaking),
then extremizing the energy per unit area yields
\begin{equation}
\pi-\theta \approx \frac{t/(e^2dn_T/2\epsilon)}
                  {\left[|\zeta|-(1-\Gamma)\right]} ,
\end{equation}
to lowest order in $t/(e^2dn_T/2\epsilon)$.
This result is a local minimum provided that it is positive,
which is true in the limit we are considering here.
It follows from Eq.~(\ref{eq:n1n2}) that $n_1/n_T\approx(\phi/2)^2$,
so that $n_1\propto t^2$.
Hence $n_1$, although small, is always nonzero for $t>0$.
In actual samples, large bias changes the tunneling matrix element $t$.
Sufficiently large bias shifts the bottom (minimum energy) of the wells
relative to one another so greatly that $t$ can be driven
(for all practical purposes) to zero.

Interlayer exchange is significant only
when the layer densities and their separation are sufficiently small.
In order for interlayer exchange to contribute, there must be
(1) more particles in one subband than another ($n_a>n_b$), and
(2) nonzero $\sin\theta$ ($\theta\ne 0,\pi$).
Thus, for example, the case of balanced layers with $\theta=\pi/2$
at very low densities ($n_{a\uparrow}=n_T$) is a situation in which interlayer
exchange contributes strongly.
We discuss this case in Sec.~\ref{sec:onecomp}, and we find there
that interlayer exchange does indeed suppress interlayer charge
transfer.
For the case of unbalanced layers at high total density,
generally $n_a>n_b$; but when $t=0$, $\theta$ is usually equal
to $0$ or $\pi$, so that in the MFA, interlayer exchange does not contribute.
Near depletion, where one of the layers empties out, the situation
is not as clear, so we now analyze that situation in some detail later
below.

Even with interlayer exchange, it turns out that the MFA model is always
unstable with respect to an abrupt exchange-driven inter{\em subband} 
charge transfer from (low-density) subband $b$ to (higher-density) 
subband $a$ when the particle density in subband $b$ gets small enough.
The abrupt intersubband charge transfer is probably an unphysical feature
of the MFA model that is not observed in real experiments.
We believe that a proper treatment of the correlation energies
(which have been entirely omitted here) would help fix this shortcoming.
Nevertheless, we can still investigate what the MFA has to say about
{\em interlayer} charge transfer, since the subbands are
in general different from the layers, when we include interlayer exchange.

\vspace{-0.1in}
\subsubsection{Subbands densities nearly equal}
\label{sec:two3comp}
\vspace{-0.1in}

We now consider the limit in which
$\Delta n \equiv (n_a-n_b) \ll n_T$,
so that the double-layer system is only slightly pseudospin-polarized.
This will be the case for the two- ($p=2a$) and four- ($p=4$) component
ground states for small $t$ and $|p_F-p_B|$.
We begin by computing the change in the ground-state energy per unit area
due to changing $\Delta n$ from zero to a small but finite value.
Expanding Eq.~(\ref{eq:etheta}) to second order in $\Delta n$ gives
\begin{eqnarray}
\label{eq:deltae}
&& \frac{\Delta{\cal{E}}_0}{L_xL_y} =
\frac{(\Delta n)^2}{p\nu_0} - t\Delta n\sin\theta
- \frac{e^2}{4\pi\epsilon} \frac{(\Delta n)^2}{\sqrt{p\pi n_T}}
\\ \nonumber
&& + \frac{e^2d}{8\epsilon} \left[\Delta n\cos\theta - (p_F-p_B)\right]^2
+ \Gamma_0 \frac{e^2d}{2\epsilon} \left(\frac{\Delta n}{2}\right)^2
  \sin^2\theta ,
\end{eqnarray}
where
\begin{equation}
\Gamma_0 \equiv \lim_{n_a\rightarrow n_b} \Gamma .
\end{equation}
The quantity $\Gamma_0$ is calculated in Sec. \ref{app:gamma0} of
the Appendix.
Extremizing Eq.~(\ref{eq:deltae}) with respect to $\theta$ gives
\begin{eqnarray}
\label{eq:delthmin}
&& -2t\cos\theta + \frac{e^2d}{2\epsilon}(p_F-p_B)\sin\theta
\\ \nonumber
&& = (1-\Gamma_0) \frac{e^2d}{2\epsilon} \Delta n \sin\theta \cos\theta .
\end{eqnarray}
By taking the second derivative of Eq.~(\ref{eq:deltae}) with respect to
$\theta$, we find that Eq.~(\ref{eq:delthmin}) is a local minimum
provided that
\begin{eqnarray}
\label{eq:dele2}
&& 2t\sin\theta + \frac{e^2d}{2\epsilon}(p_F-p_B)\cos\theta
\\ \nonumber
&& - (1-\Gamma_0) \frac{e^2d}{2\epsilon} \Delta n \cos(2\theta)
> 0 .
\end{eqnarray}

Except for $p=3$ near balance, $0<\Gamma_0<1$, so that
\begin{equation}
0 < 1-\Gamma_0 < 1 .
\end{equation}
Hence Eqs. (\ref{eq:delthmin}) and (\ref{eq:dele2}) imply that
if $t>0$ but $|p_F-p_B|=0$, then $\cos\theta=0$ so that
$\theta=\pi/2$ (except for $p=3$.)
If, however, $|p_F-p_B|>0$ but $t=0$, then $\sin\theta=0$ so that
$\theta=0,\pi$ (except for $p=1$.)
There is thus a competition between the effects of $t$ and $|p_F-p_B|$.
If neither $t$ nor $|p_F-p_B|$ is zero, then $\sin\theta\ne 0$
and $\cos\theta\ne 0$.
In the limit that $(e^2d/2\epsilon)|p_F-p_B| \ll t$, then
\begin{equation}
\cos\theta \approx \frac{(e^2d/2\epsilon)(p_F-p_B)}{2t} \ll 1
\end{equation}
for $\Delta n \rightarrow 0$.
On the other hand,
in the limit that $t \ll (e^2d/2\epsilon)|p_F-p_B|$, then
\begin{equation}
\sin\theta \approx \frac{2t}
         {(e^2d/2\epsilon)[|p_F-p_B|-(1-\Gamma_0)\Delta n]} \ll 1 .
\end{equation}
In general, $\theta$ must be solved numerically.

Extremizing Eq.~(\ref{eq:deltae}) with respect to $\Delta n$ gives
\begin{equation}
\label{eq:deltn}
\Delta n = \frac{2t\sin\theta +
                    (e^2d/\epsilon)(p_F-p_B)\cos\theta}
{\frac{4}{p\nu_0} + \frac{e^2d}{2\epsilon}(\cos^2\theta+\Gamma_0\sin^2\theta)
 - \frac{e^2}{4\pi\epsilon} \frac{4}{\sqrt{p\pi n_T}}} ,
\end{equation}
which is a local minimum provided that its denominator is positive,
as may be seen by computing the second derivative of Eq.~(\ref{eq:deltae})
with respect to $\Delta n$.
Note that both interlayer tunneling ($t$) and gate bias ($|p_F-p_B|$)
produce pseudospin polarization (increase $\Delta n$.)

When $p_F=p_B$ and $t$ is (arbitrarily) small but nonzero
so that $\theta=\pi/2$, Eq.~(\ref{eq:deltn}) yields the pseudospin
Stoner enhancement factor in Eq.~(\ref{eq:stonerfac}).
This is discussed in more detail in Sec.~\ref{sec:stoner}.

When $\sin\theta=0$ (which requires $t=0$), then Eq.~(\ref{eq:deltn}) gives
\begin{equation}
\label{eq:delratio}
\frac{\Delta n}{|p_F-p_B|}
= \frac{d/a_0}{d/a_0 + (2/p)[1-r_s/(\pi\sqrt{2/p})]} .
\end{equation}
This is the case for the $p=2a$ and $p=4$ states for
$\Delta n/n_T \ll 1$.
It follows from this that near balance ($p_F\approx p_B$),
the Eisenstein ratio for fixed $p_B$ is given by
\begin{equation}
R_E = \frac{1}{2} \left[ 1 +
\frac{(p/2)d/a_0}{1-r_s/(\pi\sqrt{2/p})} \right]^{-1} ,
\end{equation}
where we have use of Eq.~(\ref{eq:re}).
Equation (\ref{eq:delratio}) says that, when $\sin\theta=0$,
then $\Delta n < |p_F-p_B|$ for small $\Delta n/n_T$, provided that
\begin{equation}
\label{eq:rsgt}
r_s < \pi\sqrt{2/p} .
\end{equation}
Now, the GRPA estimate of the interparticle separation required for
spin polarization in a single two-dimensional layer is
$r_s=\pi/\sqrt{2}$, which is just the right-hand side of Eq.~(\ref{eq:rsgt})
for $p=4$.
We therefore expect that, as long as the ground state has four components,
it will be true that $\Delta n < |p_F-p_B|$, and this is indeed what
our MFA calculations find for small $r_s$.

According to Eq.~(\ref{eq:delratio}),
$\Delta n < |p_F-p_B|$
for small $\Delta n/n_T$ for $p=2a$ until
$r_s>\pi$.
The interparticle separation $r_s=\pi$ is also the GRPA estimate
of $r_s^{(0)}(1,2)$ required for pseudospin polarization.
Thus for $d>0$ we expect that at higher densities (small $r_s$),
$\Delta n < |p_F-p_B|$ throughout the $p=4$ state and in the
low-$r_s$ region of the $p=2a$ state,
but that $\Delta n > |p_F-p_B|$ for the high-$r_s$ region of the $p=2a$ state,
at least for small $\Delta n/n_T$.
This is in fact what our MFA calculations show.
It is also true that in the $p=3$ phase (which has $\Delta n > 0$
even when $p_F=p_B$),
$\Delta n > |p_F-p_B|$ for $\zeta\ll 1$,
although not for $\zeta$ on the order of one.
Of course, for $r_s$ sufficiently small, pseudospin polarization occurs
so that $\Delta n = n_T > |p_F-p_B|$.

\begin{figure}[h]
\epsfxsize3.5in
\centerline{\epsffile{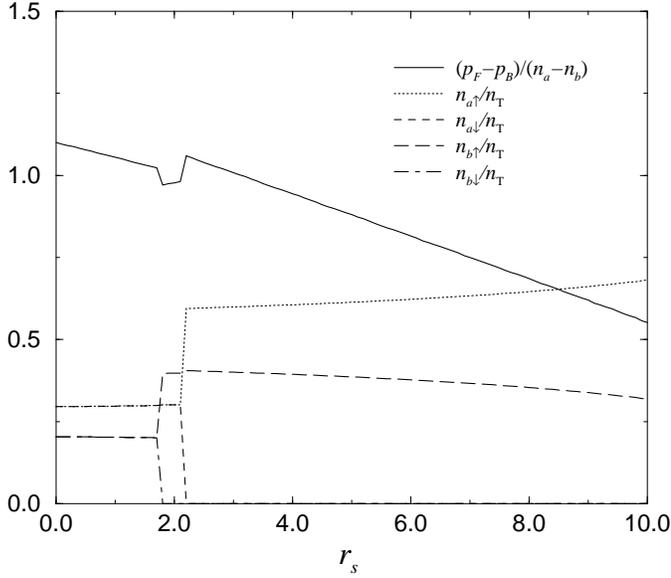}}
\caption{Normalized subband densities and the ratio $(p_F-p_B)/(n_a-n_B)$,
versus average interparticler separation $r_s$,
for fixed $d/a_0=5$ and fixed gate imbalance parameter $\zeta=0.2$.
The $p=4$, $p=3$, and $p=2a$ phases are obtained successively as $r_s$ is
increased.
}
\label{fig:zeta02}
\end{figure}

Figure \ref{fig:zeta02} shows a plot of the subband densities $n_{\alpha s}$
and the ratio $(p_F-p_B/(n_a-n_b)$ versus $r_s$ for fixed layer separation
$d/a_0=5$ and fixed layer imbalance $\zeta=0.2$.
For small $r_s$, the $p=4$ phase is obtained, and $(p_F-p_B)/(n_a-n_B)>1$.
For $r_s\approx 2.011$, the $p=4$ phase is obtained,
and $(p_F-p_B)/(n_a-n_B)<1$.
For larger $r_s$, the spin-polarized $p=2a$ phase is obtained.
Note that for $p=2a$ the ratio $(p_F-p_B)/(n_a-n_B)$ is larger than one
for smaller $r_s$, but smaller than one for larger $r_s$.
For even larger $r_s$ (not shown), the $p=1$ phase would be obtained
with $(p_F-p_B)/(n_a-n_B)=\zeta$, which in this case has $\zeta=0.2$.

If the denominator in Eq.~(\ref{eq:deltn}) is not positive, then
the global minimum for the energy per unit area occurs for $\Delta n=n_T$,
corresponding to pseudospin polarization.
Thus the GRPA condition for stability against abrupt {\em intersubband}
charge transfer is just the condition that denominator in
Eq.~(\ref{eq:deltn}) be positive, or equivalently
\begin{equation}
\label{eq:substable}
\frac{d}{a_0} +
\frac{(2/p)[1-r_s/(\pi\sqrt{2/p})]}
     {(\cos^2\theta + \Gamma_0\sin^2\theta)} > 0.
\end{equation}
When $\sin\theta=0$ (which requires $t=0$) so that the pseudospin
is Ising-like and interlayer exchange does not contribute,
then Eq.~(\ref{eq:substable}) is equivalent to the stability condition
against abrupt {\em interlayer} charge transfer given in
Eq.~(\ref{eq:stability}), when the electron lengths $s(n)$
are approximated by Eq.~(\ref{eq:sn}).
When $\cos\theta=0$ (e.g., when $p_F=p_B$ and $t$ has any finite positive
value), then the violation of the inequality in Eq.~(\ref{eq:substable})
is equivalent to the condition that the pseudospin Stoner ehancement
factor $I$, given in Eq.~(\ref{eq:stonerfac}), is equal to one, which
signals the transition to pseudospin ferromagnetism and SILC.
Because $0<\Gamma_0<1$, gate imbalance ($|p_F-p_B|>0$, so that
$\sin^2\theta$ decreases and $\cos^2\theta$ increases) makes the
double-layer system more unstable towards pseudospin polarization.

\vspace{-0.1in}
\subsubsection{Subband densities versus bias}
\label{sec:biasplots}
\vspace{-0.1in}

In this section, we show some illustrative calculations of the effect
of layer imbalance.
We plot the subband densities $n_{\alpha s}$ and the value of $\Gamma$
versus the gate imbalance parameter $\zeta = (p_F-p_B)/n_T$ at
fixed layer separation $d/a_0=5$ assuming zero interlayer tunneling ($t=0$),
for different values of $r_s$.
We find that if we fix layer density (or equivalently, $r_s$) and
vary the gate imbalance parameter $\zeta$, then (for $t=0$)
there are six distinct patterns of transitions between noncrystalline
MFA ground states.
We list the six possibilities below, in order of increasing layer imbalance
(beginning at $\zeta=0$) from left to right, and in order of increasing
average interparticle separation per layer $r_s$ from top to bottom:
\begin{enumerate}
	\item $r_s < r_s(3,4)$: \\
$(p=4) \rightarrow (p=3) \rightarrow (p=2b)$, \\
all with $\sin\theta=0$
	\item $r_s(3,4) < r_s < r_s(2,3)$: \\
$(p=3) \rightarrow (p=2b)$, \\
all with $\sin\theta=0$
	\item $r_s(2,3) < r_s < \sqrt{2}r_s(2,4)$: \\
$(p=2a) \rightarrow (p=3)
\rightarrow (p=2b)$, \\
all with $\sin\theta=0$
	\item $\sqrt{2}r_s(2,4) < r_s < r_s(1,2)$ and 
$\zeta_c>(1-\Gamma_1)$: \\
$(p=2a) \rightarrow (p=1$), \\
all with $\sin\theta=0$
	\item $\sqrt{2}r_s(2,4) < r_s < r_s(1,2)$ and 
$\zeta_c<(1-\Gamma_1)$: \\
$(p=2a) \rightarrow (p=1)$, \\
with $\sin\theta>0$ for $\zeta_c < \zeta < (1-\Gamma_1)$
	\item $r_s > r_s(1,2)$: \\
$(p=1)$ only, \\
with $\sin\theta>0$ for $0 \le \zeta < (1-\Gamma_1)$
\end{enumerate}

Here $(p=2a)$ denotes the spin-polarized two-component state,
$(p=2b)$ denotes the pseudospin-polarized two-component state,
and $\zeta_c$ is the value of the gate imbalance parameter $\zeta$
at which the  $(p=2a)\rightarrow (p=1)$ transition occurs.
The quantity $\sqrt{2}r_s(2,4)$ appearing in cases (4) and (5) above
correspond to the critical density $n_c$ for the MFA spin polarization
transition for a single layer, expressed in terms of the average
interparticle spacing per layer: $1/\sqrt{\pi(n_c/2)a_0^2}$.
(Note that for the same total density $n_T$, a double-layer system has
an average layer $r_s=1/\sqrt{\pi(n_T/2)a_0^2}$ that is $\sqrt{2}$
larger than the single-layer $r_s=1/\sqrt{\pi n_Ta_0^2}$.)
The quantity $n_c$ is discussed below Eq.~(\ref{eq:varepsn}).
We shall illustrate the first four of these possibilities in the remainder
of this section, and discuss the last two possibilities (which exhibit
SILC) in Sec.~\ref{sec:onecomp}.
It is evident from the above list that SILC, which requires $\sin\theta\ne 0$,
occurs only for $p=1$.

\begin{figure}[h]
\epsfxsize3.5in
\centerline{\epsffile{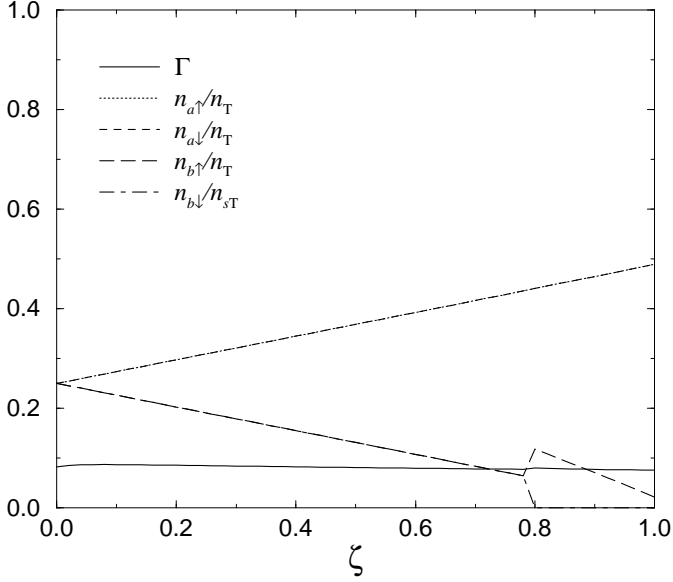}}
\caption{Normalized subband densities and interlayer exchange parameter
$\Gamma$, versus layer imbalance parameter $\zeta=(p_F-p_B)/n_T$,
for $d/a_0=5$ and $r_s=1$.
The $p=4$ state is obtained except for $\zeta$ near one, where
the $p=3$ state is obtained.
Beyond $\zeta=1$ (not shown) the $p=2b$ state is obtained.
}
\label{fig:rs1}
\end{figure}

Figure \ref{fig:rs1} is an example of case (1), with $r_s=1$.
This gives a four-component ($p=4$) phase when the gates are balanced
($\zeta=0$), and
maintains a $p=4$ state for most of the range of $\zeta$,
followed by a $p=3$ state for $\zeta$ near one.
If we were to increase $\zeta$ beyond one (not shown),
corresponding to $p_B<0$, then a pseudospin-polarized $p=2b$ state
would be obtained.

\begin{figure}[h]
\epsfxsize3.5in
\centerline{\epsffile{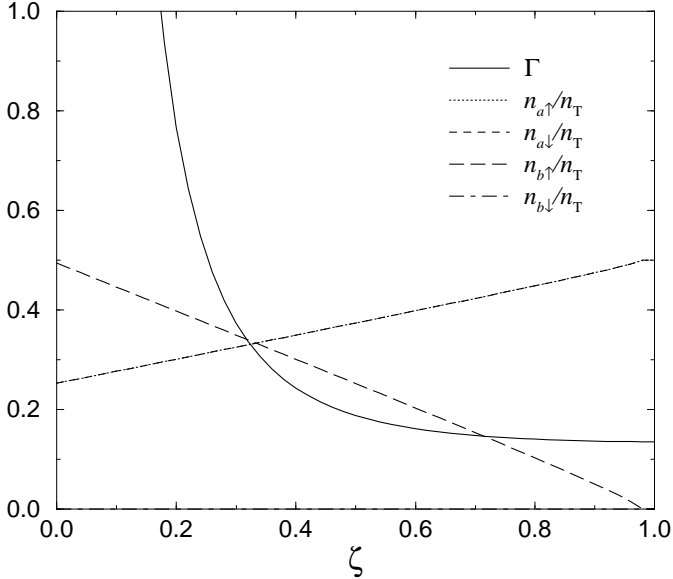}}
\caption{Normalized subband densities and interlayer exchange parameter
$\Gamma$, versus layer imbalance parameter $\zeta=(p_F-p_B)/n_T$,
for $d/a_0=5$ and $r_s=r_s(2,4)\approx 2.011$.
The $p=3$ state is obtained except for $\zeta$ near one, where
the pseudospin-polarized $p=2$ state is obtained.
}
\label{fig:rs2p011}
\end{figure}

Figure \ref{fig:rs2p011} is an example of case (2), with
$r_s=r_s(2,4)\approx 2.011$.
This gives a three-component ($p=3$) state when the gates are balanced
($\zeta=0$), and
maintains a $p=3$ state for most of the range of $\zeta$,
followed by a pseudospin-polarized $p=2b$ state for $\zeta$ near one
and beyond.
Note that for small $\zeta$, $\Gamma(p=3)>1$.

\begin{figure}[h]
\epsfxsize3.5in
\centerline{\epsffile{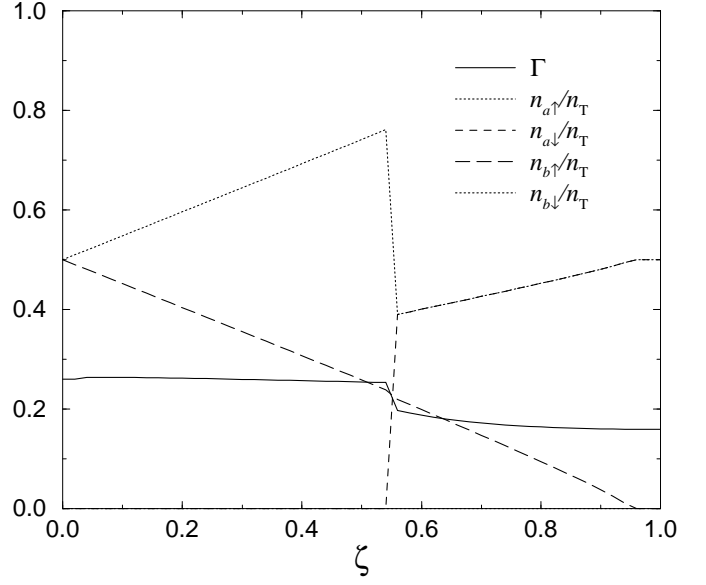}}
\caption{Normalized subband densities and interlayer exchange parameter
$\Gamma$, versus layer imbalance parameter $\zeta=(p_F-p_B)/n_T$,
for $d/a_0=5$ and $r_s=2.5$.
The $p=2a$ state is obtained for $\zeta<0.6$,
followed by a $p=3$ state for $0.6<\zeta<0.8$, and
then a pseudospin-polarized $p=2b$ state for $\zeta>0.8$.
}
\label{fig:rs2p5}
\end{figure}

Figure \ref{fig:rs2p5} is an example of case (3), with $r_s=2.5$.
This gives a spin-polarized two-component ($p=2a$) state when the gates
are balanced ($\zeta=0$), and
maintains a $p=2a$ state for $\zeta<0.55$,
followed by a $p=3$ state for $0.55<\zeta<0.95$, and
then a pseudospin-polarized $p=2b$ state for $\zeta>0.95$.

\begin{figure}[h]
\epsfxsize3.5in
\centerline{\epsffile{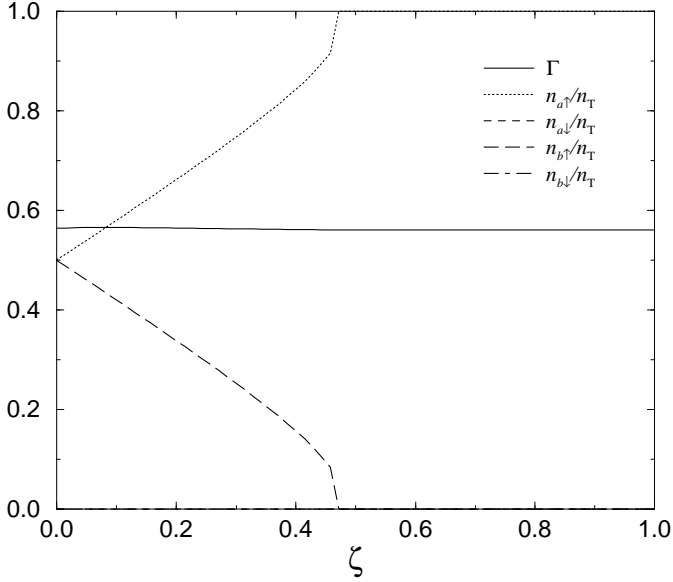}}
\caption{Normalized subband densities and interlayer exchange parameter
$\Gamma$, versus layer imbalance parameter $\zeta=(p_F-p_B)/n_T$,
for $d/a_0=5$ and $r_s=9$.
The $p=2a$ state is obtained for $\zeta<\zeta_c\approx 0.45$,
followed by a $p=1$ state (without SILC) for $\zeta>\zeta_c$.
}
\label{fig:rs9}
\end{figure}

Figure \ref{fig:rs9} is an example of case (4), with $r_s=9$.
This gives a spin-polarized two-component ($p=2a$) state when the gates
are balanced ($\zeta=0$), and
maintains a $p=2a$ state for $\zeta<\zeta_c\approx 0.45$,
followed by a  one-component ($p=1$) state for $\zeta>\zeta_c$.
Because $\zeta_c>1-\Gamma_1$, $\sin\theta=0$ throughout, and
thus no SILC is found.
In the next section, we consider values of $r_s$ large enough that
the $p=1$ state is achieved for $\zeta_c<1-\Gamma_1$,
thereby producing SILC.

\vspace{-0.1in}
\section{Pseudospin-Polarized States}
\label{sec:onecomp}
\vspace{-0.1in}

In this section, we consider the case in which all the particles
are in the lowest-energy subband ($n_a=n_T$), corresponding to full
pseudospin polarization.
There are two types of pseudospin-polarized MFA ground states:
spin-unpolarized ($p=2b$), and spin-polarized ($p=1$).
The spin-unpolarized case requires either interlayer tunneling
($t>0$) or gate imbalance $|p_F-p_B|>0$, or both.
In the absence of tunneling, the pseudospin-polarized $p=2b$ state
has $\sin\theta=0$ (Ising-like pseudospin), and
occurs whenever the total density and the layer imbalance are sufficiently
large.
If the tunneling $t$ is sufficiently large and the total density $n_T$
is not too small, then it is possible in principle to obtain a
$p=2b$ state with $\cos\theta=0$, for $p_F=p_B$.
Tunneling also reduces the value of $r_s$ required to achieve the $p=1$
state.
Because the pseudospin-polarized $p=2b$ MFA state does not occur without
bias or tunneling,
we shall focus mainly on the spin-polarized one-component ($p=1$) phase,
which can in principle arise without bias or tunneling.
The one-component phase is especially interesting because it
can occur as a broken-symmetry ground state of a double-layer system
in the absence of tunneling or layer imbalance, at very
small particle densities and layer separations.\cite{zheng}

When $n_a=n_T$, the ground-state energy (\ref{eq:etheta}) becomes
\begin{eqnarray}
\label{eq:esmalld}
&& \frac{{\cal{E}}_0}{L_xL_y} =
\frac{n_T^2}{p\nu_0} - tn_T\sin\theta -
\frac{8}{3\sqrt{p\pi}} \frac{e^2}{4\pi\epsilon} n_T^{3/2}
\\ \nonumber &+&
\frac{e^2dn_T^2}{8\epsilon} (\cos\theta-\zeta)^2 +
\frac{\sin^2\!\theta}{4} \frac{e^2dn_T^2}{2\epsilon} \Gamma_1 ,
\end{eqnarray}
where $\Gamma_1 = \Gamma(n_a=n_T)$ and $\zeta=(p_F-p_B)/n_T$.
The properties of $\Gamma_1$ are described in
Sec. \ref{app:sinteg} of the Appendix.
Equation (\ref{eq:esmalld}) includes the two cases $p=1$
($n_{a\uparrow}=n_T$) and $p=2b$ ($n_{a\uparrow}=n_{a\downarrow}=n_T/2$).
For $t=0$, the $p=2b$ case has $\cos\theta=\pm 1$, so we will focus
on the spin- and pseudospin-polarized one-component ($p=1$) ground state.

\vspace{-0.1in}
\subsection{Spontaneous interlayer coherence} 
\label{sec:sipc}
\vspace{-0.1in}

The pseudospin-polarized ground state offers the possibility of SILC.
Recall that SILC occurs when the off-diagonal (or interlayer)
density matrix $\rho_{12}$
is nonzero in the absence of interlayer tunneling:
\begin{equation}
\label{eq:sipccond}
\rho_{12} \equiv \sum_{{\bf k}s}
\langle c^\dagger_{1{\bf k}s} c_{2{\bf k}s} \rangle
= \frac{1}{2} (n_a-n_b) \sin\theta \ne 0 ,
\end{equation}
which requires both finite pseudospin polarization ($n_a>n_b$) and
$\sin\theta\ne 0$.
In the pseudospin-polarized ground state ($n_a=n_T$),
$\rho_{12}$ is just the geometric mean of the layer densities:
\begin{equation}
\label{eq:rho12}
\rho_{12}
= \frac{1}{2} n_T\sin\theta = \sqrt{n_1n_2} ,
\end{equation}
where we have made use of Eq.~(\ref{eq:ccvals}).
So if the pseudospin-polarized ground state has some density of particles
in each layer, it has interlayer phase coherence, $\rho_{12}\ne 0$.
Note that in the one-component phase,
\begin{equation}
\label{eq:sinth}
\sin\theta = (\rho_{12}+\rho_{21})/n_T
\end{equation}
measures the interlayer density matrix, normalized by the total density.

For $t=0$ and $p=1$, Eq.~(\ref{eq:thsoln}) gives
\begin{eqnarray}
\label{eq:ndiff}
&& \frac{n_1-n_2}{n_T} = \cos\theta =
\left\lbrace \begin{array}{cc}
-1, & X \le -1 \\
X, & -1 \le X \le 1  \\
1, & X \ge 1 \\
\end{array} \right.
\\ \nonumber
&& X = \frac{\zeta}{1 - \Gamma_1}
= \frac{\zeta}{1 - (32/3\pi p)[1-S(z)]/z}
\\ \nonumber
&& \qquad \qquad \qquad \rightarrow
\frac{\zeta}{(32/45\pi)z - (1/24)z^2} ,
\end{eqnarray}
where $z=2k_Fd$, and we have made use of Eq.~(\ref{eq:n1n2}).
The last line of Eq.~(\ref{eq:ndiff}) holds in the limit
that $z\rightarrow 0$.
The layer densities are equal ($n_1=n_2$) only
at exacly $p_F=p_B$;
when $p_F>p_B$, layer 1 tends to be occupied,
and when $p_F<p_B$, layer 2 tends to be occupied.
Thus, the hypothetical bistability of the one-component phase
proposed by Ruden and Wu does not exist, due to SILC.\cite{ruden}
Equation (\ref{eq:ndiff}) gives
\begin{equation}
\label{eq:sin1comp}
\sin\theta = \sqrt{1-[\zeta/(1-\Gamma_1)]^2} \Theta(1-\Gamma_1-|\zeta|)
\delta_{n_{a\uparrow},n_T} ,
\end{equation}
so that $\sin\theta$ in nonzero only for $|\zeta|<1-\Gamma_1$ and
for $n_{a\uparrow}=n_T$.
It is interesting to note that when the ground state is
pseuodspin-polarized, the dependence of the layer densities
on external parameters (e.g., layer imbalance $\zeta$) does not
involve the effective mass $m^*$ of the electrons or holes.

\begin{figure}[h]
\epsfxsize3.5in
\centerline{\epsffile{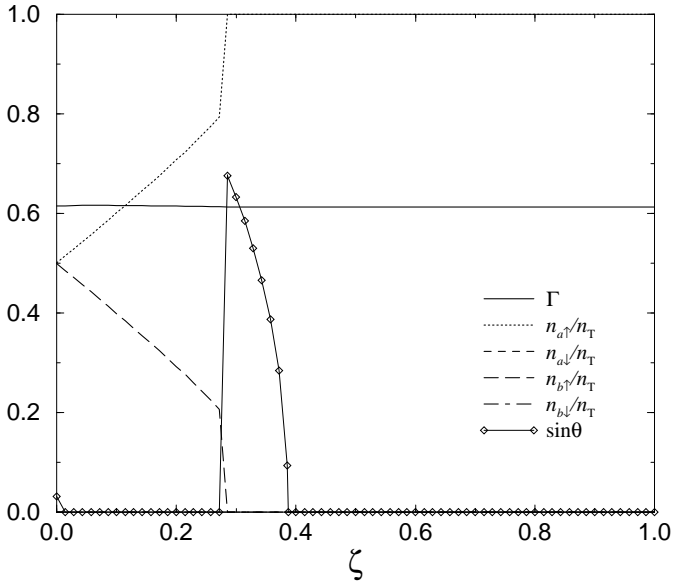}}
\caption{Normalized subband densities, interlayer exchange parameter
$\Gamma$, and normalized interlayer density matrix $\sin\theta$,
versus layer imbalance parameter $\zeta=(p_F-p_B)/n_T$,
for $d/a_0=5$ and $r_s=11$.
The $p=2a$ state is obtained for $\zeta<\zeta_c\approx 0.45$,
followed by a $p=1$ state with SILC for $\zeta_c<\zeta<1-\Gamma_1$.
SILC is lost for $\zeta>1-\Gamma_1$.
}
\label{fig:rs11}
\end{figure}

We have found that layer imbalance ($p_F\ne p_B$) can induce SILC
at higher total densities than in the balanced case.
This is illustrated for $r_s=11$ in Fig.~\ref{fig:rs11},
which is an example of case (5) introduced in Sec. \ref{sec:biasplots}.
Figure \ref{fig:rs11} shows a spin-polarized two-component ($p=2a$) state
when the gates are balanced ($\zeta=0$), and
maintains a $p=2a$ state for $\zeta<\zeta_c\approx 0.28$,
followed by a  one-component ($p=1$) state for $\zeta>\zeta_c$.
Because $\zeta_c<1-\Gamma_1$, $\sin\theta>0$ for
$\zeta_c<\zeta<1-\Gamma_1$, producing SILC in a finite region of layer
imbalance away from $\zeta=0$, at a smaller value of $r_s$ than is required
to achieve SILC for balanced layers.

\begin{figure}[h]
\epsfxsize3.5in
\centerline{\epsffile{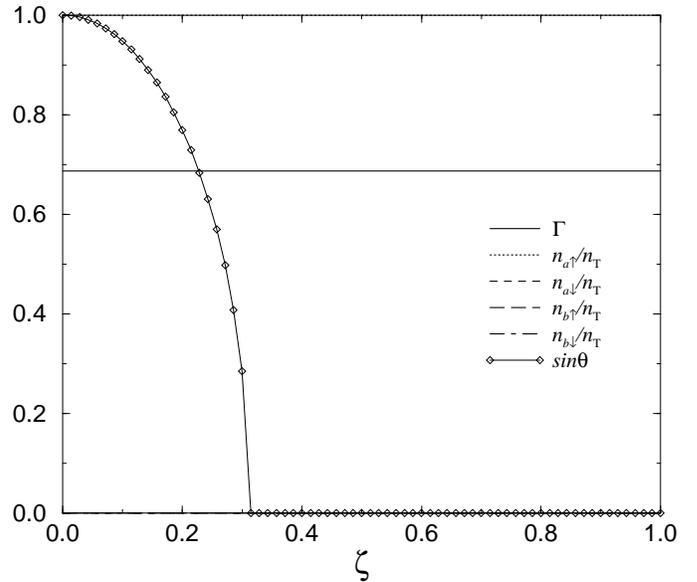}}
\caption{Normalized interlayer density matrix $\sin\theta$ and
interlayer exchange parameter $\Gamma$,
versus layer imbalance parameter $\zeta=(p_F-p_B)/n_T$
in the one-component phase ($n_{a\uparrow}=n_T$),
for $d/a_0=5$ and $r_s=15$. SILC is lost for $\zeta>1-\Gamma_1$.
}
\label{fig:rs15}
\end{figure}

When the toal density (and layer separation) are sufficiently small,
the $p=1$ phase with SILC is obtained even in the balanced case.
This is illustrated for $r_s=15$ in Fig.~\ref{fig:rs15},
which is an example of case (6) introduced in Sec. \ref{sec:biasplots}.
Figure \ref{fig:rs15} shows a one-component ($p=1$) state throughout
the range of $\zeta$,
with $\sin\theta>0$ for $\zeta<1-\Gamma_1$,
producing SILC in that region.
Note that as $r_s\rightarrow\infty$, then $z=2k_Fd\rightarrow 0$,
so that $\Gamma_1(z)\rightarrow 1$, and thus $1-\Gamma_1 \rightarrow 0$.
Therefore the maximum amount of imbalance $\zeta$ which allows SILC
decreases with the density, at very low densities.

\begin{figure}[h]
\epsfxsize3.5in
\centerline{\epsffile{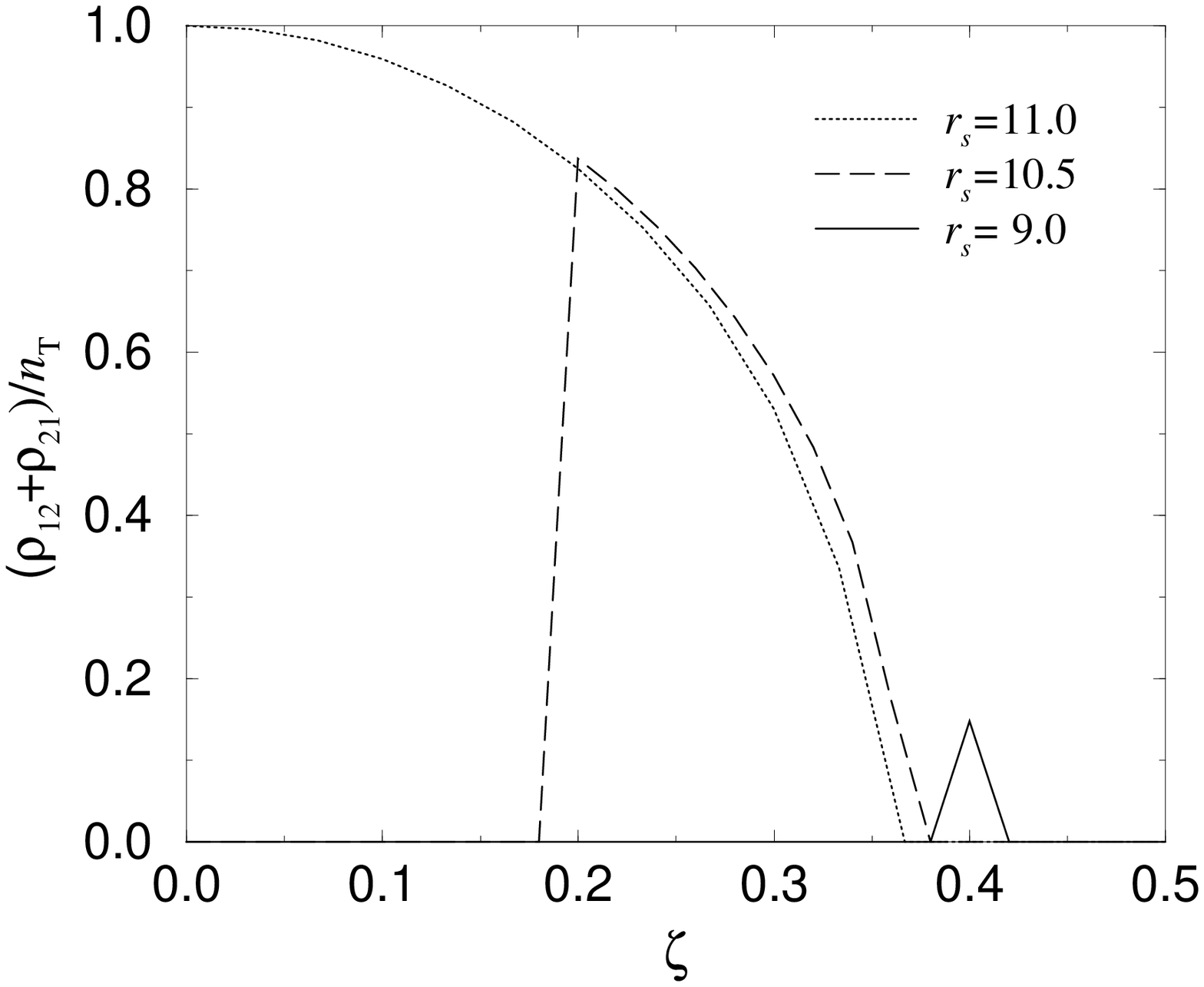}}
\caption{Interlayer exchange strength
$(\rho_{12}+\rho_{21})/n_T=\sin\theta$
as a function of the gate-imbalance parameter
$\zeta=(p_F-p_B)/n_T$ for fixed $d/a_0\approx 4.356$.
}
\label{fig:rho12}
\end{figure}

Figure \ref{fig:rho12} shows $\sin\theta$ versus $\zeta$ at
$d/a_0\approx 4.356$ for the three values of $r_s$, where
$\sin\theta$ obeys Eq.~(\ref{eq:sin1comp}).
At the lowest density shown ($r_s=11$), the system exhibits SILC
when balanced ($\zeta=0$) and under bias, until $\zeta=1-\Gamma_1$.
As the density is raised, SILC is lost for the balanced system
but appears suddenly around $\zeta\approx 0.2$ for $r_s=10.5$ when
an abrupt intersubband charge transfer produces pseudospin polarization
($p=1$).
For $r_s=9$ there is only a very small region of layer imbalance $\zeta$
that exhibits SILC, and for $r_s$ slightly smaller than this value,
SILC disappears completely.

Using Eqs. (\ref{eq:esmalld}), (\ref{eq:rho12}), and (\ref{eq:ndiff}),
we may calculate the energy per unit area $\varepsilon(n_1,n_2)$
defined in Eq.~(\ref{eq:uovera}), for the one-component phase:
\begin{equation}
\label{eq:vareps1c}
\varepsilon(n_1,n_2) =
\frac{n_T^2}{\nu_0} -
\frac{8}{3\sqrt{\pi}} \frac{e^2}{4\pi\epsilon} n_T^{3/2} +
\frac{e^2dn_1n_2}{2\epsilon}\Gamma_1 ,
\end{equation}
where $n_T=n_1+n_2$.
Recall that $\varepsilon$ does not include the electrostatic contribution
to the energy per unit area.
We may use Eqs. (\ref{eq:mui}) and (\ref{eq:vareps1c}) to calculate
$\mu_1$, the chemical potential measure relative to the energy minimum
of layer 1,
\begin{eqnarray}
\label{eq:mu1val}
\mu_1 &=&
\frac{2n_T}{\nu_0} -
\frac{4}{\sqrt{\pi}} \frac{e^2}{4\pi\epsilon} \sqrt{n_T}
\\ \nonumber && \quad +
\frac{e^2dn_2}{2\epsilon} \left(
\Gamma_1 + n_1\frac{d\Gamma_1}{dn_T}\right) ,
\end{eqnarray}
where we have used the fact that in the one-component
phase, $\Gamma_1$ depends on $n_1$ and $n_2$ only through $n_T=n_1+n_2$.
The quantity $\mu_2$ may be obtained by interchanging $n_1$ and $n_2$
in Eq.~(\ref{eq:mu1val}) and can be used to compute the front-gate voltage
$V_F$ using Eq.~(\ref{eq:vfalt}) in the Appendix.
It is straightforward to check that the difference between $\mu_1$
and $\mu_2$ satisfies the equilibrium condition in Eq.~(\ref{eq:equilib}).
Equations (\ref{eq:sij}) and (\ref{eq:mu1val})
may be used to calculate the electronic length $s_{ij}$ defined
in Sec.~\ref{sec:model}:
\begin{eqnarray}
\label{eq:s11}
\frac{s_{11}}{a_0} &=& \frac{1}{2} - \frac{r_s}{2\pi\sqrt{2}} +
\frac{1}{2}\frac{d}{a_0}n_2\left(
2\frac{\partial\Gamma_1}{\partial n_T} +
n_1 \frac{\partial^2\Gamma_1}{\partial n_T^2}
\right) .
\end{eqnarray}
The quantity $s_{22}$ may be obtained by interchanging $n_1$ and $n_2$
in Eq.~(\ref{eq:s11}).

Interlayer correlations produce a nonzero value of the electron length
$s_{12}$.  
From Eqs. (\ref{eq:sij}), (\ref{eq:s1s2}), and (\ref{eq:mu1val})
we have
\begin{equation}
\label{eq:s12}
s_{12} = s_{11} + \frac{d}{2}\left[
\Gamma_1 + (n_1-n_2) \frac{\partial\Gamma_1}{\partial n_T} \right] .
\end{equation}
Note that $s_{21}=s_{12}$, and that
\begin{equation}
\label{eq:s1val}
s_1 \equiv s_{11} - s_{12} = -\frac{d}{2}\left[
\Gamma_1 + (n_1-n_2) \frac{\partial\Gamma_1}{\partial n_T} \right] .
\end{equation}
The length $s_2$ can be obtained from $s_1$ by interchanging
$n_1$ and $n_2$.
Note that
\begin{equation}
\label{eq:sval}
s \equiv s_1 + s_2 = -\Gamma_1 d ,
\end{equation}
so that, from Eq.~(\ref{eq:regen}), the Eisenstein ratio
for fixed total density (constant $n_T$) is
\begin{equation}
\label{eq:refixnt}
R_E = \frac{s}{d+s} = \frac{-\Gamma_1}{1-\Gamma_1}
\rightarrow \frac{-1}{(32/45\pi)z} ,
\end{equation}
where $z=2k_Fd$, and the right-hand side holds in the limit $z\rightarrow 0$.
It is interesting to note that $R_E\propto -1/d$ as $d\rightarrow 0$,
just as was found in Ref.~\onlinecite{jungwirth} for the $\nu_T=1$
2LQH state.
For fixed $p_B$ (nearly equivalent to keeping the back-gate
voltage $V_B$ constant),
\begin{equation}
\label{eq:refixpb}
R_E = \frac{s_1}{d+s} = -\frac{
\left[ \Gamma_1 + (n_1-n_2) \partial\Gamma_1/\partial n_T \right]}
{2(1-\Gamma_1)} ,
\end{equation}
which in the balanced case ($p_F=p_B$ so that $n_1=n_2$) gives
Eq.~(\ref{eq:retzero}), which is exactly half of Eq.~(\ref{eq:refixnt}).

\begin{figure}[h]
\epsfxsize3.5in
\centerline{\epsffile{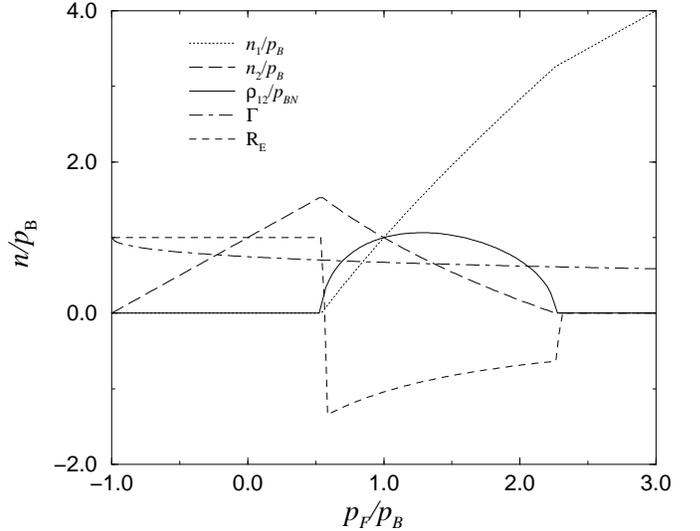}}
\caption{Layer densities $n_1, n_2$ and normalized interlayer
density matrix $2\rho_{12}/n_T=\sin\theta$, in the one-component phase,
versus front-gate density $p_F/p_B$, for fixed back-gate density $p_B$.
}
\label{fig:rhon1n2}
\end{figure}

Figure \ref{fig:rhon1n2} shows an example of the one-component phase
under bias for fixed back-gate density $p_B$ (essentially fixed
back-gate voltage $V_B$.)
The normalized layer densities $n_1$ and $n_2$ are shown, together
with the interlayer density matrix $\rho_{12}/p_B$ and $\Gamma$,
and also the Eisenstein ratio $R_E$.
For $p_F/p_B<0.5$, layer 2 contains all the charge ($n_2=n_T$) and
layer 1 is empty, so that Eq.~(\ref{eq:re}) gives $R_E=1$.
For $0.5<p_F/p_B<2.25$, both $n_1$ and $n_2$ are partially occupied,
$\rho_{12}=\sqrt{n_1n_2}$ is nonzero, and $R_E$ has dropped abruptly
and become negative, reflecting the presence of SILC, with its value
in this region given by Eq.~(\ref{eq:refixpb}).
For $p_F/p_B>2.25$, layer 2 is empty, $n_1=n_T$, and $R_E=0$.
Figure~\ref{fig:rhon1n2} illustrates an interesting hypothetical situation
in which bias and the exchange interaction have completely emptied
out layer 2, despite the fact that $p_B$ is nonzero.
It turns out that within the MFA, sufficently large $p_F$ would eventually
repopulate layer 2.
We analyze this issue further below, when we calculate the energy gap
$\Delta_{ab}$ in a pseudospin-polarized state ($n_a=n_T$)
to an otherwise empty subband ($n_b=0$).

Although Zheng and co-workers showed (within an unrestricted HFA)
that an abrupt {\em interlayer} charge transfer does not occur
when the gates are electrostatically balanced ($p_F=p_B$), one
may ask if an abrupt interlayer charge transfer can occur if the system
is biased.
According to the MFA developed here, the answer is yes, except at
sufficiently small densities.
(We are speaking here of zero tunneling; otherwise, neither layer
is strictly empty, according to the MFA.)
In the MFA, sufficiently strong bias or tunneling or sufficient lowering
of the densities eventually produces an abrupt {\em intersubband} charge
transfer, i.e., at some point, $n_a$ suddenly jumps.
Whether this translates into abrupt {\em interlayer} charge transfers
depends on what happens with the phase angle $\theta$.
If $\sin\theta\ne 0$ (SILC), then the interlayer charge 
transfer is suppressed, or at least somewhat reduced.
If the density is so high that strong bias produces the $p=2b$
pseudospin-polarized phase with $\sin\theta=0$,
then the MFA does give an abrupt interlayer charge transfer, because
then $\cos\theta=\pm 1$, and there is no difference between subband
densities and layer densities.
So, for example, a system with a density that would correspond
to $p=4$ when balanced will not exhibit SILC.
It is likely that including correlation-energy effects eliminates the
abruptness of the transition, but these effects have not been included
here.

\vspace{-0.1in}
\subsection{Intersubband gap}
\label{sec:gap}
\vspace{-0.1in}

The intersubband energy gap $\Delta_{ab}$ for the pseudospin-polarized
($p=1$ or $p=2b$) phase is defined as the energy required to move
a particle from the occupied $a$ subband with $n_a=n_T$
to the (otherwise empty) $b$ subband:
\begin{equation}
\Delta_{ab} = \frac{\delta}{\delta n}
                    \left(\frac{{\cal{E}}_0}{L_xL_y}\right) ,
\end{equation}
for
\begin{eqnarray}
\begin{array}{ll}
n_{a\uparrow} \rightarrow   (1/p)(n_T-\delta n) , &
n_{b\uparrow}   \rightarrow \delta n , \\
n_{a\downarrow}   \rightarrow (1-1/p)(n_T-\delta n) , &
n_{b\downarrow} = 0 .
\end{array}
\end{eqnarray}
An outline of the MFA calculation of $\Delta_{ab}$ is given
in Sec. \ref{app:isubgap} of the Appendix.
In units of the energy scale $v_0=e^2/4\pi\epsilon a_0$,
the MFA intersubband gap is
\begin{eqnarray}
\label{eq:abgap2}
&& \frac{\Delta_{ab}}{v_0} = \frac{2t}{v_0}\sin\theta
- \frac{4}{r_s^2}\left[\frac{1}{p} +
             \frac{d}{a_0}\cos\theta(\cos\theta-\zeta)\right]
\\ \nonumber &&
+ \sqrt{\frac{2}{p}} \frac{4}{\pi} \frac{1}{r_s}
- \sin^2\theta \sqrt{\frac{2}{p}} \frac{1}{r_s}
\left\lbrace \frac{\left[e^{-z/2}-(1-z/2)\right]}{z/2} \right.
\\ \nonumber && \qquad \qquad \qquad \qquad \qquad \left.
           + \frac{2}{\pi} \int_0^1 dx \left(1-e^{-zx}\right)
                                       \arccos(x) \right\rbrace
\end{eqnarray}
where $\zeta=(p_F-p_B)/n_T$, and $z=2k_Fd$ is the layer imbalance.

\begin{figure}[h]
\epsfxsize3.5in
\centerline{\epsffile{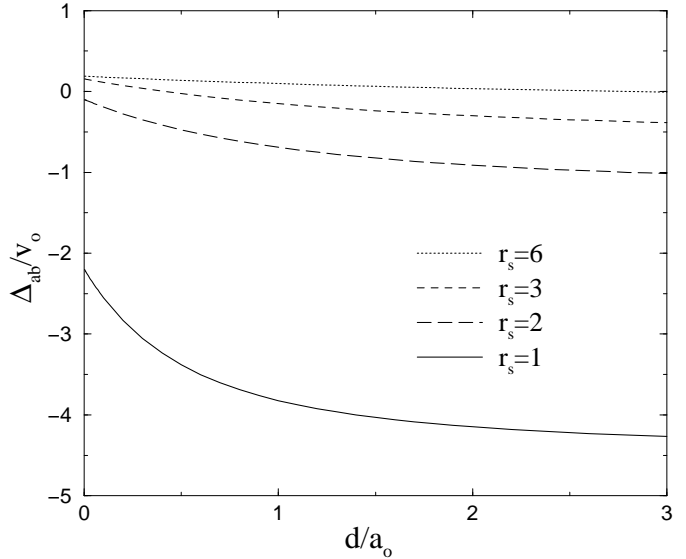}}
\caption{Normalized intersubband energy gap $\Delta_{ab}$
for transferring particles from the occupied lower-energy
subband $a$ to the (assumed empty) higher-energy subband $b$,
for $r_s=1,2,3,6,10$.
}
\label{fig:gaps}
\end{figure}

The intersubband gap $\Delta_{ab}$ is useful for at least two purposes.
First, it provides an estimate for the location of
the pseudospin-polarization transition.
The condition $\Delta_{ab}>0$ means the pseudospin-polarized ground state
is stable against intersubband charge transfers, whereas $\Delta_{ab}<0$
implies the opposite.
Thus, solving the equation $\Delta_{ab}=0$ in the MFA yields an estimate
of the location of the pseudospin-polarization transition.
It turns out that this procedure gives a lower value of $r_s$
for the $p=1$ transition than the GRPA estimate
(obtained using the pseudospin Stoner enhancement factor $I$):
at $d=0$, the MFA $\Delta_{ab}$ calculation (for $t=\zeta=0=\cos\theta=0$)
gives $r_s^{(0)}(1,2)=\pi/\sqrt{2}$, compared to $r_s^{(0)}(1,2)=\pi$
from the GRPA.
Figure \ref{fig:gaps} shows $\Delta_{ab}$ versus $d/a_0$ when
$t=\zeta=\cos\theta=0$ for $r_s=1,2,3,6$.
It is evident that $\Delta_{ab}>0$ only for sufficiently large $r_s$,
and that it decreases with layer separation $d/a_0$.
Negative values of $\Delta_{ab}$ indicate regions where the
pseudospin-polarized state is not stable.

The intersubband gap has a simple form when $t=\sin\theta=0$:
\begin{equation}
\label{eq:gapabi}
\frac{\Delta_{ab}}{v_0} =
-\frac{4}{r_s^2}\left[\frac{1}{p} + \frac{d}{a_0}(1-|\zeta|)\right]
+ \sqrt{\frac{2}{p}} \frac{4}{\pi} \frac{1}{r_s} .
\end{equation}
Equation (\ref{eq:gapabi}) provides an estimate of when one of the
layers cotains all the particles.
It tells us that for sufficently high total density (small $r_s$),
$\Delta_{ab}<0$ and both layers must be occupied, provided that
$1/2+(d/a_0)(1-|\zeta|)>0$ (which includes $\zeta=1$, corresponding to
$p_B=0$.)
This is because the kinetic energy (and the Coulomb energy, for $|\zeta|<1$),
which favors occupying both layers, dominates over the exchange energy,
which favors occupying a single layer, at higher densities.
For example, for $p=2$ and $\zeta=1$, Eq.~(\ref{eq:gapabi}) shows that
both layers will be occupied even when $p_B=0$, provided that $r_s<\pi/2$.
On the other hand, for $p=1$ and $\sin\theta=0$
(i.e., $|\zeta|>1-\Gamma_1$),
only one layer will be occupied, provided that
$r_s>(\pi/\sqrt{2})[1+(d/a_0)(1-|\zeta|)]$.
This is the situation shown in Fig.~\ref{fig:rhon1n2},
which shows that layer 2 has completely emptied for $p_F/p_B>2.25$.
At higher values of $p_F/p_B$ (e.g, $r_s<\pi/2$, not shown),
layer 2 would no longer be empty.

The second use of the MFA calculation of $\Delta_{ab}$ is as a rough
estimate of the minimum energy (thermal, or from photons) required
to excite particles from the occupied to the unoccupied subband in
the one-component state.
Perhaps this could be detected with sensitive heat-capacity measurements
or by measuring microwave absorption.

\vspace{-0.1in}
\subsection{Coulomb drag}
\label{sec:drag}
\vspace{-0.1in}

One very interesting feature of the one-component state with SILC
is that it is expected to exhibit interlayer drag
(finite dc transresistance), even at zero temperature.
Ordinarily, if the layers are not correlated in the ground state,
current in one layer can drag along particles in the other layer
(due to the Coulomb interaction between the layers) only at finite
temperature.\cite{drag}
But if $\rho_{12}\ne 0$, due to either tunneling or, more interestingly,
to SILC, then interlayer correlations present in the 2LES ground state
will produce interlayer drag even at zero temperature, as has been
predicted for interlayer-correlated 2LQH states.\cite{moon,drag2lqh}
Based on the Kubo formula with the ground state of Eq.~(\ref{eq:psimfa}),
we expect that a calculation of the zero-temperature
dc transconductivity $\sigma_d$ will give
\begin{equation}
\label{eq:sigmad}
\sigma_d \propto \frac{e^2\rho_{12}}{m^*}
         \propto \frac{e^2(n_a-n_b)\sin\theta}{m^*} .
\end{equation}
Calculations of the drag conductivity for a pseudospin-polarized ground state
are currently being carried out by other researchers.\cite{dragcalc}
According to Eqs. (\ref{eq:rho12}) and (\ref{eq:sigmad}),
we expect that in the $p=1$ phase ($n_a=n_T$),
$\sigma_d\sim\sqrt{n_1n_2}$, approximately (i.e., within an MFA
calculation of $\sigma_d$.)

It would be interesting to clarify the relationship between
$s_{12}$, $\rho_{12}$, and $\sigma_d$.
We conjecture that finite interlayer drag at zero temperature requires
$\rho_{12}\ne 0$ at zero temperature (although we have not proved this)
and that $\rho_{12}\ne 0$ (or at least $\rho_{12}\rho_{21}\ne 0$)
at zero implies finite interlayer drag at zero temperature.
It is certainly true that $\rho_{12}\ne 0$ and $\sigma_d\ne 0$
occur together at zero temperature in the interlayer correlated
2LQH effect.\cite{moon}
We also think it likely that a similar relation holds between $s_{12}$
and $\rho_{12}$, and hence between $s_{12}$ and $\sigma_d$,
although we have not proved this either.

\vspace{-0.1in}
\section{Conclusions}
\label{sec:conclude}
\vspace{-0.1in}

We investigated the effects of intralayer and interlayer exchange
in biased double-layer systems, in the absence of a magnetic field.
This was accomplished using a mean-field approximation (MFA) which,
in the limit of balanced layers (no bias),
is equivalent to the unrestricted HFA of Zheng and co-workers\cite{zheng}

\vspace{-0.1in}
\subsection{Findings}
\label{sec:findings}
\vspace{-0.1in}

We found that a balanced 2LES possesses four possible noncrystalline
MFA ground states.
The spin- and pseudospin-unpolarized four-component ($p=4$) state
is obtained at the highest densities.
In contrast to earlier work,
we found that as the density is lowered, a three-component ($p=3$)
state with slightly unequal layer densities is obtained.
Thus, we find that there is no direct four- to two-component transition.
At finite layer separation ($d>0$) and zero interlayer tunneling ($t=0$),
the $(p=4) \rightarrow (p=3)$ MFA transition involves a small but abrupt
interlayer charge transfer.
Any such abrupt interlayer charge transfer will in principle result
in a large (formally infinite) value of the Eisenstein ratio $R_E$
at the transition.
The Eisenstein ratio is
a sensitive measure of the interlayer capacitance discussed in
Sec.~\ref{sec:model}.

Like Zheng and co-workers, we found that as the total density was lowered,
a spin-polarized two-component ($p=2a$) state preceded
a low-density one-component ($p=1$) state
possessing SILC, provided that the gates were balanced ($p_F=p_B$).
This $p=1$ state is different from that of Ruden and Wu,
whose proposed one-component state occupied a single layer,
rather than a single subband consisting of a linear combination of both layers.
We obtained a MFA phase diagram for the noncrystalline phases of the
2LES, shown in Fig.~\ref{fig:phdiag}.
This phase diagram is similar to that of Ref.~\onlinecite{zheng}, except
for the presence of the $p=3$ phase between the $p=4$ and $p=2$ phases.
Only the $p=1$ phase was found to possess SILC --- i.e., a
nonzero interlayer density matrix ($\rho_{12}\ne 0$)
even with zero interlayer tunneling ($t=0$).
We also defined the pseudospin Stoner interaction parameter $I$, and
considered the linear response of the MFA ground state to interlayer
tunneling, equivalent to a GRPA calculation.
We used $I$ to obtain an alternate (GRPA) estimate of the location
of the $(p=2) \rightarrow (p=1)$ transition, shown as the dotted line at
the top of Fig.~\ref{fig:phdiag}.
Of course, in the limit of vanishing total density
($r_s\rightarrow\infty$), we expect that a Wigner crystal state is
obtained (in the absence of disorder).
We did not consider the effects of disorder here, except to note
that it limits the maximum $r_s$ for a state with mobile particles.

Under bias ($|p_F-p_B|>0$), we found that there are five possible
noncrystalline ground states.
In every case, the MFA gave subbands, which, when occupied, were
either completely spin-unpolarized or fully spin-polarized.
Including correlation-energy effects would likely produce ground
states with intermediate spin polarizations, as is apparently
the case in three dimensions.\cite{ortiz}
The additional state that can appear under sufficiently large bias and/or
interlayer tunneling is a pseudospin-polarized two-component
($n_{a\uparrow}=n_{a\downarrow}=n_T/2$) state, which we labeled
$p=2b$.
The $p=2b$ state requires bias and/or tunneling, and has $\sin\theta=0$
(and thus no SILC) for $t=0$.
In Sec.~\ref{sec:bias} we studied the effect of bias by considering
sytems at fixed total density $n_T$, for a range of values of the
layer-imbalance parameter $\zeta=(p_F-p_B)/n_T$.
We enumerated the six possible scenarios for bias-driven transitions
between the (five possible) noncrystalline MFA ground states for $t=0$.
We also showed that a very simple model that assumes no interlayer exchange
and no spin polarization is capable of fitting experimental SdH data 
quite well (see Fig.~\ref{fig:finfig}), and that a simple LDF model can
do the same in the presence of interlayer tunneling
(see Fig.~\ref{fig:yinga}).

We studied the one-component phase under applied bias,
finding that bias lowers the $r_s$ required for SILC
(see Fig.~\ref{fig:rho12}).
Within the MFA, SILC occurs only in the one-component phase:
when $t=0$, $\sin\theta$ is nonzero only when $n_{a\uparrow}=n_T$.
Perhaps including correlation-energy effects would allow SILC for
states with partial pseudospin polarization.
But $p=1$ is only a necessary condition for SILC, not a sufficient one.
We found that if the layer imbalance parameter was too large
($\zeta>1-\Gamma_1$), then SILC was lost.
When SILC occurs, the MFA gave a value of the interlayer density matrix
equal to the geometric mean of the layer densities:
$\rho_{12}=\sqrt{n_1n_2}$.
For the case that SILC is present, we calculated the layer densities ($n_i$),
local values of the chemical potential ($\mu_i$),
electronic lengths ($s_{ij}$), and Eisenstein ratio ($R_E$)
(see Fig.~\ref{fig:rhon1n2}).

Ruden and Wu originally predicted an abrupt interlayer charge tranfer
for $t=0$
at sufficiently low densities and layer separations,
in the balanced case ($p_F=p_B$).\cite{ruden}
Like Zheng and co-workers, we found that an abrupt {\em interlayer}
charge transfer does not occur in the balanced case, due to SILC.
However, the MFA (and the unrestricted HFA of Ref.~\onlinecite{zheng})
does produce abrupt {\em intersubband} charge transfers, even for
the balanced case.
This a feature of the HFA that requires correlation-energy effects to
remedy.
In the case of nonzero bias, intersubband transfers are equivalent to
interlayer transfers if $\sin\theta=0$, which is the usual case,
except possibly for $p=1$.
Interlayer subband transfers at $t=0$ are reduced or suppressed
in the MFA only for the $(p=2a) \rightarrow (p=1)$ transition,
and only if $\zeta<(1-\Gamma_1)$ in the $p=1$ phase.
So SILC, when present, does reduce or eliminate abrupt interlayer
subband transfers, but the MFA does not always eliminate them
under bias ($p_F\ne p_B$).
If the system is at sufficiently low density and layer separation
that it stays in the $p=1$ state, then there are no abrupt interlayer
charge transfers under bias in the MFA, despite the fact that
the layers can empty out as $p_F$ is changed (see Fig.~\ref{fig:rhon1n2}).

We also calculated the intersubband gap $\Delta_{ab}$ for the
pseudospin-polarized ($p=1$ or $p=2b$) phases within the MFA,
defined as the energy to move a particle from the lower energy $a$
subband to the higher energy (empty) $b$ subband.
This energy provides an estimate of the single-particle intersubband gap
in the pseudospin-polarized $p=1$ and $p=2b$ phases,
and can be used to estimate the stability of those phases.
If the $p=1$ phase can be obtained experimentally, $\Delta_{ab}$ might
be measured using heat-capacity or microwave/optical techniques.
A very interesting feature of the one-component phase with SILC is that
it should have nonzero interlayer drag, even at zero temperature,
with the size of the interlayer drag conductivity being proportional
to the interlayer density matrix $\rho_{12}$.

Pseudospin polarization can be detected by SdH measurements,
which exhibit oscillations that are periodic in
$1/H$ (where $H$ is the applied magnetic field).
The periods of the SdH oscillations are given by
\begin{equation}
\label{eq:sdhosc}
\Delta_{\alpha s}(1/H) = \frac{2\pi e}{\hbar} \frac{1}{A_{\alpha s}}
= \frac{2\pi e}{\hbar} \frac{1}{\pi k_{\alpha s}^2}
= \frac{e}{h} \frac{1}{n_{\alpha s}} ,
\end{equation}
where $A_{\alpha s}$ is the cross-sectional area of the Fermi surface
perpendicular to the applied magnetic field for electrons
in subband $\alpha$ with spin $s$.
Knowing the total density $n_T$ (e.g., from Hall measurements),
SdH measurements of the subband densities $n_{\alpha s}$ could allow
a determination of the degree of spin and pseudospin polarization.
For example, in the case of equally balanced layers ($n_1=n_2=n_T/2$)
having a $p$-component ground state ($p=1,2,4$) in the absence of
tunneling ($t=0$), there is a single ($p$-fold degenerate)
SdH oscillation period,
\begin{equation}
\label{eq:sdhp}
\Delta_p(1/H) = \frac{2\pi e}{\hbar} \frac{1}{A_p}
= \frac{2\pi e}{\hbar} \frac{1}{\pi k_F^2}
= \frac{e}{h} \frac{p}{n_T} ,
\end{equation}
which allows $p$ to be determined directly from SdH measurements.

\vspace{-0.1in}
\subsection{Can it exist?}
\label{sec:canit}
\vspace{-0.1in}

Can the one-component state be realized in the balanced case?
The $p=1$ state is a legitimate solution of in the HFA,
but we have not examined its stability here.
It is hypothetically possible that the $p=1$ state might always
be preempted by a Wigner crystal state.
Conti and Senatore\cite{conti} carried out diffusion Monte Carlo 
(DMC) simulations in the $d=0$ limit,
calculating the $p=4$ ground-state energy as a function of $r_s$
and using previous single-layer DMC results\cite{rapi} to estimate
the $p=2$ and $p=1$ energies.
They also estimated the ground-state energy for a Wigner crystal state,
and found that the $p=4$ state is obtained for $r_s<42$, and that
the Wigner crystal state is obtained for larger values of $r_s$.
In their calculation, neither the $p=2$ nor the $p=1$ states are ever 
favored energetically.

Although the DMC results in Ref.~\onlinecite{conti} show the need
to examine the existence and stability of the $p=1$ state beyond the HFA,
they do not rule out its existence.
This is because at $d=0$ the fermions possess $CP(3)$ symmetry
(spin and pseudospin fully rotateable and interchangeable),
and estimating the $p=2$ and $p=1$ energies using single-layer results
misses part of the correlation energy, which lowers the $p=1$ and $p=2$
ground-state energies.
Ideally, a DMC simulation of $CP(3)$ fermions would be most useful
to determine theoretically if the $p=1$ state can be obtained, but
such calculations might be prohibitively difficult to carry out.
As a start, allowing for the possibility of Wigner crystallization
(broken translational symmetry) within the MFA calculation would be
helpful.
Alternatively, a time-dependent MFA calculation of the collective mode
would indicate where (for what density and layer separation)
the collective mode of the 2LES goes soft ($\omega\rightarrow 0$),
signaling the onset of Wigner crystallization.
We are currently developing such a calculation of the collective mode.
Better yet would be a double-layer STLS calculation allowing for
the possibility of Wigner crystallization, or at least a determination
of when the STLS collective mode goes soft in a double-layer system.

More important is the question of whether SILC
can be achieved experimentally in the absence of a strong magnetic field,
which serves to quench the kinetic energy of the particles.
The MFA and HFA underestimate the value of $r_s$
required for the transitions, perhaps by a factor of 10.
For example, the spin-polarization transition for a single layer
has been estimate to occur for $r_s\sim 20$.\cite{rapi}
Although such high values of $r_s$ have been achieved in $p$-type
GaAs samples, even higher values of $r_s$ will be required to
achieve spontaneous pseudospin polarization.
Disorder imposes further constraints, because it limits the
maximum $r_s$ for which the particles are still mobile.
However, given the impressive progress in producing double-layer
systems with ever-lower particle densities and ever-higher mobilities,
it does not seem prudent to rule out the possibility that such a state
might someday be realized.

As pointed out by Zheng and co-workers\cite{zheng} and by
Conti and Senatore,\cite{conti}
most of the considerations presented in Ref.~\onlinecite{moon}
for SILC in the quantum Hall regime should be relevant to
the $p=1$ phase (with SILC) in the absence of a magnetic field.
In both cases, there is a ground state with broken $U(1)$ symmetry
(when $t=0$),
due to interlayer exchange at small layer separations and particle densities.
It is therefore expected that the $p=1$ state will exhibit many of the
novel features of the 2LQH state with SILC, including zero-temperature
interlayer drag, vortex excitations [of the angle $\phi$
in Eq.~(\ref{eq:etheta})] and an associated a finite-temperature
Kosterlitz-Thouless transition (for $t=0$), and interesting many-body
effects in tilted magnetic fields for finite interlayer
tunneling.\cite{yang,moon,bigyang}

\vspace{-0.1in}
\subsection{Speculations regarding the 2LQH regime}
\label{sec:speculate}
\vspace{-0.1in}
It is interesting to speculate about the applicability of these ideas
to the 2LQH effect at total filling factor unity.\cite{moon}
For sufficiently small distances ($d<d_c\approx1.2\ell$,
where $\ell=\sqrt{\hbar/eB}$ is the magnetic length\cite{yang}),
the 2LQH system exhibits a quantum Hall effect.\cite{murphy}
Theoretically, the small-$d$ 2LQH state has nonzero $\rho_{12}$
even for $t=0$ (SILC), and the 2LQH system exhibits
strong Coulomb drag.\cite{moon,drag2lqh}
At sufficiently large layer separations, it is found experimentally that
the quantum Hall effect disappears,\cite{murphy}
and it is has been proposed that there is a quantum phase transition
to a state without SILC.\cite{moon}
The nature of the ground state for $d>d_c$ is a topic of active investigation.
It has been analyzed as a system of two weakly coupled layers
of $\nu=1/2$ composite fermions (CF's).
Theoretical calculations of the drag at low temperatures
predict that the drag resistivity should scale
with temperature as $T^{4/3}$,
based on calculating the effects of gauge fluctuations
on two CF layers in the metallic state.\cite{ybkim1,sakhi,ussiprb}.
It has also been proposed that the weak coupling between the CF's
in different layers produces BCS pairing between
them at sufficiently low temperatures,\cite{bonesteel}
and that this paired state leads to a finite drag resistivity at
zero temperature.\cite{ussiprl,fzhou}

We point out here that besides the apparent BCS instability
between CF's in different layers,
double-layer CF systems might be unstable to pseudospin polarization.
In the limit $d\rightarrow\infty$, the double-layer $\nu_T=1$ system
may be regarded as a $p=2a$ phase (i.e., spin-polarized but
pseudospin-unpolarized) of CF's in zero effective magnetic field
($\nu=1/2$ per layer).
Naively, the presumably large effective mass of the CF's
would correspond to a much larger effective value of $r_s$
than for the zero-field case, perhaps producing a value of $r_s$
sufficently large to obtain a pseudospin-polarized $p=1$ phase.
Another way of saying this is that the large magnetic field experienced
in each $\nu=1/2$ layer quenches the kinetic energy of the particles,
and that this quenching might strongly enhance exchange instabilities --
in this case towards pseudospin polarization ($n_a>n_b$), presumably
with $\sin\theta=1$ (when $p_F=p_B$), since no spontaneous interlayer transfer
has been found to occur for $p_F\approx p_B$.
One appealing feature of a hypothetical $p=1$ CF state is that it would
have small or zero resistivity in the pseudospin channel (i.e., for
oppositely directed currents), the same channel in which the 2LQH state
exhibits superfluidity at sufficiently small layer separations.\cite{moon}
Such a $p=1$ state of CF's would exhibit SILC ($\rho_{12}\ne 0$ even
when $t=0$) and therefore possess zero-temperature Coulomb drag.
Recent experiments with double-layer systems corresponding to filling
factor $\nu=1/2$ in each layer provide evidence for the possibility of
zero-temperature drag.\cite{jpedrag}
We are currently investigating the possibility and consequences of
pseudospin polarization in double-layer CF systems.

We also note that a perpendicular magnetic field ${\bf B}$
will generally enhance spin and pseudospin polarization
because it tends to quench the kinetic energy.
This effect enhances the exchange and correlation effects
that lead to polarization.
We therefore expect that SILC can in principle be found
at {\em any} filling factor $\nu_T=(h/e)n_T/B$,
provided that the total density $n_T$
and the layer separation $d$ are sufficiently small.
In particular, if it exists in the one-component phase for
zero magnetic field, SILC will probably persist, and even grow stronger,
when a perpendicular magnetic field is applied.
Finally, we remark that
it may prove instructive to view the $\nu_T=1$ 2LQH state as a
Chern-Simons bosonic condensate of spin- and pseudospin-polarized
($p=1$) electrons bound to unit flux quanta.

It has recently been found
that double-layer systems in strong magnetic
fields near total filling factor unity exhibit a huge resonant enhancement
of the interlayer tunneling conductivity when SILC is present.\cite{spielman} 
It would be interesting to measure the tunneling conductivity
for a tilted sample, since many-body effects in a state with
SILC strongly suppress the interlayer tunneling amplitude when the
parallel component of the magnetic field exceeds a critical value.
This suppression is much stronger than for a system without SILC.
We expect a similar strong enhancement of the tunneling conductivity
at zero magnetic field, provided that the system possesses SILC.
Such tunneling measurements could prove very useful for measuring
the strength of SILC in double-layer systems at zero (or higher)
magnetic field.

\vspace{-0.1in}
\section{Acknowledgments}
\vspace{-0.1in}

We thank F.~David~N\'{u}\~{n}ez in deep appreciation of his 
patience and enthusiasm, and for his invaluable assistance.
We thank A.~R.~Hamilton for providing us with SdH data
for a double-layer hole system,
and for patiently answering several questions
regarding experimental measurements on double-layer systems.
Special thanks are also owed to S. Das Sarma and A.H. MacDonald for
useful discussions.
C.B.H. thanks the Institute for Theoretical Physics
(University of California, Santa Barbara),
where part of this work was carried out, and
for their support through the ITP Scholars Program.
This work was supported by a grant from the Research Corporation,
and by the National Science Foundation under grant 9972332.

\bigskip

\vspace{-0.1in}
\appendix
\section*{}
\vspace{-0.1in}

\subsection{Gate voltages}
\label{app:gatevolt}
\vspace{-0.1in}

The layer densities $(n_1,n_2)$ are determined theoretically
by minimizing the total ground-state energy per unit area
[Eq.~(\ref{eq:etheta})] for fixed gate densities $(p_F,p_B)$.
Experimentally however, it is the gate voltages which are tuned.
If needed, the gate voltages for a given value of $(p_F,p_B)$
can be calculated using
\begin{eqnarray}
\label{eq:vgate}
eV_F &=& eD_F E_F + \mu_1 + eV_F^{(0)} , \\ \nonumber
eV_B &=& eD_B E_B + \mu_2 + eV_B^{(0)} ,
\end{eqnarray}
and Eq.~(\ref{eq:mui}).
Here the gate electric fields depend on the gate sheet densities through
Gauss's law, $E_\alpha=(e/\epsilon)p_\alpha$,
and $eV_\alpha^{(0)}$ are sample-dependent constant gate-voltage shifts.
Although $eV_\alpha$ is approximately equal to $eD_\alpha E_\alpha$,
Eq.~(\ref{eq:vgate}) shows that the layer values $\mu_i$ of the
chemical potential also contribute to the gate voltages.

The layer values of the chemical potential $\mu_i$ can be
computed numerically from the variation of the equilibrium value
of the total energy per unit area (regarded as a function of
the front- and back-gate densities $p_F$ and $p_B$)
with respect to infinitesimal changes in gate densities:
\begin{equation}
\label{eq:muequil}
\delta \bar{\cal{E}}_0/L_xL_y = \mu_1 \delta p_F + \mu_2 \delta p_B .
\end{equation}
For the typical case in which the back-gate voltage $V_B$ is kept constant
and the back-gate distance $D_B$ is much larger than the interparticle
and interlayer separation so that $p_B$ is nearly constant,
it is convenient to use Eq.~(\ref{eq:equilib}) and write
\begin{equation}
\label{eq:vfalt}
eV_F = eD_F E_F + edE_{12} + \mu_2 + eV_F^{(0)} ,
\end{equation}
where Eq.~(\ref{eq:muequil}) gives
\begin{equation}
\label{eq:mu1equil}
\mu_2 \approx \frac{\partial}{\partial p_B}
\left(\frac{\bar{\cal{E}}_0}{L_xL_y}\right)_{p_F} .
\end{equation}
Equation (\ref{eq:vfalt}) has the advantage of being applicable
even when layer 1 empties out.
Equations (\ref{eq:vfalt}) and (\ref{eq:mu1equil}) can be used to calculate
theoretically the front-gate voltage.
In the limit where the interlayer separation is larger than the intralayer
particle separation and (which usually amounts to the same thing)
interlayer correlations can be neglected,
then variations of $\mu_2$ with $p_F$ are small in comparison with
$edE_{12}$, so that the effects of $\mu_2$ can be absorbed into the
voltage shift $eV_F^{(0)}$, thus giving
\begin{equation}
V_F \approx E_FD_F + E_{12}d + V_F^{(0)} .
\end{equation}

\vspace{-0.1in}
\subsection{Hartree-Fock approximation}
\label{app:hfa}
\vspace{-0.1in}

In the Hartree-Fock Approximation (HFA) the two-body interaction
is factored so that the ground-state energy per unit area is
\begin{eqnarray}
\label{eq:elxly}
\frac{{\cal{E}}_{HF}}{L_xL_y} &=&
\frac{1}{L_xL_y} \sum_{{\bf k}s} \varepsilon_k 
      \langle c_{1{\bf k}s}^\dagger c_{1{\bf k}s} +
              c_{2{\bf k}s}^\dagger c_{2{\bf k}s} \rangle \\ \nonumber
  &-& \frac{t}{L_xL_y} \sum_{{\bf k}s}
     \langle c_{1{\bf k}s}^\dagger c_{2{\bf k}s}
         +   c_{2{\bf k}s}^\dagger c_{1{\bf k}s} \rangle \\ \nonumber
  &-& \frac{1}{2(L_xL_y)^2} \sum_{j_1{\bf k}_1s_1}
                        \sum_{j_2{\bf k}_2s_2}
                          V_{j_1j_2}(|{\bf k_2-k_1}|) \\ \nonumber
&&\qquad\qquad\times \langle c_{j_1{\bf k_2}s_1}^\dagger
                             c_{j_2{\bf k_2}s_2} \rangle
     \langle c_{j_2{\bf k_1}s_2}^\dagger c_{j_1{\bf k_1}s_1}
                                                   \rangle \\ \nonumber
  &+& \frac{1}{2}  \sum_{j_1} \sum_{j_2}
                   V_{j_1j_2}(q=0) n_{j_2} n_{j_1} \\ \nonumber
&-& \sum_{j}\sum_{\alpha} V_{j\alpha}(q=0)  p_\alpha n_j \\ \nonumber
&+& \frac{1}{2}\sum_{\alpha\beta} V_{\alpha\beta}(q=0) p_\alpha
                                                       p_\beta .
\end{eqnarray}
The effect last three (the Hartree) terms of Eq.~(\ref{eq:elxly})
may be calculated by noting that
\begin{eqnarray}
\label{eq:vqeq0}
&& \lim_{q\rightarrow 0} V_{ij}(q) =
\lim_{q\rightarrow 0} \frac{e^2}{2\epsilon q}
\left[ 1 - \left(1 - e^{-qd_{ij}}\right) \right] \\ \nonumber
&=& \left( \lim_{q\rightarrow 0} \frac{e^2}{2\epsilon q} \right)
    - \frac{e^2}{2\epsilon} d_{ij}
\equiv (\infty) - \frac{e^2}{2\epsilon} d_{ij} ,
\end{eqnarray}
where $(\infty)$ denotes the formally divergent part in the last line
of Eq.~(\ref{eq:vqeq0}).
The last three (the Hartree) terms of Eq.~(\ref{eq:elxly}) become
\begin{eqnarray}
\label{eq:inftyeq}
&(\infty)& \frac{1}{2} (n_1 + n_2 - p_F - p_B)^2
+ \frac{e^2d}{2\epsilon} (p_F-n_1)(n_2-p_B) \\ \nonumber
&+& \frac{e^2D_F}{2\epsilon} p_F (n_1 +n_2-p_B) +
    \frac{e^2D_B}{2\epsilon} p_B (n_1 +n_2-p_F) .
\end{eqnarray}
Requiring the first term of Eq.~(\ref{eq:inftyeq}) to not diverge
imposes charge neutrality: $n_1+n_2=p_F+p_B$.
From Gauss's law, the Hartree energy [Eq.~(\ref{eq:inftyeq})]
may therefore be written, up to an overall constant, as
\begin{equation}
\label{eq:efields}
\frac{\epsilon}{2} \left[
E_{12}^2 d + E_F^2 D_F + E_B^2 D_B \right] ,
\end{equation}
where $E_{12}$ is the electric field between layers 1 and 2,
$E_F$ is the electric field between the front gate and layer 1, and
$E_B$ is the electric field between the back gate and layer 2.

Equation~(\ref{eq:efields}) is just the electric field energy per
unit area for the sample; we drop the last two terms
since they may be regarded as constants for fixed $p_F$ and $p_B$.
The ground-state energy per unit area may thus be written as
\begin{eqnarray}
\label{eq:elxly2}
\frac{{\cal{E}}_0}{L_xL_y} &=&
\frac{1}{L_xL_y} \sum_{{\bf k}s} \varepsilon_k 
      \langle c_{1{\bf k}s}^\dagger c_{1{\bf k}s} +
              c_{2{\bf k}s}^\dagger c_{2{\bf k}s} \rangle \\ \nonumber
  &-& \frac{t}{L_xL_y} \sum_{{\bf k}s}
     \langle c_{1{\bf k}s}^\dagger c_{2{\bf k}s}
         +   c_{2{\bf k}s}^\dagger c_{1{\bf k}s} \rangle
                   + \frac{e^2d}{2\epsilon} (n_1-p_F)^2 \\ \nonumber
  &-& \frac{1}{2(L_xL_y)^2} \sum_{j_1{\bf k}_1s_1}
                        \sum_{j_2{\bf k}_2s_2}
                           V_{j_1j_2}(|{\bf k}_2-{\bf k}_1|) \\ \nonumber
&& \qquad\qquad \times \langle c_{j_1{\bf k_2}s_1}^\dagger
                               c_{j_2{\bf k_2}s_2} \rangle
     \langle c_{j_2{\bf k_1}s_2}^\dagger c_{j_1{\bf k_1}s_1}
                                                   \rangle .
\end{eqnarray}

\vspace{-0.1in}
\subsection{Exchange integrals}
\label{app:exint}
\vspace{-0.1in}

\begin{figure}[h]
\epsfxsize3.5in
\centerline{\epsffile{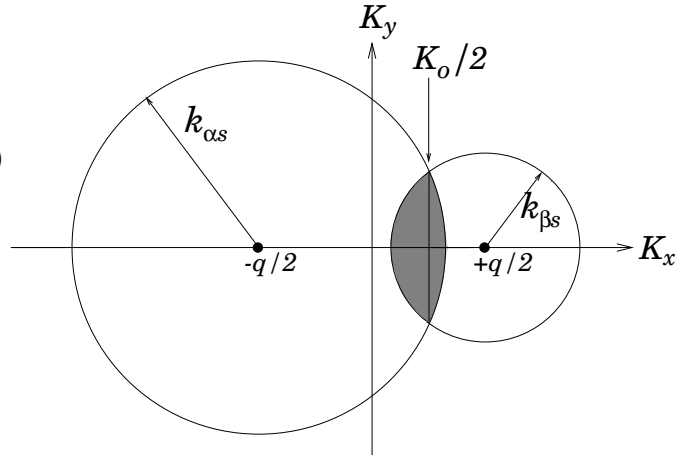}}
\caption{
The quantity $I_{\alpha\beta s}(q)$ is proportional to the area of
the shaded region of overlap between two circular Fermi surfaces of radii
$k_{\alpha s}$ and $k_{\beta s}$, which are centered at $K_x=\pm q/2$.
The circular Fermi surfaces intersect at $K_x\equiv K_0/2$.
}
\label{fig:fermicircs}
\end{figure}

The exchange integral $I_{\alpha\beta s}(q)$ is defined as
\begin{eqnarray}
\label{eq:iabapp}
I_{\alpha\beta s}(q) =
\frac{1}{L_xL_y} \sum_{\bf K} &&
\Theta(k_{\alpha s}-|{\bf K+q}/2|) \\ \nonumber
&& \times \Theta(k_{\beta s}-|{\bf K-q}/2|) ,
\end{eqnarray}
where $\alpha$ and $\beta$ can be either $a$ or $b$,
and where $k_{\alpha s}, k_{\beta s}$ denote the Fermi wave vectors
for particles of spin $s=\uparrow,\downarrow$ in subbands $a$ or $b$.
Note that Eq.~(\ref{eq:iabapp}) implies that $I_{\alpha\beta s}$ is
$1/(2\pi)^2$ times the shaded area shown in Fig.~\ref{fig:fermicircs},
where for concreteness ${\bf q}$ is taken be in the ${\bf \hat{x}}$
direction.
Let the quantity $K_0/2$ equal the value of $K_x$ at which the
Fermi circles of radius $k_{\alpha s}$ and $k_{\beta s}$ intersect:
\begin{equation}
\label{eq:k0}
K_0 = (k_{\alpha s}^2 - k_{\beta s}^2)/q =
      4\pi (n_{\alpha s} - n_{\beta s})/q .
\end{equation}
Then
\begin{eqnarray}
\label{eq:iab2}
I_{\alpha\beta s}(q) &\equiv&
\left[ n_{\beta s}  \Theta(k_{\alpha s} - k_{\beta s} - q) \right.
\\ \nonumber && \qquad +
\left. n_{\alpha s} \Theta(k_{\beta s} - k_{\alpha s} - q) \right]
\\ \nonumber &+&
\Theta(k_{\alpha s} + k_{\beta s} - q)
\Theta(q - |k_{\alpha s} - k_{\beta s}|)
\\ \nonumber && \times \frac{1}{\pi} \lbrace
n_{\alpha s} \lbrack \cos^{-1}\left(\frac{q+K_0}{2k_{\alpha s}}\right)
\\ \nonumber && \qquad -
            \left(\frac{q+K_0}{2k_{\alpha s}}\right)
            \sqrt{1-\left(\frac{q+K_0}{2k_{\alpha s}}\right)^2} \rbrack
\\ \nonumber && \quad +
n_\beta \lbrack \cos^{-1}\left(\frac{q-K_0}{2k_{\beta s}}\right)
\\ \nonumber && \qquad -
            \left(\frac{q-K_0}{2k_{\beta s}}\right)
            \sqrt{1-\left(\frac{q-K_0}{2k_{\beta s}}\right)^2} \rbrack \rbrace .
\end{eqnarray}
When $\beta=\alpha$, then $k_{\beta s}=k_{\alpha s}$, $K_0=0$, and
Eq.~(\ref{eq:iab2}) becomes
\begin{eqnarray}
\label{eq:iaa}
&& I_{\alpha\alpha s}(q) \equiv n_{\alpha s} \Theta(2k_{\alpha s} - q)
\\ \nonumber && \times \frac{2}{\pi}
\left[ \cos^{-1}\left(\frac{q}{2k_{\alpha s}}\right) -
            \left(\frac{q}{2k_{\alpha s}}\right)
            \sqrt{1-\left(\frac{q}{2k_{\alpha s}}\right)^2} \right] ,
\end{eqnarray}
and the first exchange integral in Eq.~(\ref{eq:etheta})
may be carried out explicitly:
\begin{equation}
\label{eq:v11termapp}
-\frac{1}{2L_xL_y} \sum_{\bf q} V_{11}(q) I_{\alpha\alpha s}(q) =
-\frac{8}{3\sqrt{\pi}} \frac{e^2}{4\pi\epsilon} n_{\alpha s}^{3/2} .
\end{equation}
Equation~(\ref{eq:v11termapp}) is just the exchange energy per unit area
for a uniform spin-polarized two-dimensional electron gas
of areal density $n_{\alpha s}$.

\vspace{-0.1in}
\subsection{Interlayer exchange parameter}
\label{app:sinteg}
\vspace{-0.1in}

A key quantity in our discussion of the effects of interlayer exchange
in double-layer systems is the interlayer exchange parameter $\Gamma$,
defined by
\begin{eqnarray}
\label{eq:appgamma}
\Gamma &\equiv& \frac{1}{L_xL_y} \sum_{{\bf q}s}
\left[ \frac{V_{11}(q)-V_{12}(q)}{e^2d/2\epsilon} \right] \\ \nonumber
&& \qquad \qquad \times
\left[ \frac{I_{aas}(q)+I_{bbs}(q)-2I_{abs}(q)}{(n_a-n_b)^2} \right] ,
\end{eqnarray}
which is positive and a monotonically decreasing function
of the interlayer separation $d$:
\begin{equation}
\label{eq:gamineq}
0 < \Gamma(d>0) < \Gamma(d\rightarrow 0) ,
\end{equation}
when the subband densities are regarded as fixed.
We note that in the three-component phase ($p=3$),
unlike the ($p=1,2,4$) phases,
the equilibrium subband densities change with $d$,
so that according to Eq.~(\ref{eq:xdinf}),
\begin{equation}
\label{eq:gammap3}
\lim_{d\rightarrow\infty} \Gamma_{\rm eq}(p=3)
\propto \frac{1}{(n_a-n_b)^2d}
\propto \frac{d}{a_0} \rightarrow \infty ,
\end{equation}
when $p_F=p_B$,
in apparent disagreement with Eq.~(\ref{eq:gamineq}).
We stress that the inequality in Eq.~(\ref{eq:gamineq}) 
is true only when the subband densities $n_{\alpha s}$ are regarded
as fixed, which is not the case in Eq.~(\ref{eq:gammap3}).

The interlayer exchange parameter $\Gamma$ is important because it determines
when SILC is possible (for $\zeta<1-\Gamma_1$ and $p=1$) --
i.e., when $\sin\theta\ne 0$.
It also determines
the value of the pseudospin Stoner enhancement factor $I$
(which depends on $\Gamma_0$).
$\Gamma$ affects the state of the system (e.g., layer densities
and Eisenstein ratio $R_E$) whenever $\sin\theta\ne 0$.

\vspace{-0.1in}
\subsubsection{Inequality}
\label{app:sinequal}
\vspace{-0.1in}

Using the inequality
\begin{equation}
\label{eq:appinequal}
e^2d/2\epsilon>V_{11}(q)-V_{12}(q) ,
\end{equation}
which is true for $d>0$,
it follows from Eq.~(\ref{eq:appgamma}) that for $d>0$,
\begin{eqnarray}
\label{eq:appgineq}
\Gamma &<& \frac{1}{L_xL_y} \sum_{{\bf q}s}
\left[ \frac{I_{aas}(q)+I_{bbs}(q)-2I_{abs}(q)}{(n_a-n_b)^2} \right]
\\ \nonumber && \qquad =
\sum_s \frac{(n_{as}-n_{bs})^2}{(n_a-n_b)^2} = \Gamma(d\rightarrow 0) ,
\end{eqnarray}
where we have used the fact that
\begin{equation}
\lim_{d\rightarrow 0} \frac{[V_{11}(q)-V_{12}(q)]}{e^2d/2\epsilon} = 1
\end{equation}
and
\begin{eqnarray}
\label{eq:iabsum}
&& \frac{1}{L_xL_y} \sum_{{\bf q}s} \left[
I_{aas}(q)+I_{bbs}(q)-2I_{abs}(q)\right] \\ \nonumber
&& = \frac{1}{(L_xL_y)^2} \sum_{{\bf qK}s} \left[
\Theta(k_{as}-|{\bf K+q/2}|) - \Theta(k_{bs}-|{\bf K+q}/2|) \right]
\\ \nonumber
&& \qquad \qquad \times \left[
\Theta(k_{as}-|{\bf K-q/2}|) - \Theta(k_{bs}-|{\bf K-q}/2|) \right]
\\ \nonumber
&& = \sum_s \left\{ \frac{1}{L_xL_y} \sum_{\bf k} \left[
\Theta(k_{as}-k) - \Theta(k_{bs}-k) \right] \right\}^2
\\ \nonumber
&& = \sum_s (n_{as}-n_{bs})^2 .
\end{eqnarray}
Thus the condition
\begin{equation}
\label{eq:apppi2cond}
(n_{a\uparrow}-n_{b\uparrow})(n_{a\downarrow}-n_{b\downarrow}) \ge 0
\end{equation}
is sufficient to guarantee that $\Gamma<1$.

Equation~(\ref{eq:iabsum}) is true at finite temperature when
$I_{\alpha\beta s}(q)$ is generalized appropriately.
This is because
\begin{equation}
\label{eq:dsmallv}
\lim_{d\rightarrow 0} \left[ V_{11}(q) - V_{12}(q) \right] =
\frac{e^2 d}{2\epsilon} 
\end{equation}
is independent of the wave vector $q$.
We may write
\begin{eqnarray}
&& \Gamma = \frac{1}{(L_xL_y)^2}
\sum_{{\bf k}_1,{\bf k}_2,s}
\left[ \frac{V_{11}(|{\bf k}_2-{\bf k}_1|)-V_{12}(|{\bf k}_2-{\bf k}_1|)}
            {e^2d/2\epsilon}\right]
\\ \nonumber
&& \times
\left\{ \frac{\left(\langle a_{{\bf k}_2s}^\dagger a_{{\bf k}_2s} \rangle -
\langle b_{{\bf k}_2s}^\dagger b_{{\bf k}_2s} \rangle\right)
        \left(\langle a_{{\bf k}_1s}^\dagger a_{{\bf k}_1s} \rangle -
\langle b_{{\bf k}_1s}^\dagger b_{{\bf k}_1s} \rangle\right)}
            {(n_a-n_b)^2} \right\} ,
\end{eqnarray}
which generalizes $\Gamma$ to finite temperatures.
In the limit $d\rightarrow 0$, Eq.~(\ref{eq:dsmallv}) shows that
$\Gamma$ approaches
\begin{eqnarray}
&& \frac{1}{(n_a-n_b)^2} \sum_s
\left[ \frac{1}{L_xL_y} \sum_{\bf k}
\left(\langle a_{{\bf k}s}^\dagger a_{{\bf k}s} \rangle -
\langle b_{{\bf k}s}^\dagger b_{{\bf k}s} \rangle\right) \right]^2
\\ \nonumber
&& \qquad \quad = \frac{1}{(n_a-n_b)^2} \sum_s
\left(n_{as}-n_{bs}\right)^2
\\ \nonumber
&& \rightarrow \left\{
\begin{array}{ll}
1 ,   & p=1 \quad (n_{a\uparrow}=n_T) \\
1 ,   & p=2a \quad (n_{a\uparrow}=n_{b\uparrow}=n_T/2) \\
1 ,   & p=3 \quad (n_{a\uparrow}=n_{a\downarrow}=n_{b\uparrow}=n_T/3) \\
1/2 , & p=2b \quad (n_{a\uparrow}=n_{a\downarrow}=n_T/2) \\
1/2 , & p=4 \quad (n_{a\uparrow}=n_{a\downarrow}
                                =n_{b\uparrow}=n_{b\downarrow}=n_T/4)
\end{array} .
\right.
\end{eqnarray}

Empirically, we find that within the MFA, the spins in subbands $a$ and $b$
are either completely unpolarized (at higher densities) or fully polarized
(at sufficiently low densities.)
Therefore the only possible MFA configurations of spin and pseudospin
that would not satisfy the inequality in Eq.~(\ref{eq:apppi2cond}) and
that might therefore have $\Gamma>1$ would be three-component ($p=3$)
states in which subband $a$ (the majority subband) is spin-unpolarized
and subband $b$ (the minority subband) is spin-unpolarized:
\begin{equation}
n_{a\uparrow}=n_{a\downarrow}, n_a>n_{b\uparrow}>n_{b\downarrow}=0.
\end{equation}

\vspace{-0.1in}
\subsubsection{Pseudospin-unpolarized $\Gamma$}
\label{app:gamma0}
\vspace{-0.1in}

It is useful to define and evaluate
\begin{equation}
\Gamma_0 \equiv \lim_{n_a\rightarrow n_b} \Gamma .
\end{equation}
We begin our calculation of $\Gamma_0$ by noting that
\begin{eqnarray}
\label{eq:idiff}
&& I_{aas}(q) + I_{bbs}(q)-2I_{abs}(q) = \\ \nonumber
&& \frac{1}{L_xL_y} \sum_{\bf K} \left[
\Theta(k_{as}-|{\bf K+q}/2|) -
\Theta(k_{bs}-|{\bf K+q}/2|) \right] \\ \nonumber
&& \qquad \qquad \times \left[
\Theta(k_{as}-|{\bf K-q}/2|) -
\Theta(k_{bs}-|{\bf K-q}/2|) \right] \\ \nonumber
&& \approx
\frac{(k_F \Delta n/n_T)^2}{L_xL_y} \sum_{\bf K}
\delta(k_F-|{\bf K+q}/2|) \\ \nonumber
&& \qquad \qquad \qquad \qquad \quad \times \delta(k_F-|{\bf K-q}/2|)
\\ \nonumber
&& \qquad \qquad \qquad =
\left( \frac{k_F}{2\pi} \frac{\Delta n}{n_T} \right)^2
\frac{\Theta(1-x)}{x\sqrt{1-x^2}} ,
\end{eqnarray}
where $\Delta n \equiv (n_a-n_b) \rightarrow 0$,
$k_F=\sqrt{4\pi n_T/p}$ is the Fermi wave vector per layer
for the state with $p$ components ($p=2,4$),
and \mbox{$x\equiv q/2k_F$}.
Using Eq.~(\ref{eq:idiff}), we obtain
\begin{eqnarray}
\label{eq:gamma0}
\Gamma_0 &=& \frac{2}{p} \frac{2/\pi}{e^2d/2\epsilon} \int_0^1 dx
\frac{[V_{11}(2k_Fx)-V_{12}(2k_Fx)]}{\sqrt{1-x^2}} \\ \nonumber
&=& \frac{2}{p} \frac{2}{\pi z} \int_0^{\pi/2} d\theta
\frac{\left(1-e^{-z\sin\theta}\right)}{\sin\theta} ,
\end{eqnarray}
where $z=2k_Fd$, and the second line is obtained by the substitution
of variables $x=\sin\theta$.
It is straightforward to obtain $\Gamma_0(z)$ for small $z\rightarrow 0$
by expanding the last line of Eq.~(\ref{eq:gamma0}) in powers of $z$,
\begin{equation}
\lim_{z\rightarrow 0} \Gamma_0(z) = \frac{2}{p}
\left[ 1 - \frac{1}{\pi}z + \frac{1}{12}z^2 \right] ,
\end{equation}
up to second order in $z$.

Obtaining $\Gamma_0$ for $z\rightarrow\infty$ is more cumbersome.
One way to proceed is to define a cutoff $\epsilon$ that satisfies
\begin{equation}
\frac{1}{z} \ll \epsilon \ll 1
\end{equation}
so that $\sin\theta\approx\theta$ for $\theta<\epsilon$ and
$z\sin\theta \gg 1$ for $\theta>\epsilon$.
Then
\begin{eqnarray}
\label{eq:cutoff}
\Gamma_0 &\approx& \frac{2}{p} \frac{2}{\pi z} \left[
\int_0^\epsilon d\theta \frac{\left(1-e^{-z\theta}\right)}{\theta}
\right.
\\ \nonumber
&& \qquad \qquad \qquad \left. +
\int_\epsilon^{\pi/2} d\theta \frac{1}{\sin\theta} \right] .
\end{eqnarray}
The first integral in Eq.~(\ref{eq:cutoff}) may be carried out
using the identity\cite{gradshteyn,greix}
\begin{equation}
\label{eq:greix}
\int_0^R dt \frac{\left(1-e^{-t}\right)}{t} =
\ln(R) + \gamma + \int_R^\infty dt \frac{e^{-t}}{t} ,
\end{equation}
where $\gamma\approx 0.5772$ is Euler's constant, and
$R\equiv z\epsilon \rightarrow \infty$, so that
the last term of Eq.~(\ref{eq:greix}) can be dropped for large $R$.
The second integral in Eq.~(\ref{eq:cutoff}) is well known:\cite{grinvsin}
\begin{equation}
\int d\theta \frac{1}{\sin\theta} = \ln[\tan(\theta/2)] .
\end{equation} 
Now, $\tan(\epsilon/2)\approx\epsilon/2$ for $\epsilon \ll 1$,
so that the logarithmically divergent ($\sim\ln\epsilon$) parts
of the two integrals in Eq.~(\ref{eq:cutoff}) cancel each other, leaving
\begin{equation}
\lim_{z\rightarrow\infty} \Gamma_0 =
\frac{2}{p} \frac{2}{\pi z} \left[\ln(2z) + \gamma\right] .
\end{equation}

\vspace{-0.1in}
\subsubsection{Pseudospin-polarized $\Gamma$}
\label{app:onecomp}
\vspace{-0.1in}

We now compute
\begin{equation}
\Gamma_1 \equiv \lim_{n_a\rightarrow n_T} \Gamma .
\end{equation}
When the double-layer system is pseudospin polarized so that
$n_a=n_T$, then $I_{bbs}(q)=I_{abs}(q)=0$, and it follows from
Eqs. (\ref{eq:iaa}) and (\ref{eq:appgamma}) that
\begin{eqnarray}
\label{eq:gamma1}
\Gamma_1 &=& \frac{1}{p} \frac{16}{\pi z} \int_0^1 dx
\left(1-e^{-zx}\right) \left[\arccos(x)-x\sqrt{1-x^2}\right]
\\ \nonumber
&=& \frac{1}{p} \frac{32}{3\pi z} \left[1-S(z)\right] ,
\end{eqnarray}
where $z\equiv 2k_Fd$, and
\begin{eqnarray}
\label{eq:appsinteg}
S(z) &=& \frac{3}{2} \int_0^1 dx e^{-zx}
                     \left[\arccos(x) - x\sqrt{1-x^2}\right]
\\ \nonumber &=&
\frac{3\pi}{4z} \left\{ 1 - \frac{2}{z} \left[I_1(z)-L_1(z)\right] \right\}
\\ \nonumber &\rightarrow& \left\{
\begin{array}{ll}
1 - (3\pi/32)z + (1/15)z^2 - (\pi/256)z^3 , & z \rightarrow 0 \\
3\pi/4z , & z \rightarrow \infty
\end{array}
\right.
\end{eqnarray}
so that
\begin{eqnarray}
\label{eq:appgamlims}
\Gamma_1(z) &\rightarrow& \frac{1}{p} \left\{
\begin{array}{ll}
1 - (32/45\pi)z + (1/24)z^2 , & z \rightarrow 0 \\
(32/3\pi)/ z - 8/z^2 , & z \rightarrow \infty
\end{array}
\right.
\end{eqnarray}
Here $I_1$ and $L_1$ are modified Bessel and modified Struve functions
of the first kind, respectively:\cite{grinjn,grlnhn}
\begin{equation}
\label{eq:inln}
I_n(z) = i^{-n} J_n(iz) , \qquad
L_n(z) = i^{-(n+1)} H_n(iz) ,
\end{equation}
where $J_n$ is the ordinary Bessel function of order $n$, and
$H_n$ is the Hankel function of order $n$.

Obtaining Eq.~(\ref{eq:appsinteg}) is somewhat involved.
The first part of the integral in the first line of Eq.~(\ref{eq:appsinteg})
can be obtained by writing
\begin{equation}
\int_0^1 dx e^{-zx} \arccos(x)
= \int_0^1 dx e^{-zx} \left[\frac{\pi}{2} - \arcsin(x)\right] ,
\end{equation}
and using the identity\cite{grarcsin}
\begin{eqnarray}
\label{eq:correction}
&& \int_0^1 dx e^{-zx} \arcsin(x) \\ \nonumber
&& = \frac{\pi}{2z} \left[ I_0(z) - L_0(z) - e^{-z} \right] ,
\end{eqnarray}
which corrects a misprint in Eq.~(4.551.1) of Ref.~\onlinecite{gradshteyn}.
One then obtains
\begin{eqnarray}
\label{eq:ident1}
&& \int_0^1 dx e^{-zx} \arccos(x) \\ \nonumber
&& = \frac{\pi}{2z} \left[ 1 +  L_0(z) - I_0(z) \right] ,
\end{eqnarray}
The second part of the integral in the first line of Eq.~(\ref{eq:appsinteg})
can be obtained by writing
\begin{eqnarray}
&& \int_0^1 dx e^{-zx} x\sqrt{1-x^2}
= - \frac{\partial}{\partial z} \int_0^1 dx e^{-zx} \sqrt{1-x^2}
\\ \nonumber
&& = - \frac{\partial}{\partial z} \left[ 1 -
\int_0^1 dx \frac{ze^{-zx}}{\sqrt{1-x^2}} \right]
\\ \nonumber
&& = \frac{\pi}{2} \frac{\partial}{\partial z} \left[
\frac{L_1(z)-I_1(z)}{z} \right] ,
\end{eqnarray}
and by using the identities\cite{gridents}
\begin{eqnarray}
\frac{\partial}{\partial z} \left[ \frac{I_1(z)}{z} \right]
&=& \frac{I_2(z)}{z} \\ \nonumber
\frac{\partial}{\partial z} \left[ \frac{L_1(z)}{z} \right]
&=& \frac{L_2(z)}{z} + \frac{2}{3\pi}
\end{eqnarray}
to obtain
\begin{equation}
\label{eq:ident2}
\int_0^1 dx e^{-zx} x\sqrt{1-x^2}
= \frac{1}{3} + \frac{\pi}{2z} \left[ L_2(z)-I_2(z) \right]
\end{equation}
Combining Eqs. (\ref{eq:ident1}) and (\ref{eq:ident2})
and using the identities\cite{gr021}
\begin{eqnarray}
I_0(z) - I_2(z) &=& \frac{2I_1(z)}{z}
\\ \nonumber
L_0(z) - L_2(z) &=& \frac{2L_1(z)}{z} + \frac{2z}{3\pi}
\end{eqnarray}
gives the second line of Eq.~(\ref{eq:appsinteg}).
The third line of Eq.~(\ref{eq:appsinteg}) follows from power series
(for small $z$) and asymptotic (for large $z$) expansions of $I_1(z)$
and $L_1(z)$.\cite{grseries}

\begin{figure}[h]
\epsfxsize3.5in
\centerline{\epsffile{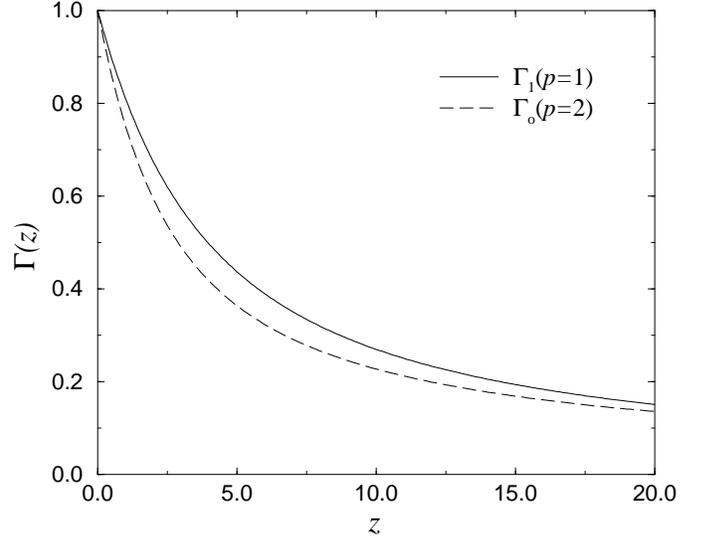}}
\caption{
$\Gamma_1(p=1)$ (solid curve) and $\Gamma_0(p=2)$ (dashed curve) as a
function of $z=2k_Fd$.
For $0<z<z_c\approx 44.09$, $\Gamma_1(p=1) > \Gamma_0(p=2)$.
}
\label{fig:gamma10}
\end{figure}

We note that for the same value of $p$,
$\Gamma_1(z) \le \Gamma_0(z)$ for all $z$.
This may be seen by the subsitution of variables
$x=\sin\theta$ in Eq.~(\ref{eq:gamma1}):
\begin{eqnarray}
\Gamma_1(z) &=&
\frac{2}{p} \frac{2}{\pi} \int_0^{\pi/2} d\theta
\frac{\left(1-e^{-z\sin\theta}\right)}{z\sin\theta} \\ \nonumber
&& \qquad \times \left\{
\sin(2\theta) \left[(\pi-2\theta)-\sin(2\theta)\right] \right\}
\le \Gamma_0(z) ,
\end{eqnarray}
where $\Gamma_0$ is expressed as an integral over $\theta$ in the last
line of Eq.~(\ref{eq:gamma0}).
However, for differing values of $p$,
$\Gamma_1(p=1) > \Gamma_0(p=2)$ for $0<z<z_c\approx 44.09$.
Both $\Gamma_1(p=1)$ and $\Gamma_0(p=2)$ are plotted in
Fig.~\ref{fig:gamma10}.
For $z>z_c$ (not shown), $\Gamma_1(p=1) < \Gamma_0(p=2)$.

\vspace{-0.1in}
\subsection{Stoner enhancement factor}
\label{app:stoner}
\vspace{-0.1in}

We now outline our calculation of the Stoner interaction parameter for
$t\rightarrow 0$.
Consider a two- or four-component state with
equal subband densities (pseudospin unpolarized), $n_a=n_b=n_T/2$.
For small $t\rightarrow 0$,
imagine moving a small amount of charge $\Delta n/2$
from subband $b$ to subband $a$,
\mbox{$n_a\rightarrow n_T/2 + \Delta n/2,
       n_b\rightarrow n_T/2 - \Delta n/2$},
so that $(n_a-n_b)=\Delta n$,
and calculate the change in the total energy [Eq.~(\ref{eq:etheta})]
as $\Delta n\rightarrow 0$.
The effect of the change of densities on the last term of
Eq.~(\ref{eq:etheta}) may be calculating by using
Eq.~(\ref{eq:gamma0}).
The change in the energy per unit area due to $\Delta n$ as
\mbox{$\Delta n \rightarrow 0$} is then given by
\begin{eqnarray}
\label{eq:delen}
\frac{\Delta {\cal{E}}_0}{L_xL_y} &=&
\frac{(\Delta n)^2}{p\nu_0} - t\Delta n
- \frac{1}{\sqrt{\pi p}} \frac{e^2}{4\pi\epsilon}
                          \frac{(\Delta n)^2}{\sqrt{n_T}} \\ \nonumber
&+& \frac{e^2d}{2\epsilon} \left(\frac{\Delta n}{2}\right)^2 \Gamma_0 .
\end{eqnarray}
Minimizing $\Delta {\cal{E}}_0/L_xL_y$ with respect to $\Delta n$
to solve for $\Delta n$ and using the definition Eq.~(\ref{eq:istoner})
of the Stoner enhancement $I$ gives
\begin{eqnarray}
\label{eq:appival}
I &=& \frac{\nu_0e^2}{2\pi\epsilon k_F} \left( 1 -
p \frac{\pi}{4} k_Fd \Gamma_0 \right)
\\ \nonumber
  &=& \frac{\nu_0}{\pi} \left\lbrace 2V_{11}(2k_F) -
\int_0^1 dx \frac{[V_{11}(2k_Fx)-V_{12}(2k_Fx)]}{\sqrt{1-x^2}} \right\rbrace ,
\end{eqnarray}
which is just the first line of Eq.~(\ref{eq:stonerfac}).
Equation (\ref{eq:appival}) is equal to the $t\rightarrow 0$ limit
of the Stoner interaction parameter calculated in the GRPA,
given in Eq.~(14) of Ref.~\onlinecite{swier}.

\vspace{-0.1in}
\subsection{Intersubband gap}
\label{app:isubgap}
\vspace{-0.1in}
The subband transfer energy $\Delta_{ab}$ for the
pseudospin-polarized ($n_a=n_T$) $p=1$ or $p=2b$ phase
is defined as the MFA energy required to move a particle from the
occupied $a$ subband to the (otherwise empty) $b$ subband:
\begin{equation}
\Delta_{ab} = \frac{\delta}{\delta n}
                    \left(\frac{{\cal{E}}_0}{L_xL_y}\right) ,
\end{equation}
where $\delta {\cal{E}}_0$ denotes the change of ${\cal{E}}_0$ under
\begin{eqnarray}
\label{eq:gaptrans}
\begin{array}{ll}
n_{a\uparrow} \rightarrow   (1/p)(n_T-\delta n) , &
n_{b\uparrow}  \rightarrow \delta n , \\
n_{a\downarrow} \rightarrow (1-1/p)(n_T-\delta n) , &
n_{b\downarrow} = 0 ,
\end{array}
\end{eqnarray}
where $\delta n \ll n_T$, and $n_{b\downarrow} = 0$ reflects the
fact that at low densities ($n_b=\delta n\rightarrow 0$), subband
$b$ will be spin-polarized.
In order to calculate $\Delta_{ab}$ we first compute
\begin{eqnarray}
\label{eq:abgap1}
&& \frac{\delta}{\delta n} \sum_s
\left[ I_{aas}(q) + I_{bbs}(q)  - 2I_{abs}(q) \right]
\\ \nonumber
&& = \frac{\delta}{\delta n} \frac{1}{L_xL_y} \sum_{\bf K}
\left[ \Theta(k_{as} - |{\bf K} + {\bf q}/2|) -
       \Theta(k_{bs} - |{\bf K} + {\bf q}/2|) \right]
\\ \nonumber && \qquad \qquad \qquad \times
\left[ \Theta(k_{as} - |{\bf K} - {\bf q}/2|) -
       \Theta(k_{bs} - |{\bf K} - {\bf q}/2|) \right]
\\ \nonumber
&& \approx \frac{\delta}{\delta n} \frac{2}{L_xL_y} \sum_{{\bf K}s}
\Theta(k_{as} - |{\bf K} + {\bf q}/2|) \left[
\delta k_{as} \delta(k_{as} - |{\bf K} - {\bf q}/2|)
\right. \\ \nonumber &&
\qquad \qquad \qquad \qquad \qquad \qquad \qquad \qquad \left. -
\Theta(\delta k_{bs} - |{\bf K} - {\bf q}/2|) \right]
\\ \nonumber
&& \approx \frac{\delta}{\delta n} \frac{1}{2\pi^2} \sum_s \left[
\delta k_{as} \int d^2K \Theta(k_{as} - |{\bf K} + {\bf q}|)
                     \delta(k_{as} - |{\bf K}|) \right. \\ \nonumber
&& \qquad \qquad \quad \quad \quad  - \left.
\Theta(\delta k_{bs} - |{\bf K}|) \right]
\\ \nonumber
&& = \frac{\delta}{\delta n} \frac{1}{2\pi^2} \sum_s \left[
2k_{as}\delta k_{as} \arccos(q/2k_{as}) \Theta(2k_{as}-q) \right.
\\ \nonumber
&& \qquad \qquad \quad \quad \quad  - \left.
\pi (\delta k_{bs})^2 \Theta(k_{as} - q) \right]
\\ \nonumber
&& = - \left[ \frac{2}{\pi} \Theta(1-x) \arccos(x) +
              2 \Theta(1/2 - x) \right] ,
\end{eqnarray}
where $k_{as} = \sqrt{4\pi n_{as}}$ so that
\begin{equation}
\delta k_{as} = -(2\pi/k_F)\delta n_{as} , \qquad
\delta k_{b\uparrow} = -\sqrt{4\pi \delta n} ,
\end{equation}
and $k_F=\sqrt{4\pi n_T/p}$.
In order to obtain the last line of Eq.~(\ref{eq:abgap1}), we
expanded the bracketed terms to first order with respect to $\delta n$,
and defined $x \equiv q/2k_F$.

The subband transfer gap is then given by
\begin{eqnarray}
\label{eq:appabgap2}
\Delta_{ab} &=& -\frac{2n_T}{p\nu_0} + 2t\sin\theta
+ \frac{4}{\sqrt{\pi}} \frac{e^2}{4\pi\epsilon} \sqrt{n_T/p}
\\ \nonumber && 
- \frac{e^2dn_T}{2\epsilon} \cos\theta \left(\cos\theta - \zeta\right)
\\ \nonumber &&
- \frac{\sin^2\!\theta}{4} \frac{e^2k_F}{2\pi\epsilon}
\left\lbrace \frac{\left[e^{-z/2}-(1-z/2)\right]}{z/2} \right.
\\ \nonumber && \qquad \qquad \left.
           + \frac{2}{\pi} \int_0^1 dx \left(1-e^{-zx}\right)
                                       \arccos(x) \right\rbrace ,
\end{eqnarray}
where $\zeta=(p_F-p_B)/n_T$ and $z=2k_Fd$.
Equation (\ref{eq:appabgap2}) is expressed in dimensionless form
in Eq.~(\ref{eq:abgap2}).

\end{document}